\documentclass[11pt,a4paper]{article}

\usepackage[margin=2.2cm]{geometry}
\usepackage[utf8]{inputenc}
\usepackage[T1]{fontenc}
\usepackage{lmodern}
\usepackage{amsmath,amssymb}
\usepackage{booktabs}
\usepackage{tabularx}
\usepackage{multirow}
\usepackage{makecell}
\usepackage{float}
\usepackage{graphicx}
\usepackage[dvipsnames,table]{xcolor}
\usepackage{tikz}
\usetikzlibrary{shapes.geometric, arrows.meta, positioning, fit, calc, backgrounds, decorations.pathreplacing, patterns}
\usepackage{listings}
\usepackage{enumitem}
\usepackage[colorlinks=true,linkcolor=MidnightBlue,citecolor=OliveGreen,urlcolor=BrickRed]{hyperref}
\usepackage[numbers,sort&compress]{natbib}
\usepackage{subcaption}
\usepackage{authblk}
\usepackage{xspace}
\usepackage{pifont}
\usepackage{longtable}

\definecolor{openhands}{HTML}{4A90D9}
\definecolor{aider}{HTML}{2ECC71}
\definecolor{claudecode}{HTML}{D4A574}
\definecolor{codex}{HTML}{E74C3C}
\definecolor{miniswe}{HTML}{9B59B6}
\definecolor{openclaw}{HTML}{F39C12}
\definecolor{vibe}{HTML}{FF7000}
\definecolor{gemini}{HTML}{1A73E8}
\definecolor{hermes}{HTML}{16A085}
\definecolor{pi}{HTML}{34495E}
\definecolor{opencode}{HTML}{C0392B}
\definecolor{omnigent}{HTML}{8E44AD}
\definecolor{lightgray}{HTML}{F5F5F5}
\definecolor{codebg}{HTML}{F8F8F8}

\lstdefinestyle{pseudocode}{
  basicstyle=\ttfamily\scriptsize,
  backgroundcolor=\color{codebg},
  frame=single,
  rulecolor=\color{gray!30},
  breaklines=true,
  captionpos=b,
  tabsize=2,
  showstringspaces=false,
  numbers=left,
  numberstyle=\tiny\color{gray},
  numbersep=5pt,
  xleftmargin=12pt,
  framexleftmargin=10pt,
  keywords={function, if, else, for, while, return, await, spawn, emit, yield, break, continue, match, let, async, fn, loop, new, true, false, null, try, catch, raise, class, def, self, import, from, in, not, and, or},
  keywordstyle=\bfseries\color{MidnightBlue},
  commentstyle=\itshape\color{gray},
  stringstyle=\color{OliveGreen},
  morecomment=[l]{\#},
  morecomment=[l]{//},
}
\newcolumntype{L}[1]{>{\raggedright\arraybackslash}p{#1}}
\newcolumntype{C}[1]{>{\centering\arraybackslash}p{#1}}
\newcolumntype{R}[1]{>{\raggedleft\arraybackslash}p{#1}}

\newcounter{observationctr}
\newcommand{\observation}[1]{\vspace{0.4em}\noindent\refstepcounter{observationctr}\fbox{\parbox{0.97\linewidth}{\textbf{Observation~\theobservationctr.} #1}}\vspace{0.4em}}
\newcommand{\recommendation}[2]{\vspace{0.4em}\noindent\fcolorbox{OliveGreen!60}{OliveGreen!5}{\parbox{0.965\linewidth}{\textbf{Recommendation~#1.} #2}}\vspace{0.4em}}
\newcommand{\clcode}{\textsc{Claude~Code}\xspace}
\newcommand{\cdx}{\textsc{Codex}\xspace}
\newcommand{\oh}{\textsc{OpenHands}\xspace}
\newcommand{\adr}{\textsc{Aider}\xspace}
\newcommand{\mswe}{\textsc{Mini-SWE-Agent}\xspace}
\newcommand{\ocl}{\textsc{OpenClaw}\xspace}
\newcommand{\vibe}{\textsc{Mistral~Vibe}\xspace}
\newcommand{\gemini}{\textsc{Gemini~CLI}\xspace}
\newcommand{\hrm}{\textsc{Hermes}\xspace}
\newcommand{\pib}{\textsc{Pi}\xspace}
\newcommand{\ocd}{\textsc{OpenCode}\xspace}
\newcommand{\omni}{\textsc{Omnigent}\xspace}
\newcommand{\cmark}{\ding{51}}
\newcommand{\xmark}{\ding{55}}

\title{\textbf{Harness Engineering:\\Anatomy, Architecture, and Evolution of Coding Agents}\\[0.35em]\large A Source-Code Study of Eleven Systems}

\author[1,2]{Paul Barbaste\thanks{Lead and corresponding author.}}
\author[2]{Tristan Darrigol}
\author[2]{Germain Vu}
\author[2]{Tom Wiltberger}
\affil[1]{Inclusive Brains}
\affil[2]{Wavestone AI Lab}
\date{July 2026}

\begin{document}
\maketitle

\begin{abstract}
An agent is a model plus a harness: the runtime that couples an LLM to the world through a loop, tools, context management, safety controls, orchestration, and extension surfaces.
\emph{Harness engineering}, named as a discipline in early 2026, is the design and evolution of that runtime.
This paper gives the young discipline its most comprehensive empirical foundation to date.
It is a source-code anatomy of eleven production coding harnesses: \textbf{Claude~Code} (Anthropic), \textbf{Codex~CLI} (OpenAI), \textbf{Gemini~CLI} (Google), \textbf{Mistral~Vibe} (Mistral), \textbf{OpenHands}, \textbf{Aider}, \textbf{Mini-SWE-Agent}, \textbf{Hermes} (Nous Research), \textbf{Pi}, \textbf{OpenCode}, and \textbf{OpenClaw}, plus \textbf{Omnigent} (Databricks), the first meta-harness we are aware of, analyzed as a contrast point.
The paper defines what a harness is, maps its seven canonical subsystems with the minimal and maximal implementation of each, and dissects all eleven systems along those subsystems.
It does not benchmark or rank; it describes and compares how the systems are built.

The audit yields 13 cross-cutting observations and a catalog of 29 recurring design patterns.
Two absences survive a threefold corpus expansion: across roughly four million lines of Python, TypeScript, and Rust, no agent runtime imports a general-purpose agentic framework (LangChain, LangGraph, AutoGen, or a dozen others; Gemini~CLI uses neither of Google's own), and none retrieves code with vector embeddings; the field runs on hand-rolled async loops and deterministic retrieval (\texttt{ripgrep}, tree-sitter, glob, auto-discovered Markdown context files).
The extensibility standards resolved: SKILL.md skills lead MCP in adoption (9/11 vs.\ 8/11), with registries, trust tiers, and the corpus's first agent-authored skills; ACP ships in six systems and acquired a third role, \emph{harness hosting}, with OpenHands running Claude~Code, Codex, or Gemini~CLI as interchangeable backends.

Because the original eight systems were re-pinned rather than replaced, the study also contains a controlled longitudinal sample: the same harnesses, source-diffed across one quarter.
The diff shows convergence becoming \emph{imitation} (Codex adopts Claude~Code's hook vocabulary verbatim and ships an importer for its sessions and settings; OpenHands reads Claude~Code's plugin format), while behavioral policy migrates from prompt prose to configuration, and three of the April edition's observations required substantive revision in place.

These threads converge on the paper's thesis: in the first half of 2026 the coding harness completed a turn from tool to \emph{platform}.
Harnesses became importable SDKs while framework vendors shipped harnesses; marketplaces, switching-cost tooling, and enterprise governance layers appeared; the agent became addressable as a model behind an OpenAI-compatible endpoint; and a meta-harness now orchestrates eleven vendor harnesses (half this corpus among them) behind one API, re-implementing the expensive parts and arbitraging the proprietary ones.
The paper closes with 18 design recommendations pinned to the observed code, and a 90-line minimum-viable-harness scaffold that implements ten of them.
\end{abstract}

\vspace{0.3em}

\section{Introduction}
\label{sec:intro}

An agent is a model plus a harness.
The model supplies the intelligence; the harness turns that intelligence into work through a loop, tools, context management, safety controls, orchestration, and extension surfaces.
In the span of five months, \emph{harness engineering} went from a phrase coined in a vendor blog post to a discipline with practitioner guides, formal definitions, automated-evolution systems, and its own arXiv genealogy (Section~\ref{sec:genealogy}).
What the discipline does not yet have is a reference: an account, grounded in production source code, of what a harness actually is, what it is made of, and how the field's leading implementations differ.
This paper is that reference, and it arrives with a thesis: the harness has stopped being a tool and become a \emph{platform}.

The existing literature reaches neither goal.
Benchmark surveys report scores (typically SWE-Bench~\cite{jimenez2024swebench}) without touching architecture; conceptual taxonomies~\cite{weng2023llmagents,wang2024survey} describe patterns without implementations; the one concurrent source-level taxonomy~\cite{rombaut2026} excludes the provider-native systems that define the field's frontier.
None reaches the engineering decisions that determine whether a harness is reliable, extensible, or deployable: how the loop is structured, how tools are defined and sandboxed, how safety is enforced, how context is rationed.
Nor does any of it ask where the category itself is heading.

We study eleven systems at source level, pinned to July~2026 releases.
Four are the flagship products of the major commercial LLM providers: \textbf{Claude~Code} (Anthropic, TypeScript, via a circulated source snapshot), \textbf{Codex~CLI} (OpenAI, Rust), \textbf{Gemini~CLI} (Google, TypeScript), and \textbf{Mistral~Vibe} (Mistral AI, Python).
Seven are open source, spanning the field's full spread: \oh{} (event-sourced SDK, since mid-2026 also a host for rival harnesses), \adr{} (thirteen polymorphic edit formats), \mswe{} (the 100-line research floor), \hrm{} (the fastest-growing harness on GitHub, with a self-improving skill loop), \pib{} (the minimal-core countertrend), \ocd{} (the most-starred dedicated coding agent, with the corpus's most thoroughgoing client/server architecture), and \ocl{} (a multi-channel gateway, our non-SWE contrast point).
A twelfth system, \omni{} (Databricks), is the first \emph{meta}-harness, an orchestration layer over the others, and is analyzed as a second contrast point.
Because eight of these systems were already audited in this study's April~2026 edition, their retained snapshots yield a controlled longitudinal sample: the same harnesses, source-diffed across one quarter.

The paper makes seven contributions.
The first is the discipline's reference anatomy (Section~\ref{sec:whatisaharness}): a definition of the harness with its genealogy and boundary cases, and a component map of the seven canonical subsystems, each with its minimal and maximal observed implementation, that organizes the rest of the paper.
On that map the paper builds a systematic source-level comparison of eleven harnesses, the first corpus to cover all four provider-native systems, the two most-adopted open-source harnesses, and each system's \emph{signature} capability (Table~\ref{tab:overview}).
The audit yields 13 cross-cutting observations and a catalog of 29 recurring patterns, several documented here for the first time: agent-maintained memory pipelines, verify-on-stop guards, lineage compaction, session-tree version control, log-as-queue loops, syntax-aware command permissioning, cache-dialect fanout, and harness mimicry.

Because the original eight systems were re-pinned rather than replaced, the study also carries a longitudinal analysis (Section~\ref{sec:evolution_window}): ninety days of harness evolution measured in source, in which convergence becomes imitation (hook vocabularies copied verbatim, plugin formats adopted, session importers shipped), patterns diffuse down the corpus, and behavioral policy migrates from prompt prose to configuration.
Two empirical absences are established at eleven-system scale and survive a threefold corpus expansion: no agent runtime uses a general-purpose agentic framework, and none uses embedding-based retrieval over code; their historical resolution lies in the harness--framework merger.
That merger opens onto the platform-turn thesis (Section~\ref{sec:platformturn}), documented in named artifacts: harness SDKs and framework-built harnesses converging on one shape, plugin and skill marketplaces with supply-chain security, cross-vendor session importers, mobile-device-management (MDM) governance, the agent-as-a-model gateway, and a source-level analysis of the first meta-harness.
The analysis condenses, finally, into a practitioner's guide of 18 design recommendations, each anchored in a cited observation and a system that implements it, plus a 90-line minimum-viable-harness scaffold that realizes ten of them directly and is compatible with the rest.

The rest of the paper proceeds as follows.
Section~\ref{sec:whatisaharness} defines the harness and its seven subsystems.
Section~\ref{sec:background} surveys related work and the 2026 landscape.
Section~\ref{sec:method} sets out the methodology; Section~\ref{sec:overview} gives the system overview.
Sections~\ref{sec:agentloop}--\ref{sec:extensibility} dissect the seven subsystems across the corpus.
Section~\ref{sec:synthesis} ties the cross-cutting observations together.
Section~\ref{sec:platformturn} develops the platform-turn thesis, including the meta-harness and the longitudinal analysis.
Section~\ref{sec:discussion} discusses implications and limits.
Section~\ref{sec:recommendations} distills the analysis into 18 design recommendations and the 90-line scaffold.
Section~\ref{sec:conclusion} closes.

\section{What Is a Harness?}
\label{sec:whatisaharness}

Before comparing implementations, we fix the object of study.
This section defines the harness, traces the short history of the term, separates it from the concepts it is routinely confused with, and lays out the seven subsystems that every harness in our corpus implements in some form.
A reader new to the field should leave this section knowing exactly what a harness is and what it is made of; the rest of the paper substantiates each cell of the map against production source code.

\subsection{Definition and Genealogy}
\label{sec:genealogy}

\begin{quote}
\emph{An agent is a model plus a harness. The harness is everything except the model: the runtime that couples an LLM to the world---its loop, its tools, its context, its safety controls, its orchestration, and its extension surfaces. Harness engineering is the discipline of designing and evolving that runtime.}
\end{quote}

The one-line algebra \textbf{Agent = Model + Harness} was popularized by Trivedy's LangChain engineering series in early 2026~\cite{trivedy2026anatomy}, and the term \emph{harness engineering} entered circulation in February~2026: first used in passing by Mitchell Hashimoto, then defined and elaborated by Trivedy in the context of LangChain's Deep Agents work~\cite{deepfeed2026}.\footnote{There is a pleasing irony in the provenance: the term was named and defined from within LangChain, the framework vendor whose libraries are absent from every harness runtime in this corpus (Observation~\ref{obs:absences}). LangChain's response to that absence was not to lobby for adoption but to ship a harness of its own---Deep Agents~\cite{deepagents2026}, built on LangGraph---which independently converges on the very conventions this corpus documents: \texttt{SKILL.md} skills with progressive disclosure, \texttt{AGENTS.md} memory, sub-agent spawning, and a todo-planning tool. Section~\ref{sec:merger} returns to this two-way traffic between frameworks and harnesses.}
The discipline acquired its practitioner literature within weeks: B{\"o}ckeler's guide to harness engineering \emph{for coding-agent users}~\cite{bockeler2026} (April), O'Reilly and independent syntheses (June), and community-curated pattern catalogs.
An academic wave followed the same arc: Lin et~al.'s \emph{Agentic Harness Engineering}~\cite{lin2026ahe} automates harness evolution; HARBOR~\cite{harbor2026} optimizes harnesses against benchmarks; Wang et~al.\ recast code itself as the harness substrate~\cite{codeharness2026}; Rombaut's source-code taxonomy of thirteen open-source scaffolds~\cite{rombaut2026} is the closest methodological relative of the present study (Section~\ref{sec:related_taxonomies} maps his twelve dimensions onto our seven subsystems); and Macedo~\cite{macedo2026} gives the concept its first operational definition, deriving necessary and sufficient conditions for what counts as a harness.
We adopt Macedo's definitional core---\emph{the layer that wraps a language model and turns it into an agent able to act on a repository}---and extend it with the subsystem decomposition of Section~\ref{sec:subsystems}, which the definitional literature gestures at but has not yet grounded in production source.

\subsection{What a Harness Is Not}
\label{sec:terminology}

Usage of the term is loose, and four boundary cases account for most of the confusion~\cite{macedo2026}.

\begin{itemize}[itemsep=2pt]
  \item \textbf{A harness is not a scaffold---quite.} The two words are near-synonyms in practice, and this paper inherits ``scaffold'' from its own earlier vocabulary. Where a distinction is useful, \emph{scaffold} names the structural code (the loop, the registries) and \emph{harness} names the shipped runtime artifact that embeds it: \mswe{}'s scaffold is 100 lines of Python; \clcode{}'s harness is a product with a terminal UI, a permission system, and a plugin ecosystem.
  \item \textbf{A harness is not an agentic framework.} A framework (LangChain, AutoGen, CrewAI) is a library the developer imports to \emph{build} an agent; a harness is a runtime the developer works \emph{inside of}. The distinction was clean in 2025 and is dissolving in 2026---from both directions---which is the subject of Section~\ref{sec:merger}.
  \item \textbf{A harness is not an evaluation harness.} The SWE-bench ``harness'' wraps an \emph{agent} to run it against tasks; an agent harness wraps a \emph{model} to make it act. Same word, opposite direction of wrapping.
  \item \textbf{A harness is not an orchestrator.} An orchestrator (or meta-harness, Section~\ref{sec:omnigent}) coordinates one or more harnesses from above and implements no editing loop of its own. \omni{} orchestrates eleven vendor harnesses---five of them systems in this paper---and implements no editing loop of its own; it is evidence about harnesses, not one of them.
\end{itemize}

\subsection{The Seven Subsystems}
\label{sec:subsystems}

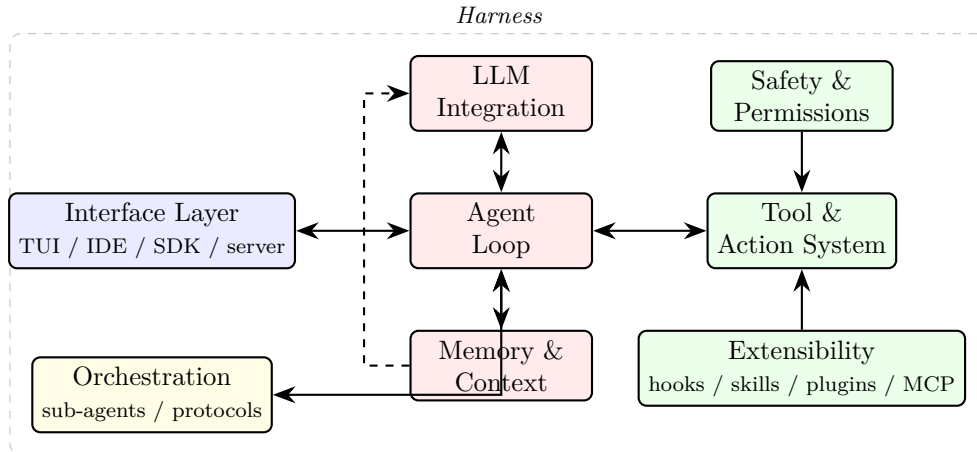
\begin{figure}[H]
\centering
\begin{tikzpicture}[
  node distance=0.8cm and 1.2cm,
  box/.style={rectangle, draw, rounded corners=3pt, minimum height=0.9cm, minimum width=2.4cm, font=\small, align=center, thick},
  io/.style={box, fill=blue!8},
  core/.style={box, fill=red!8},
  ext/.style={box, fill=green!8},
  arr/.style={-{Stealth[length=3mm]}, thick},
  darr/.style={{Stealth[length=3mm]}-{Stealth[length=3mm]}, thick},
]
  \node[io] (user) {Interface Layer\\\scriptsize TUI / IDE / SDK / server};
  \node[core, right=1.5cm of user] (loop) {Agent\\Loop};
  \node[core, above=of loop] (llm) {LLM\\Integration};
  \node[core, below=of loop] (mem) {Memory \&\\Context};
  \node[ext, right=1.5cm of loop] (tools) {Tool \&\\Action System};
  \node[ext, above=of tools] (safety) {Safety \&\\Permissions};
  \node[ext, below=of tools] (config) {Extensibility\\\scriptsize hooks / skills / plugins / MCP};
  \node[box, fill=yellow!12, below=1.15cm of user, minimum width=2.4cm] (orch) {Orchestration\\\scriptsize sub-agents / protocols};

  \draw[darr] (user) -- (loop);
  \draw[darr] (loop) -- (llm);
  \draw[darr] (loop) -- (mem);
  \draw[darr] (loop) -- (tools);
  \draw[arr] (safety) -- (tools);
  \draw[arr] (config) -- (tools);
  \draw[arr, dashed] (mem.west) -- ++(-0.6,0) |- (llm.west);
  \draw[darr] (orch) -| (loop);

  \begin{scope}[on background layer]
    \node[draw=gray!50, dashed, rounded corners, fit=(loop)(llm)(mem)(tools)(safety)(config)(orch), inner sep=8pt, label={[font=\footnotesize\itshape]above:Harness}] {};
  \end{scope}
\end{tikzpicture}
\caption{The canonical anatomy of a coding-agent harness: seven subsystems around an agent loop, plus the interface layer through which humans and programs drive it. All eleven systems in the corpus implement these subsystems with varying complexity; Table~\ref{tab:componentmap} gives the observed range per subsystem.}
\label{fig:architecture_overview}
\end{figure}

Every system in our corpus, from the 100-line research baseline to the million-line production CLI, must take a position on the same seven subsystems (Figure~\ref{fig:architecture_overview})---even when that position is deliberate absence.
They are the analysis dimensions D1--D7 of Section~\ref{sec:method}, but they are also, we argue, the canonical anatomy of the artifact itself: the minimal implementations show how small each subsystem can shrink---in one case (orchestration) to a deliberate absence---and the maximal implementations show where a decade of engineering headroom lies in each.
Table~\ref{tab:componentmap} maps the anatomy: what each subsystem does, its minimal and maximal observed forms, and where this paper dissects it.

\begin{table}[H]
\centering
\caption{The component map of a coding-agent harness: the seven subsystems, with the minimal and maximal implementations observed in the corpus.}
\label{tab:componentmap}
\small
\renewcommand{\arraystretch}{1.25}
\begin{tabularx}{\textwidth}{@{}L{2.5cm} L{3.6cm} L{3.2cm} L{4.4cm} c@{}}
\toprule
\textbf{Subsystem} & \textbf{Role} & \textbf{Minimal form} & \textbf{Maximal form} & \textbf{\S} \\
\midrule
Agent loop & Alternates inference with action execution; owns stop conditions and failure recovery & \mswe{}: a linear \texttt{while} over one bash tool & \oh{}: event-sourced conversation over a persistent event log, with parallel action batches & \ref{sec:agentloop} \\
LLM integration & Speaks provider protocols; assembles the prompt; manages caching, thinking, routing & \mswe{}: one LiteLLM call, one Jinja template & \hrm{}: five owned transports, 29 provider profiles; \cdx{}: server-delivered model catalog & \ref{sec:llm} \\
Tools \& actions & Defines and executes what the agent can do, file editing above all & \mswe{}: bash only & \clcode{}: 43 typed tools with deferred loading; \cdx{}: tool calls as V8-executed code & \ref{sec:tools} \\
Memory \& context & Rations the context window; persists knowledge across turns and sessions & \mswe{}: unbounded linear history & \cdx{}: agent-maintained cross-session memory pipeline; \gemini{}: graph-based context distillation & \ref{sec:memory} \\
Safety \& permissions & Decides what runs, what asks, what is forbidden; isolates execution & \mswe{}: cost and step limits & \cdx{}: policy rules + LLM approval reviewer + three-platform OS sandbox & \ref{sec:safety} \\
Orchestration & Spawns and coordinates sub-agents; connects to other agents & \adr{}: none (single-agent by design) & \clcode{}: recursive composition; \omni{}: cross-vendor coordination (meta-layer) & \ref{sec:multiagent} \\
Extensibility & Lets users and ecosystems add capability: config, hooks, skills, plugins, MCP & \mswe{}: structural typing (Python protocols) & \pib{}: everything-is-an-extension runtime; \cdx{}: marketplace-distributed plugins & \ref{sec:extensibility} \\
\bottomrule
\end{tabularx}
\end{table}

Two cross-cutting surfaces sit alongside the seven: the \emph{interface layer} (TUI, CLI flags, IDE protocol, HTTP server, SDK) through which humans and programs drive the harness, and the \emph{session substrate} (transcripts, persistence, resume/fork) that several subsystems share.
Both recur throughout the corpus analysis, and the interface layer in particular carries the platformization argument of Section~\ref{sec:platformturn}.

\subsection{The Minimal Harness}
\label{sec:minimalharness}

The decomposition above says what a harness is made of; it does not say how much of each part is needed.
The corpus contains an existence proof at the floor: \mswe{} implements all seven subsystems in roughly 100 lines---a \texttt{while} loop, one template, one tool, a message list, two limits, no orchestration, and structural typing as its entire extension story---and reports results on SWE-bench Verified in the same range as systems three orders of magnitude larger.\footnote{Self-reported figures, different models and dates; see the methodology caveats in Section~\ref{sec:discussion}. The point survives the caveats: the floor is low.}
Section~\ref{sec:recommendations} closes the loop with a 90-line minimum-viable harness that a practitioner can copy and specialize.
What separates the floor from the production systems is not task completion but everything else this paper documents: safety, recovery, cost management, extensibility, and---increasingly---the platform surfaces of Section~\ref{sec:platformturn}.

\section{Background and Related Work}
\label{sec:background}

\subsection{LLM-Powered Code Agents}

The concept of using LLMs as autonomous coding agents gained traction with SWE-Bench~\cite{jimenez2024swebench}, a benchmark of real-world GitHub issues requiring end-to-end resolution.
SWE-Agent~\cite{yang2024sweagent} introduced the \emph{Agent-Computer Interface} (ACI), providing LLMs with specialized file viewing and editing tools.
The CodeAct paradigm~\cite{wang2024codeact} proposed unifying agent actions into executable code, arguing that this subsumes discrete tool calling.
ReAct~\cite{yao2023react} established the reasoning-action loop that underpins most current agent architectures.

A second wave of SWE-specific agents has appeared since.
AutoCodeRover~\cite{zhang2024autocoderover} replaces text-based search with AST-aware retrieval and spectrum-based fault localization.
MASAI~\cite{arora2024masai} decomposes the issue-resolution pipeline into modular sub-agents and reports cost-aware results on SWE-Bench Lite.
MAGIS~\cite{tao2024magis} and CodeR~\cite{chen2024coder} both formalize role-based multi-agent coordination for issue resolution (Manager/Custodian/Developer/QA in MAGIS; explicit task graphs in CodeR).
Lingma SWE-GPT~\cite{ma2024lingmaswegpt} trains a base model jointly with the SE process loop, an instance of the model--agent co-design we examine in Section~\ref{sec:llm}.
SWE-Search~\cite{antoniades2025sweseach} layers Monte Carlo tree search over agent trajectories (continuing the line of search-based reasoning developed in Tree of Thoughts~\cite{yao2023tot} and LATS~\cite{zhou2024lats}, with Plan-and-Solve~\cite{wang2023planandsolve} as the simpler precursor), and Diversity-Empowers-Intelligence~\cite{zhang2024dei} ensembles heterogeneous agents under a meta-policy.
At the other end of the design spectrum, Agentless~\cite{xia2024agentless} reports that a fixed localize/repair/validate pipeline is competitive with agent loops on SWE-Bench Lite, a useful counterweight when discussing whether agentic complexity is necessary.
The OpenHands platform paper~\cite{wang2025openhands} provides peer-reviewed grounding for the system whose source we audit in Section~\ref{sec:agentloop}.
Concurrent with this study, Lin et~al.~\cite{lin2026ahe} propose \emph{Agentic Harness Engineering} (AHE), which automatically evolves agent harnesses (tools, middleware, memory, sub-agent configuration) using observability-driven feedback.
Starting from a bash-only seed comparable to \mswe{}, their system reaches 71.9\% on SWE-Bench Verified, and the evolved harness transfers across model families.
Their component-level ablation provides quantitative evidence for the architectural emphases we arrive at independently through source-code analysis: tools, middleware, and long-term memory carry the improvement, while the system prompt alone does not.

\subsection{Agent Architecture Surveys}

Prior work on agent architectures~\cite{weng2023llmagents,wang2024survey,xi2023agents} has identified common patterns including tool use, memory management, and planning.
The multi-agent dimension is covered in detail by Guo et~al.~\cite{guo2024multiagent}; the SWE-specific subset is mapped by Liu et~al.~\cite{liu2025seagents}; AgentBench~\cite{liu2024agentbench} provides the standard cross-domain evaluation harness.
These surveys operate at a conceptual level, describing what components exist rather than how they are implemented.
The present work complements them by grounding architectural analysis in concrete implementations, showing that the ``same'' pattern (e.g., tool use) admits considerable variation in practice.

\subsection{Concurrent Source-Code Taxonomies}
\label{sec:related_taxonomies}

The closest methodological relative of this study is Rombaut's \emph{Inside the Scaffold}~\cite{rombaut2026}, developed concurrently with our April edition: a source-code taxonomy of thirteen open-source coding-agent scaffolds at pinned commits, organized as twelve dimensions in three layers (control architecture; tool/environment interface; resource management), with the useful finding that most agents \emph{compose} loop primitives (ReAct, generate-test-repair, plan-execute, retry, tree search) rather than implementing a single one.
The two studies are complementary rather than redundant.
Rombaut's corpus is open-source-only, skews toward benchmark-oriented pipelines (AutoCodeRover, Agentless, Moatless Tools, DARS, Prometheus, SWE-agent), and explicitly excludes \clcode{} for lack of published source.
Ours covers all four provider-native harnesses at source level (including the circulated \clcode{} snapshot), the gateway and meta-harness contrast points, and the newest mass-adoption systems (\hrm{}, \pib{}), and it extends the analysis to dimensions his taxonomy does not reach: permission and safety architecture beyond execution isolation, multi-agent orchestration, extensibility surfaces (hooks, skills, plugins, MCP), prompt content, inter-agent protocols, and the longitudinal and platform analyses of Sections~\ref{sec:evolution_window} and~\ref{sec:platformturn}.
His twelve dimensions map onto our seven subsystems as follows: control-loop strategies, loop driver, and control-flow implementation refine our D1; tool-set design, edit/patch format, and tool-discovery strategy refine D3; context-retrieval paradigm, state management, context compaction, and persistent memory refine D4; execution isolation is the isolation half of our D5; multi-model routing is one cell of our D2.
Where his vocabulary is sharper than ours (loop-primitive composition), we adopt it.

\subsection{Multi-Agent Systems for Software Engineering}

Recent work has explored multi-agent collaboration for software engineering.
MetaGPT~\cite{hong2024metagpt} assigns specialized roles (product manager, architect, engineer) to different agents.
ChatDev~\cite{qian2024chatdev}, developed independently, organizes a role-playing development pipeline around chat-chain communication.
AutoGen~\cite{wu2023autogen} provides a framework for multi-agent conversations.
Magentic-One~\cite{fourney2024magneticone} implements an orchestrator pattern with specialized sub-agents.
For coding specifically, MAGIS~\cite{tao2024magis} and CodeR~\cite{chen2024coder} pair role decomposition with task graphs, and Chain-of-Agents~\cite{zhang2024chainofagents} sequences sub-agents over long inputs.
Several systems in the present study incorporate multi-agent patterns, enabling direct comparison of these approaches within a single domain.

\subsection{Tool Use in Language Models}

Toolformer~\cite{schick2023toolformer} demonstrated that LLMs can learn to use external tools.
Gorilla~\cite{patil2023gorilla} showed that LLMs can generate accurate API calls when provided with documentation.
ToolLLM~\cite{qin2024toolllm} scaled tool-use evaluation to 16,000+ real-world APIs, and AnyTool~\cite{du2024anytool} demonstrated hierarchical tool routing for large API surfaces.
The Model Context Protocol (MCP)~\cite{mcp2024} proposes a standard interface for connecting LLMs to external tools and data sources.
The analysis here finds that MCP has emerged as a cross-system extensibility standard, adopted by eight of the eleven systems studied (all four provider-native agents plus \oh{}, \hrm{}, \ocd{}, and \ocl{}).

\subsection{Vendor-Authored Engineering Guidance}
\label{sec:bg_anthropic_series}

A complementary thread to the academic surveys is engineering guidance published directly by major LLM vendors.
Between December~2024 and September~2025, Anthropic released a series of four articles that collectively articulate a coherent design philosophy for agents:
(i)~\emph{Building Effective Agents}~\cite{schluntz2024agents}, which contrasts ``workflows'' (predefined code paths) with ``agents'' (LLM-directed dynamic processes), enumerates five composable patterns (prompt chaining, routing, parallelization, orchestrator-workers, evaluator-optimizer), and warns explicitly against general-purpose agentic frameworks;
(ii)~\emph{Effective Context Engineering for AI Agents}~\cite{rajasekaran2025context}, which reframes ``prompt engineering'' as ``context engineering,'' documents \emph{context rot} in long windows, and recommends a hybrid approach that favors just-in-time (JIT) retrieval via \texttt{grep}/\texttt{tail}/file-system while acknowledging that pre-indexed retrieval can complement JIT methods for specific use cases;
(iii)~\emph{Writing Effective Tools for AI Agents}~\cite{aizawa2025tools}, which builds on the agent-computer-interface (ACI) concept introduced in \emph{Building Effective Agents}~\cite{schluntz2024agents}, recommends consolidating tools into ``few thoughtful'' high-signal operations, and reports SWE-Bench gains from prompt-engineered tool descriptions;
(iv)~\emph{How We Built Our Multi-Agent Research System}~\cite{hadfield2025multiagent}, a case study reporting that orchestrator-worker architectures with parallel sub-agents outperformed single-agent baselines by 90.2\% on internal evaluation while consuming $\sim$15$\times$ more tokens than a single chat interaction (the comparison is against a simple chat baseline, not against an optimized single-agent pipeline).

The four articles, read as a set, sketch a prescriptive framework that anticipates many of the patterns we later observe across vendors.
We use them as a reference baseline against which to evaluate the source-code-level decisions of the studied systems (see especially Section~\ref{sec:anthropic_validation}).

\subsection{The 2026 Harness Landscape}
\label{sec:landscape}

The corpus studied in this paper is a deliberate sample from a field that has grown considerably faster than the literature tracking it.
A market sweep conducted for this revision (July 2026) identified over two dozen actively maintained coding-agent harnesses beyond the systems analyzed here; Table~\ref{tab:landscape} lists the most significant.
Three developments since the first version of this study (April 2026) reshape the context in which the corpus sits.\footnote{Landscape events in this subsection are drawn from vendor announcements, release notes, and repository metadata as of 2026-07-10; unlike the corpus claims, they are not source-verified.}

\paragraph{Every vendor now ships a harness, and the market is consolidating.}
Beyond the four provider-native systems we analyze in depth, GitHub's Copilot~CLI reached general availability in February~2026, Amazon folded its Q~Developer CLI into the spec-driven \emph{Kiro} brand, and xAI shipped \emph{Grok~Build}~\cite{grokbuild2026} in May~2026---notably, a closed-source binary (Rust, per its release artifacts) whose documentation describes wholesale adoption of the open ecosystem's conventions: \texttt{AGENTS.md} context files, MCP servers, skills, hooks, and an ACP server mode for editor embedding.
Consolidation followed: SpaceX acquired xAI in February~2026 and announced a \$60B acquisition of Cursor (Anysphere) in June; Google announced in May that Gemini~CLI would transition toward an \emph{Antigravity~CLI}, then ended Gemini~CLI service for its consumer tiers on June~18 (enterprise Gemini~Code~Assist licenses retain access and updates).
The Google transition is a licensing turn as much as a rebrand: Gemini~CLI is Apache-2.0, developed in the open to ${\sim}$105k stars, while Antigravity~CLI ships as a closed-source Go binary with no published source---so of the three corpus systems whose vendors develop them in the open (\cdx{}, \gemini{}, \vibe{}), one now has a closed successor, and our v0.50.0 pin captures the last major open-source state of a provider-native harness lineage.
The harness layer, in other words, is no longer a greenfield: it is strategic infrastructure that vendors buy, rebrand, and gate.

\paragraph{The open-source field has its own centers of gravity.}
\ocd{}~\cite{opencode2026} is now the most-starred dedicated coding agent on GitHub ($\sim$184k stars), and \hrm{}~\cite{hermes2026} the fastest-growing by star-accrual rate ($\sim$212k stars within five months of release; counts as of 2026-07-10, as in Table~\ref{tab:landscape}); both enter our corpus in this revision, together with \pib{}~\cite{pi2026}, the poster child of the opposite, minimal-core countertrend.
Block donated its Rust harness \emph{Goose} to the Linux Foundation's Agentic AI Foundation in April~2026, the first vendor-neutral governance story we identified in the space.
A community project, \emph{Claw~Code}~\cite{clawcode2026}, reimplemented Claude~Code's architecture in Python and Rust as a self-described clean-room derivative of the same March~2026 source exposure our \clcode{} analysis draws on, reaching $\sim$100k stars within days---suggesting that harness \emph{architecture} itself, not just model access, commands community interest in its own right.
A parallel wave of Chinese-vendor harnesses (Alibaba's Qwen~Code, a Gemini~CLI fork; Moonshot's Kimi~CLI; ByteDance's Trae~Agent) ties open-weight model releases to their own CLIs.
Meanwhile \adr{}, one of the field's pioneering systems, entered de-facto community-maintenance mode (last stable release February~2026), with architectural development continuing in a community fork (\texttt{aider-ce}).

\paragraph{A meta-layer has appeared.}
Databricks open-sourced \omni{}~\cite{omnigent2026} in June~2026: an orchestration layer that treats entire harnesses (\clcode{}, \cdx{}, \ocd{}, \hrm{}, \pib{}, Cursor) as interchangeable components behind a common API, adding cross-harness policies, a uniform OS sandbox, and shareable multi-device sessions.
Section~\ref{sec:omnigent} analyzes it as this paper's second contrast point; its existence is itself a datum for the platformization argument of Section~\ref{sec:cli_as_framework}.

\begin{table}[H]
\centering
\caption{The broader coding-agent harness landscape, July~2026 (systems not in the study corpus). Star counts are approximate GitHub stars as of 2026-07-10; closed-source systems marked ``---''.}
\label{tab:landscape}
\scriptsize
\renewcommand{\arraystretch}{1.15}
\begin{tabularx}{\textwidth}{@{}l l c c L{6.2cm}@{}}
\toprule
\textbf{Harness} & \textbf{Vendor} & \textbf{OSS} & \textbf{Stars} & \textbf{Note} \\
\midrule
Antigravity CLI & Google & \xmark & --- & Announced May~2026 as \gemini{}'s successor; closed-source Go binary; consumer service cut over June~18 \\
Copilot CLI & GitHub/Microsoft & \xmark & --- & GA Feb.~2026; autopilot + fleet modes; deep GitHub integration \\
Grok Build & xAI (SpaceX) & \xmark & --- & May~2026; Rust; plan-first loop; ACP server; AGENTS.md/MCP/skills \\
Cursor CLI & Anysphere & \xmark & --- & IDE-harness parity in the terminal; \$60B SpaceX acquisition pending \\
Kiro CLI & Amazon/AWS & \xmark & --- & ex-Q Developer; spec-driven development identity \\
Cline & Cline Bot Inc. & \cmark & $\sim$64k & 5M+ IDE installs; plan/act workflow; Roo/Kilo fork family \\
Goose & Block / Linux Found. & \cmark & $\sim$29k & Rust; native MCP; donated to Agentic AI Foundation Apr.~2026 \\
Claw Code & Community & \cmark & $\sim$100k+ & Self-described clean-room Python/Rust reimplementation of Claude Code \\
Crush & Charmbracelet & \cmark & $\sim$26k & Go-native, LSP-aware TUI; ex-opencode (Go) lineage \\
Qwen Code & Alibaba & \cmark & $\sim$26k & Gemini CLI fork for Qwen coder models \\
Kimi CLI & Moonshot AI & \cmark & $\sim$9k & Skills, MCP, ``agent swarm'' parallelism \\
Trae Agent & ByteDance & \cmark & $\sim$12k & Research-friendly Python CLI agent \\
Droid & Factory AI & \xmark & --- & Terminal-Bench SOTA Sept.~2025; enterprise CI focus \\
Amp & Sourcegraph & \xmark & --- & Ad-supported free tier; deep-mode research workflows \\
Continue CLI & Continue.dev & \cmark & $\sim$30k & Multi-surface (IDE+CLI); local/privacy focus \\
\bottomrule
\end{tabularx}
\end{table}

\section{Methodology}
\label{sec:method}

\subsection{System Selection}

The study covers eleven harnesses plus one meta-harness contrast point, chosen for their spread along three axes: design philosophy, maturity, and market position (Table~\ref{tab:selection}); Section~\ref{sec:landscape} situates the corpus in the broader July-2026 field.
The set includes one agent per major commercial LLM provider (\clcode{} for Anthropic, \cdx{} for OpenAI, \gemini{}~\cite{geminicli2026} for Google, \vibe{}~\cite{mistralvibe2026} for Mistral) plus seven open-source projects that occupy different research and production niches---among them the three systems added in this revision: \hrm{} (the fastest-growing open-source harness), \pib{} (the minimal-core countertrend), and \ocd{} (the most-starred dedicated coding agent, with a distinctive client/server split); the carried-over open-source systems are \oh{}~\cite{openhands2024}, \adr{}~\cite{aider2024}, \mswe{}~\cite{miniswe2024}, and \ocl{}.
It spans three programming languages (Python, TypeScript, Rust), three orders of magnitude in code size, and design philosophies that range from a research baseline to fully productized CLI tools.
All snapshots were re-pinned to the latest available releases in July~2026 (the April-2026 snapshots of the original systems are retained for longitudinal comparison, and Section~\ref{sec:evolution_window} draws on both).

\begin{table}[H]
\centering
\caption{System selection criteria. The top eleven rows are the study corpus; \omni{} is analyzed as a meta-harness contrast point only.}
\label{tab:selection}
\small
\renewcommand{\arraystretch}{1.15}
\begin{tabularx}{\textwidth}{@{}l L{2.6cm} L{1.7cm} L{3.4cm} L{3.9cm}@{}}
\toprule
\textbf{System} & \textbf{Provider} & \textbf{Language} & \textbf{Version} & \textbf{Selection Rationale} \\
\midrule
\oh{}        & OpenHands (ex-All-Hands-AI) & Python & V1~SDK v1.34.0 (Jul.~2026) & Mature modular framework \\
\adr{}       & Community (P.~Gauthier) & Python & v0.86.3.dev (May~2026; maintenance mode) & Polymorphic edit strategies \\
\clcode{}    & Anthropic        & TypeScript & source snapshot Mar.~2026 (binary 2.1.206, Jul.~2026) & Anthropic's flagship agent \\
\cdx{}       & OpenAI           & Rust       & rust-v0.144.1 (Jul.~2026) & OpenAI's flagship agent \\
\gemini{}    & Google           & TypeScript & v0.50.0 (Jul.~2026)  & Google's flagship agent \\
\vibe{}      & Mistral AI       & Python     & v2.19.1 (Jul.~2026)  & Mistral's flagship agent \\
\mswe{}      & Princeton/Stanford & Python   & v2.4.5 (Jul.~2026)   & Minimalist research baseline \\
\hrm{}       & Nous Research    & Python     & 0.18.2 (rel.\ 2026.7.7.2, Jul.~2026) & Self-improving skills; hybrid personal/SWE harness \\
\pib{}       & earendil-works (M.~Zechner) & TypeScript & v0.80.6 (Jul.~2026) & Minimal core, maximal extensibility \\
\ocd{}       & Anomaly (ex-SST) & TypeScript & v1.17.18 (Jul.~2026) & Client/server architecture; most-starred dedicated coding agent \\
\ocl{}       & Community        & TypeScript & v2026.6.11 (Jul.~2026) & Contrast: multi-channel assistant \\
\midrule
\omni{}      & Databricks       & Python     & v0.4.0 (Jul.~2026)   & Contrast: meta-harness above harnesses \\
\bottomrule
\end{tabularx}
\end{table}

\paragraph{\ocl{} as an External Contrast Point.}
\ocl{} is an outlier in the set.
Its README describes it as a \emph{personal AI assistant gateway}, not a coding agent; it spans 20+ messaging platforms (WhatsApp, Slack, Discord, Signal, iMessage, Matrix, and so on) and ships no native code-editing tool (see Table~\ref{tab:editing}, where its entry is \emph{N/A}).
Instead of writing code itself, \ocl{} \emph{delegates} coding tasks to dedicated SWE agents through plugins (\texttt{opencode}, \texttt{kimi-coding}, \texttt{github-copilot}, \texttt{copilot-proxy}).
It is included as an external contrast point for three reasons.
First, its plugin architecture (260 SDK files, manifest-driven discovery, strict import boundaries) isolates extensibility patterns that are SWE-specific from those that belong to a broader agentic-platform category.
Second, it adopts the same cross-cutting standards as SWE-first agents (ACP session spawn, MCP, agentskills.io Skills), providing a non-SWE reference to test convergence claims against.
Third, it illustrates the gateway-delegation pattern: a multi-channel assistant that consumes rather than implements coding capabilities.
\hrm{} sits halfway between the two categories: like \ocl{} it is a multi-channel personal assistant gateway (Telegram, Discord, Slack, WhatsApp, Signal, CLI), but unlike \ocl{} it ships a full native coding toolset, so we treat it as a corpus member rather than a contrast point and flag the hybrid where it matters.
Where adoption counts are sensitive to \ocl{}'s inclusion, both the 11-system count and the 10-coding-harness count are reported.
Readers who interpret the convergence claims as strictly about coding agents should use the 10-system figures.

\paragraph{\omni{} as a Meta-Harness Contrast Point.}
\omni{} (Databricks, open-sourced June~2026~\cite{omnigent2026}) is not a coding harness but an orchestration layer \emph{above} harnesses: it wraps \clcode{}, \cdx{}, Cursor, \ocd{}, \hrm{}, \pib{}, and custom YAML-defined agents behind a common API, adding cross-harness policies, a uniform OS sandbox, and shareable multi-device sessions.
Because it does not implement its own editing loop, scoring it on the seven dimensions alongside the corpus would be a category error; instead, Section~\ref{sec:omnigent} analyzes it as evidence about which harness capabilities are \emph{migrating up} into a meta-layer.

\subsection{Dimensions of Analysis}

Each system is analyzed along seven dimensions:
\textbf{(D1)}~Agent loop design;
\textbf{(D2)}~LLM integration and model--agent co-design;
\textbf{(D3)}~Tool and action systems;
\textbf{(D4)}~Memory and context management;
\textbf{(D5)}~Safety and permission models;
\textbf{(D6)}~Multi-agent orchestration;
\textbf{(D7)}~Extensibility mechanisms.
All versions are as of July~2026 (exact pins in Table~\ref{tab:selection}); for the eight systems carried over from the April~2026 edition of this study, the April snapshots were retained and diffed against the July ones, and Section~\ref{sec:evolution_window} reports what three months of harness evolution look like at source level.
Code-size figures are approximate line counts measured on the pinned snapshots, reported to position systems by order of magnitude; they are not reproducible metrics across counting conventions.

\section{System Overview}
\label{sec:overview}

Table~\ref{tab:overview} presents a high-level comparison of the eleven systems.
The corpus spans three orders of magnitude in code size, three programming languages, and fundamentally different design philosophies; the final column names each system's \emph{signature}---the capability that, as of July~2026, no other system in the corpus implements in the same form.\footnote{The April edition of this study tabulated SWE-Bench Verified scores per system. We have removed them from the comparison table: the available figures are self-reported, obtained on different model generations and configurations, and several predate the systems' current defaults. For the record, the spring-2026 self-reported figures were: \oh{} 77.6\%, \mswe{} 74\%+, \clcode{} 72.7\%, \cdx{} 69.1\% (all SWE-Bench Verified); the remaining systems publish no comparable number---\pib{} explicitly solicits real-world session data ``instead of toy benchmarks.''}

\begin{table}[H]
\centering
\caption{High-level comparison of the eleven systems, with each system's signature capability. \clcode{} cells here and in the other July-dated tables reflect the March-2026 source snapshot, the newest source available to us; the shipping binary (2.1.206, July~2026) may differ.}
\label{tab:overview}
\scriptsize
\renewcommand{\arraystretch}{1.3}
\setlength{\tabcolsep}{3.5pt}
\begin{tabularx}{\textwidth}{@{}l l l L{1.9cm} C{0.75cm} C{0.95cm} L{1.55cm} L{1.7cm} X@{}}
\toprule
\textbf{System} & \textbf{Lang.} & \textbf{Scale} & \textbf{Provider} & \textbf{Tools} & \textbf{Multi-ag.} & \textbf{Sandbox} & \textbf{UI} & \textbf{Signature} \\
\midrule
\oh{}     & Python & Medium   & Multi    & 25+  & Yes & Docker/ Apptainer/ remote & Web+CLI +SDK & Resource-locked parallel tools; hosts rival harnesses as ACP backends \\
\adr{}    & Python & Small    & Multi    & 13~fmt & No & None & CLI & 13 polymorphic edit formats; the corpus's only ranked repo map \\
\clcode{} & TS     & Large    & Anthropic& 43   & Yes & Opt-in runtime + worktree & CLI & Deferred tool loading; prompt-cache-sharing forks; the reference others copy \\
\cdx{}    & Rust   & Very lg. & OpenAI   & 25--30 & Yes & Native (3~OS) & CLI/TUI & Agent-maintained cross-session memory; tool calls as V8-executed code \\
\gemini{} & TS     & Very lg. & Google   & 35+  & Yes & Native (3~OS) & CLI/TUI & Model routing as runtime scheduling; the corpus's only A2A server \\
\vibe{}   & Python & Small    & Mistral+ & 12+  & Yes & Worktree opt. & CLI/TUI & Middleware-pipeline loop; rewind exposed over ACP \\
\mswe{}   & Python & Tiny     & Multi    & 1    & No  & Docker+ & CLI & The 100-line floor: all seven subsystems, minimally \\
\hrm{}    & Python & Very lg. & Multi (owned) & 69 & Yes & 6 pluggable backends & CLI/TUI/ Web+28 ch. & Self-improving skill loop; lineage compaction; verify-on-stop \\
\pib{}    & TS     & Medium   & Multi, 35 prov. & 7 & Ext. & None (by design) & CLI/TUI & Everything-is-an-extension core; session-tree version control \\
\ocd{}    & TS     & Large    & Multi    & 17   & Yes & None (policy) & Client/ server & Client/server harness; syntax-aware command permissioning \\
\ocl{}    & TS     & Large    & Multi    & 109+ & Yes & None & Multi-ch. & Gateway delegation; Active-Memory sub-agent before each reply \\
\bottomrule
\end{tabularx}

\vspace{0.4em}
{\scriptsize ``Scale'' is qualitative and reflects order-of-magnitude codebase size in the system's primary source tree: Tiny ($\sim$5\,K LoC), Small ($\sim$20--40\,K), Medium ($\sim$80--110\,K), Large ($\sim$500\,K+), Very large ($\sim$600\,K+, with \cdx{} at $\sim$1.1\,M lines of Rust in the July snapshot). These figures are not directly comparable across languages and counting conventions; they are reported only to position systems on the simplicity-vs-capability axis.}
\end{table}

\observation{\label{obs:size}The eleven systems span three orders of magnitude in code size while targeting similar tasks, yet loop sophistication does not predict benchmark performance. \mswe{}'s minimal linear loop achieves reported results in the same range as \oh{}'s event-sourced conversation engine, despite the latter committing substantially more code to loop orchestration alone. Most of the mass in the production systems addresses concerns orthogonal to task completion: safety, user experience, extensibility---and increasingly \emph{clients and transport}: roughly three-fifths of \ocd{}'s non-test source is its TUI, web, desktop, and SDK clients rather than the harness proper, and \cdx{}'s July tree devotes six-figure line counts to app-server transports, plugins, and a realtime voice layer that no benchmark will ever measure.}

\section{Agent Loop Design}
\label{sec:agentloop}

The agent loop, the central control flow that alternates between LLM inference and action execution, is the architectural backbone of every SWE agent.
We identify three distinct paradigms across the studied systems.

\subsection{Taxonomy of Agent Loop Paradigms}

All eleven systems implement variations of the ReAct~\cite{yao2023react} pattern, but the implementations differ radically in complexity, concurrency, and state management.
We identify three paradigms: \emph{iterative action-observation} (nine systems), \emph{reflection-augmented} (\adr{}, with \hrm{} relocating the same function into the loop's stop condition), and \emph{coordinator-worker} (an overlay on the iterative loop, prescriptive in \clcode{} and \cdx{}, configuration-gated in \hrm{}).
Rombaut's concurrent taxonomy~\cite{rombaut2026} makes the useful finer-grained point that most production loops \emph{compose} primitives (ReAct, generate-test-repair, plan-execute, retry) rather than implementing one; the plan modes that all four provider-native systems now ship (Section~\ref{sec:safety}) are plan-execute grafted onto ReAct.

The field's vocabulary is still being minted around this subsystem.
\emph{Loop engineering} entered circulation in June~2026, between this study's two snapshots: Steinberger (\ocl{}'s original author) condensed the shift into a widely circulated dictum---stop prompting coding agents, ``design loops that prompt your agents''~\cite{steinberger2026loops}---and Osmani named and structured the practice days later, giving the loop an anatomy of triggers, topology, verifiers, and stop rules~\cite{osmani2026loop}.
The coinage is complementary to this section's subject rather than a synonym for it: harness engineering builds the \emph{inner} action--observation cycle dissected below, while loop engineering composes that cycle, from the outside, into self-sustaining outer loops that prompt, verify, and re-run the agent without a human authoring each turn.
The seam between the two disciplines is already visible in the corpus as the \emph{outer verification loop} pattern (Table~\ref{tab:patterns2}): \oh{}'s \texttt{/goal} endpoint runs an LLM judge over each finished run and either re-prompts or stops, and \hrm{}'s verify-on-stop guard vetoes the inner loop's own exit---harness features whose only purpose is to close an outer loop around the inner one.

\subsection{Iterative Action-Observation Loop}

The most common pattern follows a cycle of prompt construction, LLM inference, tool execution, and observation collection.
Nine systems implement this pattern, each with distinctive characteristics.

\paragraph{OpenHands: Event-Sourced Conversation Engine.}
\oh{}'s loop---since the mid-2026 restructuring, in the \texttt{software-agent-sdk} repository rather than the application repo---is an event-sourced conversation engine: a \texttt{LocalConversation} drives \texttt{Agent.step()}, every event is appended to a persistent \texttt{EventLog} (one file per event through a \texttt{FileStore}, with flock-based locking and secret redaction via a \texttt{SecretRegistry}), and the LLM's view of history is a cached projection of the active branch.
Conversation state is a \emph{tree} with a movable head, so replay, fork, and branch navigation are first-class.
Each \texttt{step()} makes one LLM call but executes \emph{all} tool calls of the response as an action batch---optionally in parallel through a \texttt{ParallelToolExecutor} governed by a \texttt{tool\_concurrency\_limit} and a resource-lock manager keyed on each tool's declared resources (files, terminal session, browser), so only same-resource calls serialize (the \clcode{} paragraph below and Table~\ref{tab:pipeline_detail} contrast this with \clcode{}'s boolean partitioning).
The \texttt{StuckDetector} survives the rearchitecture intact: the same five failure scenarios as in the V0 codebase (repeating action-observation pairs, repeating actions with errors, monologue loops, alternating patterns, and context-window error loops), now with configurable thresholds and enabled by default.

\paragraph{Claude Code: Streaming Loop with Concurrent Tool Batching.}
\clcode{}'s loop responses stream via Server-Sent Events (SSE) with fine-grained event handling (\texttt{content\_block\_start}, \texttt{content\_block\_delta}, \texttt{content\_block\_stop}).
The most distinctive feature is \emph{intelligent tool batching}: tool calls are partitioned into batches by concurrency safety via a single-pass reduce algorithm.
Tools default to \texttt{isConcurrencySafe = false} (the safe default), and must explicitly opt in to parallel execution.
Read-only tools (grep, glob, file read) are safe; write tools (edit, bash) are not.
This design enables significant latency reduction when the LLM requests multiple independent reads simultaneously.

\paragraph{Codex: Tokio Async State Machine.}
\cdx{}'s loop is implemented in Rust as a Tokio-based async state machine.
A \texttt{Session} struct orchestrates turns via streaming \texttt{ResponseItem} events from the OpenAI Responses API (deserialized from SSE or WebSocket stream frames), with tool invocations processed through \texttt{FuturesOrdered} for ordered parallel execution, now factored through a dedicated \texttt{ToolCallRuntime}.
Key entry points: \texttt{Codex::spawn()} creates a new session, \texttt{submit\_with\_id()} submits operations, and \texttt{next\_event()} provides non-blocking event reading for streaming responses (the once-monolithic \texttt{codex.rs} has been split into \texttt{session/} modules and a \texttt{CodexThread} abstraction, but the entry points survive).

\paragraph{Mini-SWE-Agent: Minimal Linear Loop.}
\mswe{} implements the simplest variant (Listing~\ref{lst:miniswe}):

\begin{lstlisting}[caption={\mswe{} agent loop (condensed; omits format-error handling, cost/time accounting, and trajectory saving present in v2.4.5).},label={lst:miniswe}]
def run(self, task):
    self.add_messages(model.format_message(
        "system", render_template(system_template)))
    self.add_messages(model.format_message(
        "user", render_template(instance_template, task=task)))
    while True:
        result = self.step()
        if result.role == "exit":
            return result
def step(self):
    self.query()
    return self.execute_actions()
def query(self):
    self.check_limits()
    self.n_calls += 1
    msg = self.model.query(self.messages)
    self.add_messages(msg)
def execute_actions(self):
    for action in msg["extra"]["actions"]:
        obs = self.env.execute(action)  # subprocess.run()
        self.add_messages(
            model.format_observation_messages(obs))
\end{lstlisting}

There is no state machine, no event sourcing, and no concurrency mechanism.
The message list grows linearly and unboundedly, relying entirely on the LLM's native context window.
The minimalism is deliberate: the design intentionally isolates LLM capability from agent scaffolding complexity, to see how much of one can substitute for the other.
Even the floor drifts upward, though: the v2.4 line added a wall-clock time limit on runs and a cap on \emph{consecutive malformed-action responses}---the loop now aborts after $N$ successive format errors instead of cycling indefinitely---so the corpus's minimal system has acquired a minimal stuck detector.

\paragraph{Mistral Vibe: Middleware-Pipeline Loop.}
\vibe{}'s loop is a Python async/await iterative loop in which turn-level policies are factored out into a composable \emph{middleware pipeline}.
The core conversation loop drives a standard prompt~$\to$~LLM~$\to$~tools cycle, but each iteration first walks a stack of middleware pre-turn checks and then yields events to the caller as they complete.
Six middlewares ship out of the box: \texttt{TurnLimitMiddleware}, \texttt{PriceLimitMiddleware} (cost cap per session), \texttt{TokenLimitMiddleware} (session-total token cap, added in the 2.9 line), \texttt{AutoCompactMiddleware} (triggers summarization at a token threshold), \texttt{ContextWarningMiddleware} (warns when the conversation approaches the context-window limit), and \texttt{ReadOnlyAgentMiddleware} (gates write tools for read-only agent profiles); user-cancellation is handled inline via an \texttt{is\_user\_cancellation\_event()} check rather than as a middleware.
Tool calls within a single LLM response are executed concurrently by spawning each tool as an \texttt{asyncio.create\_task()} and yielding events as they finish, a finer-grained variant of \cdx{}'s \texttt{FuturesOrdered}.
The middleware design is unique in the corpus.
New turn-level policies can be added without modifying the loop body, and the same loop powers six registered agent profiles (\texttt{default}, \texttt{plan}, \texttt{accept-edits}, \texttt{auto-approve}, \texttt{explore}, \texttt{lean}) simply by swapping the middleware composition; a seventh \texttt{chat} profile is not registered for CLI sessions but is dynamically registered by the ACP layer for IDE integrations.
Plan mode has meanwhile become structural rather than prompt-only: a plan file under a dedicated plans directory is the profile's only writable target, and an \texttt{exit\_plan\_mode} tool emits a \texttt{PlanReviewRequestedEvent} that gates the transition back to write-capable profiles.

\paragraph{Gemini CLI: Async-Generator Loop with Hybrid Loop Detection.}
\gemini{}'s loop is implemented in TypeScript as an async generator.
Each iteration yields typed \texttt{ServerGeminiStreamEvent} tuples to the UI, runs the active model via a \texttt{ModelRouterService}, streams the response via the \texttt{@google/genai} SDK, and dispatches function calls through a state-machine \texttt{Scheduler} that walks each tool through \emph{Validating~$\to$~Executing~$\to$~Completed/Errored}.
A distinctive feature is the \texttt{LoopDetectionService}, which is hybrid rather than purely deterministic: SHA-256 hashing catches the common failure modes cheaply (five identical tool calls or ten identical content chunks abort the loop), and after thirty turns in a single prompt an LLM-based self-check runs at an adaptive interval---a two-tier design that contrasts with \oh{}'s five hand-enumerated scenarios.
A session turn limit (default 100) provides a hard ceiling, and a set of eleven lifecycle hook events (\texttt{BeforeAgent}/\texttt{AfterAgent}, \texttt{BeforeModel}/\texttt{AfterModel}, \texttt{BeforeToolSelection}, \texttt{PreCompress}, and others) lets extensions inspect or veto the loop.
The scheduler executes tool calls in parallel and, since v0.45, partitions by concurrency safety: file-mutating tools (\texttt{edit}, \texttt{write\_file}) and \texttt{update\_topic} are hard-forced to sequential execution, and---uniquely in the corpus---the model itself is given a concurrency knob, an auto-injected \texttt{wait\_for\_previous} boolean on every tool schema through which it can serialize any call.
The scheduler streams live tool output to the UI via an \texttt{outputUpdateHandler} callback so the user sees long-running commands progress.

\paragraph{Hermes: Budgeted Loop with Stop-Guards.}
\hrm{} runs a single iterative tool-calling loop per user turn, bounded by a thread-safe \texttt{IterationBudget} (default 90, with \texttt{execute\_code} iterations refunded and a final grace call that lets the model summarize when the budget expires).
What distinguishes it is not the loop body but its \emph{exits}: eleven enumerated turn-exit reasons, and two stop-guards that can veto a premature final answer.
The \emph{verify-on-stop} guard rewrites a text-only response into a continuation whenever the turn mutated code files without producing fresh verification evidence---the reflection loop's function, relocated from inside the loop to its stop condition, at a fraction of the cost.
Stuck detection hashes tool name plus sorted-JSON arguments (SHA-256), warning after two identical failures and halting after eight---but the hard stop ships \emph{disabled}: the default posture is warnings appended to tool results, trusting the model to self-correct.
Tool batches parallelize on a pool of eight workers only when every call is read-only-safe or path-scoped with non-overlapping prefixes; mid-turn user steering is spliced into the last tool result behind an anti-injection marker.

\paragraph{Pi: Functional Core with Steering Queues.}
\pib{}'s loop is the purest iterative implementation in the corpus after \mswe{}: a $\sim$790-line functional core with no planner, no reflection step, no turn cap, no stuck detection, and no cost kill-switch---the loop runs until the model stops calling tools, and all of those absences are documented design refusals.
Two primitives stand out.
Mid-run user input is first-class: a \texttt{steer()} queue injects messages after the current turn's tool calls, while \texttt{followUp()} drains only when the agent would otherwise stop.
And tool calls execute in parallel by default via \texttt{Promise.all}, with a realpath-keyed per-file mutation queue serializing edits and a distinctive \emph{truncation-poisoning guard}: when the assistant message was cut off at the length limit, \emph{all} of its tool calls are failed unexecuted, because salvage-parsed streaming arguments can validate while being silently incomplete.
Recovery is layered instead of enumerated: a $\sim$40-pattern transient-error retry allowlist and a one-shot compact-and-retry on context overflow.

\paragraph{OpenCode: Log-as-Queue Loop.}
\ocd{}'s loop is a literal \texttt{while(true)} with an unusual twist: the message log doubles as the work queue.
Pending sub-agent spawns and compactions are not control-flow branches but persisted \emph{message parts} that the loop dequeues one per iteration---which makes the loop trivially resumable across process restarts, since the queue \emph{is} the transcript.
There is no default step cap (\texttt{maxSteps} defaults to infinity); when a configured cap is reached, the harness appends an assistant message declaring tools disabled rather than raising an error.
Stuck detection is a single heuristic routed, characteristically, through the permission system: three consecutive byte-identical tool calls raise a \texttt{doom\_loop} permission \emph{ask} rather than an automated abort---the user, not the harness, decides whether the repetition is a bug.
Parallel tool calls execute concurrently with no harness cap, and retry is harness-level with retry-after-aware exponential backoff; context overflow is never retried---it flips to compaction.

\subsection{Reflection-Augmented Loop}

\textbf{Aider} extends the basic loop with a \emph{reflection} mechanism.
The \texttt{run\_one()} method implements a nested loop: after each LLM response, the system applies edits, then checks for lint errors (via a three-stage Python pipeline: syntax check $\to$ compile check $\to$ flake8), test failures, and unresolved file mentions.
If issues are detected, a \texttt{reflected\_message} triggers re-invocation of the LLM with corrective context, up to a configurable maximum (default: 3 reflections).

The linter integration is particularly notable: it uses \texttt{TreeContext} from \texttt{grep\_ast} to show code context around error lines, providing the LLM with precise locality information for its corrections.
\adr{} remains the only system whose \emph{main loop} is reflection-shaped, but the corpus has grown three structural cousins: \gemini{}'s edit tool ends in an LLM ``edit fixer'' subcall that repairs a failed match (Section~\ref{sec:editing}), \ocd{} feeds LSP diagnostics back into every edit result, and \hrm{}'s verify-on-stop guard performs the reflection check once, at turn exit, instead of per iteration.
The pattern is the in-the-wild instantiation of a line of work on language-model self-correction, Reflexion~\cite{shinn2023reflexion}, Self-Refine~\cite{madaan2023selfrefine}, CRITIC~\cite{gou2024critic}, and Self-Debug~\cite{chen2024selfdebug}, that has otherwise stayed largely in the academic literature.
\adr{}'s contribution is the integration: lint and test signals are routed back into the loop as the corrective feedback those papers theorize about.
A note on its stop condition: \adr{} has no tool-call loop at all---a turn ends when the completion stream finishes, and \texttt{run\_one} exits when no reflection is pending or the cap is reached; it never inspects an \texttt{end\_turn} finish reason.

\subsection{Coordinator-Worker Pattern}

\clcode{} and \cdx{} support a \emph{coordinator mode} where a parent agent orchestrates multiple worker agents; \hrm{} gates the same shape behind configuration (an \texttt{orchestrator} role plus a spawn-depth setting unlocks nested delegation trees, and a separate Kanban-swarm mode runs planning-root~$\to$~workers~$\to$~verifier as subprocesses over a SQLite blackboard).
This is an overlay on the iterative loop, adding a hierarchical dispatch layer.

In \clcode{}, the coordinator spawns sub-agents via \texttt{AgentTool}, each receiving a forked context with isolated \texttt{AbortController}, cloned file state cache, and suppressed permission dialogs.
Workers communicate results back via \texttt{<task-notification>} XML blocks, and the coordinator synthesizes findings before delegating the next phase.
The workflow is structured: Research $\to$ Synthesis $\to$ Implementation $\to$ Verification.

In \cdx{}, sub-agents are spawned via \texttt{AgentControl::spawn\_agent()}, receiving dedicated \texttt{ThreadId}s and communicating through typed \texttt{Mailbox} channels.
Each thread maintains its own history, with \texttt{SpawnAgentForkMode} controlling context inheritance (\texttt{FullHistory} or \texttt{LastNTurns(N)}).

We defer the detailed analysis of multi-agent orchestration to Section~\ref{sec:multiagent}.

\subsection{Comparative Analysis}

\begin{figure}[H]
\centering
\begin{tikzpicture}[
  node distance=0.5cm,
  stepbox/.style={rectangle, draw, rounded corners=2pt, minimum width=1.8cm, minimum height=0.6cm, font=\scriptsize, align=center, thick},
  arr/.style={-{Stealth[length=2mm]}, thick},
]
  \begin{scope}[local bounding box=simple]
    \node[stepbox, fill=blue!10] (s1) {Prompt};
    \node[stepbox, fill=red!10, below=of s1] (s2) {LLM Call};
    \node[stepbox, fill=green!10, below=of s2] (s3) {Execute};
    \node[stepbox, fill=yellow!10, below=of s3] (s4) {Observe};
    \draw[arr] (s1) -- (s2);
    \draw[arr] (s2) -- (s3);
    \draw[arr] (s3) -- (s4);
    \draw[arr] (s4.east) -- ++(0.5,0) |- (s1.east);
    \node[font=\scriptsize\bfseries, above=0.3cm of s1] {(a) Iterative};
  \end{scope}

  \begin{scope}[xshift=4.5cm, local bounding box=reflect]
    \node[stepbox, fill=blue!10] (r1) {Prompt};
    \node[stepbox, fill=red!10, below=of r1] (r2) {LLM Call};
    \node[stepbox, fill=green!10, below=of r2] (r3) {Apply Edits};
    \node[stepbox, fill=orange!15, below=of r3] (r4) {Lint/Test};
    \node[stepbox, fill=purple!10, below=of r4] (r5) {Reflect?};
    \draw[arr] (r1) -- (r2);
    \draw[arr] (r2) -- (r3);
    \draw[arr] (r3) -- (r4);
    \draw[arr] (r4) -- (r5);
    \draw[arr] (r5.west) -- ++(-0.6,0) node[font=\tiny, left]{yes} |- (r1.west);
    \draw[arr] (r5.east) -- ++(0.5,0) node[font=\tiny, right]{no};
    \node[font=\scriptsize\bfseries, above=0.3cm of r1] {(b) Reflection};
  \end{scope}

  \begin{scope}[xshift=9cm, local bounding box=coord]
    \node[stepbox, fill=red!10] (c1) {Coordinator};
    \node[stepbox, fill=blue!10, below left=0.6cm and -0.2cm of c1] (w1) {Worker 1};
    \node[stepbox, fill=blue!10, below=0.6cm of c1] (w2) {Worker 2};
    \node[stepbox, fill=blue!10, below right=0.6cm and -0.2cm of c1] (w3) {Worker 3};
    \node[stepbox, fill=green!10, below=0.8cm of w2] (syn) {Synthesize};
    \draw[arr] (c1) -- (w1);
    \draw[arr] (c1) -- (w2);
    \draw[arr] (c1) -- (w3);
    \draw[arr] (w1) |- (syn);
    \draw[arr] (w2) -- (syn);
    \draw[arr] (w3) |- (syn);
    \node[font=\scriptsize\bfseries, above=0.3cm of c1] {(c) Coordinator};
  \end{scope}
\end{tikzpicture}
\caption{Three agent loop paradigms: (a)~iterative action-observation (\oh{}, \clcode{}, \cdx{}, \gemini{}, \vibe{}, \mswe{}, \hrm{}, \pib{}, \ocd{}; event-driven variant in \ocl{}), (b)~reflection-augmented (\adr{}), (c)~coordinator-worker overlay on (a) (\clcode{}, \cdx{}; configuration-gated in \hrm{}).}
\label{fig:loop_comparison}
\end{figure}
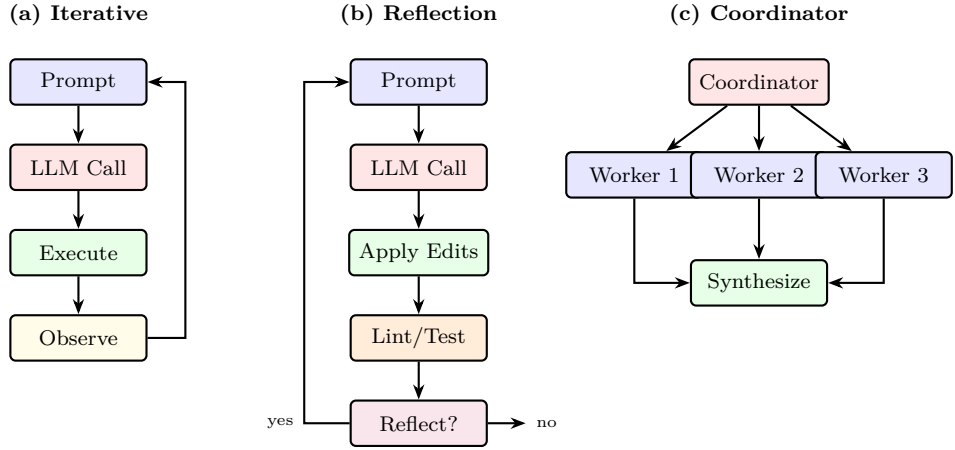

Figure~\ref{fig:loop_comparison} contrasts the three paradigms.

Table~\ref{tab:loop_detail} provides a detailed comparison of loop characteristics.

\begin{table}[H]
\centering
\caption{Detailed agent loop characteristics across the eleven systems.}
\label{tab:loop_detail}
\small
\renewcommand{\arraystretch}{1.2}
\begin{tabularx}{\textwidth}{@{}l L{2.9cm} L{3.1cm} L{3.1cm} L{2.9cm}@{}}
\toprule
\textbf{System} & \textbf{Loop Type} & \textbf{Concurrency} & \textbf{Stuck Detect.} & \textbf{Stop Cond.} \\
\midrule
\oh{}     & Event-sourced conversation & Parallel batches (resource locks; limit default 1) & 5 scenarios & State=FINISHED  \\
\adr{}    & Generator + reflect   & Sequential          & Max reflections & stream end, no reflection pending \\
\clcode{} & Streaming + batch     & Safety-partitioned  & None (manual)   & end\_turn        \\
\cdx{}    & Tokio async SM        & FuturesOrdered      & None            & end\_turn        \\
\gemini{} & Async-gen + scheduler & Parallel; edits forced sequential; model-visible \texttt{wait\_for\_previous} & Hybrid: SHA-256 + LLM check after 30 turns & end\_turn / 100  \\
\vibe{}   & Async + middleware    & \texttt{asyncio.create\_task} & Middleware (turn/price/token/compact) & end\_turn \\
\mswe{}   & Linear while          & Sequential          & step limit; format-error cap; wall clock & exit message     \\
\hrm{}    & Budgeted while + stop-guards & Parallel batches (8 workers, safety-gated) & SHA-256 call signatures (warn-first) & text response unless verify-on-stop vetoes; 90-iter budget \\
\pib{}    & Functional while + steer/followUp queues & \texttt{Promise.all}; per-file mutation queue & None (by design) & end\_turn (no cap) \\
\ocd{}    & Log-as-queue \texttt{while(true)} & Parallel, uncapped & Doom-loop $\to$ permission ask & no tool calls; \texttt{steps ?? Infinity} \\
\ocl{}    & Event-driven ACP      & RPC isolation       & Rate limit      & Session end      \\
\bottomrule
\end{tabularx}
\end{table}

\section{LLM Integration and Model--Agent Co-design}
\label{sec:llm}

How an agent scaffold relates to its underlying LLM is one of the most consequential design decisions in the system.
The corpus spans the full range from tight single-provider coupling to complete provider agnosticism, and the position on that spectrum has direct implications for the optimizations available to the scaffold.

\subsection{Provider Abstraction Spectrum}

The eleven systems adopt five distinct strategies:

\paragraph{Single-provider tight coupling (Anthropic, OpenAI, Google).}
\clcode{} exclusively uses the Anthropic SDK (\texttt{@anthropic-ai/sdk}); \cdx{} uses the OpenAI Responses API with WebSocket streaming; \gemini{} uses the \texttt{@google/genai} SDK and additionally couples to Vertex AI for enterprise deployments.
These systems cannot easily switch providers: \clcode{} depends on Claude-specific features (extended thinking, prompt caching with static/dynamic boundaries, \texttt{tool\_use} block format); \cdx{} depends on OpenAI's Responses API wire format and maintains model-specific prompts per GPT generation, delivered as of mid-2026 as \emph{server-side model-catalog data} refreshed at runtime rather than compiled-in templates (Section~\ref{sec:prompt_arch}); \gemini{} runs a \texttt{ModelRouterService} that, via a chain of pluggable strategies (fallback, override, approval-mode, three classifier strategies---including one that can run against a \emph{local} Gemma model through a managed LiteRT-LM runtime---and default), dispatches each request to a Gemini variant.
The router's target set shifted a full model generation between our two snapshots: \texttt{gemini-3(.1)-pro-preview} and GA \texttt{gemini-3.1-flash-lite} are now the default resolution targets, with Gemma~4 models routable and the 2.5 family surviving only as a no-preview-access fallback---evidence that routing layers exist precisely to absorb model churn so the loop does not.

\paragraph{Provider-first with generic fallback (Mistral).}
\vibe{} sits between tight coupling and full abstraction: a factory selects between a \texttt{MistralBackend} (using the official \texttt{mistralai} SDK) and a \texttt{GenericBackend} that speaks Anthropic, Vertex, OpenAI-compatible, and---since v2.9---OpenAI Responses endpoints.
The Mistral backend exploits Mistral-specific features (the \texttt{reasoning\_effort} enum mapped from a model's five-level \texttt{thinking} field, and the \texttt{ThinkChunk} stream block parsed into a separate \texttt{reasoning\_content} field), while the generic backend provides a portable escape hatch.
This is a different philosophy from the LiteLLM-based systems: rather than treating all providers symmetrically, \vibe{} privileges its home provider with a deeper integration and treats others as a fallback.

\paragraph{Multi-provider via abstraction layer.}
\oh{}, \adr{}, and \mswe{} all use LiteLLM as a universal abstraction, supporting 100+ models across providers.
This enables rapid model switching but introduces a dependency on LiteLLM's provider mappings.
\adr{} goes furthest with a model registry containing per-model metadata for 350+ models: edit format, weak model name, cache control, extra API parameters, and reasoning tag handling.
\oh{} has meanwhile layered its own routing on top of the abstraction: an \texttt{LLMRegistry} keyed by usage, a separate \texttt{LLMProfileStore} for named profiles, pluggable \texttt{RouterLLM} implementations, and a fallback strategy---so ``LiteLLM-based'' no longer implies feature-minimal.

\paragraph{Multi-provider via owned transports.}
The three systems added in this revision occupy a position the April edition's four-way spectrum did not contain: full multi-provider support with \emph{no} abstraction library, the provider matrix hand-built and exercised to the same depth as the vendor-native integrations.
\hrm{} ships five transport implementations (chat-completions, Anthropic Messages, Bedrock Converse, Codex Responses, and an out-of-process Codex app-server) behind 29 declarative \texttt{ProviderProfile} plugins---the profile ``is read by the transport instead of receiving 20+ boolean flags''---with model metadata for 3{,}800+ models pulled from models.dev.
\pib{} hand-rolls nine wire-protocol implementations across 35 built-in providers, with $\sim$20 per-model \texttt{compat} quirk flags (including ten \texttt{thinkingFormat} dialects for OpenAI-compatible hosts) paying the per-provider conditional cost once, centrally---and it exercises the corpus's boldest coupling trick, \emph{harness mimicry}: on a Claude Pro/Max OAuth token it presents Claude Code's identity wholesale (the ``You are Claude Code'' system-prompt opening, the beta headers, even the canonical tool-name casing, re-cased on the wire) to ride the consumer subscription, and it speaks the ChatGPT-plan Codex backend the same way.
\ocd{} delegates streaming and tool dispatch to Vercel's AI SDK and the models.dev registry---the only corpus system whose inner LLM plumbing is a third-party SDK---then concentrates its co-design in a $\sim$1{,}400-line per-vendor transform matrix (release-date-gated reasoning tiers, per-model temperature defaults, six simultaneous prompt-cache dialects), and can even install \emph{new} providers at runtime from a registry field.

\paragraph{Hybrid plugin-based.}
\ocl{} implements a pluggable provider system with first-party adapters for 10+ providers, featuring auth profile rotation, last-good tracking for failover, and cooldown expiry for degraded keys.

\observation{\label{obs:coupling}Provider-native optimizations---cache boundaries, extended thinking, reasoning effort, model-specific prompts---are not gated on tight coupling; they are gated on \emph{who pays the per-provider conditional-code cost}. \clcode{} splits its prompt at a Blake2b-hashed cache boundary whose static prefix is cached with global scope across sessions (fork-mode sub-agents additionally inherit the parent prompt byte-exactly); \cdx{} keeps per-generation prompts (now delivered as server-side model-catalog data) and a \texttt{thread\_id}-derived prompt-cache key; \gemini{}'s \texttt{ModelRouterService} dispatches each request to the cheapest sufficient Gemini variant. But three multi-provider systems now exercise the same menu from the other end of the spectrum, by paying the conditional-code cost deliberately and centrally: \hrm{} hand-rolls five transports with provider-agnostic cache markers and bit-perfect prefix normalization that hits even local llama.cpp KV caches; \pib{} places explicit \texttt{cache\_control} breakpoints with TTL tiers and audits per-turn cache-miss dollar waste as a first-class metric; \ocd{} emits six providers' cache dialects simultaneously and shapes its system prompt to at most two messages (one by default) to match the cache slots. The LiteLLM-based systems (\oh{}, \adr{}, \mswe{}) exercise the menu partially (Table~\ref{tab:api_features}), with \oh{} closing the gap through cache-tiered prompt assembly and pluggable routing. What tight coupling still buys, uniquely, is server-side co-evolution: \cdx{}'s scaffold behavior---prompts, reasoning tiers, tool modes, even the multi-agent tool generation---is re-tuned per model release from the provider's catalog endpoint without shipping a client. Vendor coupling has become less about \emph{capability} and more about \emph{who controls the update loop}.}

\subsection{Prompt Engineering Architectures}
\label{sec:prompt_arch}

Prompt construction strategies vary dramatically across the eleven systems.
Table~\ref{tab:prompts} summarizes the approaches.

\begin{table}[H]
\centering
\caption{Prompt engineering architectures across the eleven systems.}
\label{tab:prompts}
\scriptsize
\renewcommand{\arraystretch}{1.25}
\begin{tabularx}{\textwidth}{@{}l L{3.2cm} L{3.4cm} L{3.4cm} L{2cm}@{}}
\toprule
\textbf{System} & \textbf{Architecture} & \textbf{Sections} & \textbf{Caching} & \textbf{Template} \\
\midrule
\oh{}     & PromptRegistry with cache tiers & 18 named static sections (Soul, Role, Security, \ldots) + dynamic block & STATIC/DYNAMIC two-block assembly; \texttt{cache\_control} markers & Python; .j2 escape hatch \\
\adr{}    & Polymorphic CoderPrompts & Per-edit-format prompts, repo-map, examples & Anthropic markers & Python classes \\
\clcode{} & Modular sections with boundary & 12--15 named sections, static/dynamic split & Blake2b hash, boundary marker & TypeScript \\
\cdx{}    & Server-delivered per-model prompts & Capabilities, personality (user-templated), AGENTS.md, planning & \texttt{thread\_id} cache key & Model-catalog data (bundled fallback) \\
\gemini{} & PromptProvider + capability gating & Base, tools, GEMINI.md (global/proj/ext), IDE ctx, memory, skills & None (rebuilt on model change) & TS snippets (modern/legacy) \\
\vibe{}   & Per-agent prompts + A/B variants & Base, tools, git status, date, subagent roster, scratchpad, AGENTS.md contents, agent override, skills & Git status cached & Markdown files (GrowthBook-selected variant) \\
\mswe{}   & Jinja2 one-shot & system\_template, instance\_template & Anthropic markers & YAML-embedded \\
\hrm{}    & Three-tier assembly (stable/context/volatile), built once per session & $\sim$15 guidance blocks, model-family-gated & \texttt{system\_and\_3} markers + bit-perfect prefix normalization & Python constants \\
\pib{}    & Single $\sim$170-line builder; tool snippets co-vary with active toolset & Persona, tools, deduplicated guidelines, project context, skills XML, date, cwd & \texttt{cache\_control} breakpoints + TTL tiers + waste audit & TS concatenation \\
\ocd{}    & Model-family prompt matrix (9 base prompts by model-id substring) & Base, env block, AGENTS.md, MCP instructions, skills catalog & Six-dialect breakpoints; \texttt{promptCacheKey}=sessionID; system capped at 2 messages & .txt files \\
\ocl{}    & ACP translator & Session config, tool list, thinking level & Per-provider & Runtime config \\
\bottomrule
\end{tabularx}
\end{table}

\paragraph{Gemini CLI: Capability-Gated Snippets with Hierarchical Memory.}
\gemini{}'s \texttt{PromptProvider} assembles the system prompt from base instructions chosen by model capability (\texttt{isModernModel ? snippets : snippets.legacy}), tool/function declarations from the registry, and a hierarchically merged set of \texttt{GEMINI.md} files discovered at three scopes (global \texttt{\textasciitilde/.gemini/}, extension-provided, project-local \texttt{.gemini/}).
IDE diagnostics and open-file diffs are injected mid-conversation via the IDE companion package; persistent facts are written by the model editing \texttt{GEMINI.md} or the private per-project \texttt{MEMORY.md} index directly with ordinary edit tools---the system prompt states that no \texttt{save\_memory} tool exists (Section~\ref{sec:memory_pipelines}).
Unlike in \clcode{}, the prompt is rebuilt on each model change but not split by a static/dynamic cache boundary---Gemini does not currently expose request-level prompt caching, so cache amortization is left to the API server.

\paragraph{Mistral Vibe: Per-Agent Markdown Prompts.}
\vibe{}'s prompts live as plain Markdown files; only the \texttt{explore} and \texttt{lean} profiles ship a dedicated system prompt, while \texttt{plan}, \texttt{accept-edits}, and \texttt{auto-approve} reuse the default \texttt{cli.md} via overrides.
A universal system-prompt builder composes a base prompt with auto-generated tool descriptions, a Git context block (output of \texttt{git log --oneline -N --decorate} and \texttt{git status --porcelain}, cached), and an agent-specific overlay, so the \texttt{plan} and \texttt{explore} subagents receive entirely different system prompts from the default agent without any code branching.
Skill descriptions are appended dynamically.
This per-agent specialization is more granular than \clcode{}'s shared prompt with sub-agent prompt overrides, and structurally similar to \cdx{}'s model-specific templates, except keyed on \emph{agent profile} rather than on model generation.

\paragraph{Claude Code: Modular Assembly with Cache Boundary.}
\clcode{}'s prompt is assembled from a dozen-plus named sections, divided by a \texttt{SYSTEM\_PROMPT\_DYNAMIC\_BOUNDARY} marker.
Sections before the boundary (identity, system rules, task guidance, tool patterns, tone) are \emph{static} and cached via Blake2b hashing with \texttt{scope: 'global'}.
Sections after the boundary (session guidance, memory, environment, MCP instructions, language, scratchpad) are \emph{dynamic} and computed per-turn via \texttt{systemPromptSection()} with memoization.
This design minimizes prompt cache invalidation: the static prefix (which constitutes the majority of tokens) achieves high cache hit rates across turns and even across sub-agents (via \texttt{renderedSystemPrompt} sharing at fork time).

\paragraph{Codex: Per-Model Prompts as Server-Delivered Data.}
\cdx{} maintains separate prompts per model generation, and the delivery mechanism has itself co-evolved: the compiled-in Markdown templates of early versions still sit in the tree but are no longer referenced by code---the operative prompts ship as \texttt{base\_instructions} fields of a model manifest (\texttt{models.json}, bundled as fallback and refreshed from a remote \texttt{/models} endpoint with ETag caching).
The July manifest spans GPT-5.2 through a GPT-5.6 family, and the prompts are \emph{personality-templated}: a \texttt{\{\{ personality \}\}} variable is filled with user-selectable variants (friendly/pragmatic).
Prompt content drifts measurably across generations: GPT-5.2's prompt opens with version identity (``You are GPT-5.2 running in the Codex CLI\ldots'') and carries explicit no-commit and citation-ban rules; the GPT-5.5 prompt opens ``You are Codex, a coding agent based on GPT-5\ldots'' (further shortened by the 5.6 family to ``You are Codex, an agent based on GPT-5''), expands a rich Personality section, and \emph{drops} the no-commit rule and the anti-gold-plating directives entirely. Behavioral policy is migrating from prompt prose to feature flags: a \texttt{codex\_git\_commit} flag now governs commit behavior.
Section~\ref{sec:prompt-content} returns to this thinning.

\paragraph{OpenHands: Prompt Registry with Cache Tiers.}
\oh{} replaced its Jinja2 \texttt{PromptManager} with a \texttt{PromptRegistry} of guarded, ordered \texttt{PromptSection} objects bucketed into \texttt{CacheTier.STATIC} and \texttt{DYNAMIC}, rendered as a two-block system message whose static prefix stays byte-stable for provider prompt caching---the same static/dynamic cache-boundary design as \clcode{}'s, arrived at independently from within a LiteLLM-based multi-provider system.
Eighteen named static sections ship (Soul, Role, Memory, Security, SecurityRiskAssessment, VersionControl, and others), with Jinja templates surviving only as an escape hatch.

\paragraph{Hermes: Three-Tier Assembly with Bit-Perfect Prefixes.}
\hrm{} builds its prompt once per session in three tiers---stable, context, volatile---with $\sim$15 guidance blocks gated by model family (tool-use enforcement blocks are sent to GPT/Codex/Gemini/Grok/Qwen/DeepSeek models and never to Claude, with code comments citing observed per-model failures).
Cache economics drive the details: a single \texttt{system\_and\_3} strategy places four \texttt{cache\_control} breakpoints, and the prompt prefix is \emph{bit-perfect normalized} (sorted compact re-serialization of tool-call JSON, a date-only timestamp, a frozen memory snapshot) so that even local llama.cpp and vLLM KV caches hit across turns.

\paragraph{Pi: One Small Builder, Co-Varying with the Toolset.}
\pib{}'s entire system prompt is a $\sim$170-line builder emitting 30--40 lines: persona, tool list, deduplicated per-tool guideline snippets (each tool contributes \texttt{promptSnippet}/\texttt{promptGuidelines}, so the prompt co-varies with the active toolset), \texttt{<project\_context>} from AGENTS.md, an \texttt{<available\_skills>} XML index, date and cwd.
Almost all behavioral policy is deliberately absent, delegated to user-supplied context files and extensions.

\paragraph{OpenCode: A Model-Family Prompt Matrix.}
\ocd{} dispatches one of nine base prompts by model-id substring---\texttt{claude}, \texttt{gpt-}, \texttt{gemini}, and others each get dedicated .txt prompts with different registers (the Claude prompt opens ``You are OpenCode, the best coding agent on the planet.'')---generalizing \cdx{}'s per-generation templates across vendors.
Assembly deliberately caps the system prompt at two messages (one by default; plugin-expanded prompts are collapsed back), matching the two leading cache-breakpoint slots it emits in six provider dialects simultaneously.

\paragraph{Aider: Polymorphic Prompts.}
\adr{}'s prompt system is unique: each of its 13 registered edit formats (\texttt{architect}, \texttt{ask}, \texttt{context}, \texttt{diff}, \texttt{diff-fenced}, \texttt{editor-diff}, \texttt{editor-diff-fenced}, \texttt{editor-whole}, \texttt{help}, \texttt{patch}, \texttt{udiff}, \texttt{udiff-simple}, \texttt{whole}) has a dedicated \texttt{CoderPrompts} subclass defining system prompts, examples, and reminders tailored to that format.\footnote{Three function-calling coders remain in the tree unregistered---one commented out of \texttt{coders/\_\_init\_\_.py}, two simply never imported---so \texttt{Coder.create()} resolves exactly 13 formats; the April edition inconsistently reported 14.}
A factory pattern (\texttt{Coder.create()}) selects the format based on the model's metadata.
The prompt dynamically includes: file contents, a token-constrained repository map (generated via tree-sitter symbol extraction), chat history, and format-specific editing instructions.

\paragraph{Mini-SWE-Agent: One-Shot Templates.}
\mswe{} uses Jinja2 templates embedded in YAML configuration files, rendered once at session start.
Template variables include system information (\texttt{platform.uname()}), model statistics (\texttt{n\_model\_calls}, \texttt{model\_cost}), and the task description.
A notable detail: the default template includes OS-specific instructions (e.g., \texttt{sed -i ''} for macOS).

\subsection{Prompt Content and Rhetorical Style}
\label{sec:prompt-content}

The previous subsection examined \emph{how} prompts are assembled (cache boundaries, templates, hierarchical merges).
We now turn to \emph{what they actually say}: the persona each scaffold projects, the explicit directives it issues to the model, and the rhetorical devices it uses to make those directives stick.
We read the canonical system prompt of each system and extracted recurring axes.
Several patterns prove convergent across otherwise unrelated codebases.

\paragraph{Identity and persona.}
Persona openings range from terse to elaborate.
\mswe{} is the most minimal: ``You are a helpful assistant that can interact with a computer''---though \pib{}'s \emph{inner} agent library beats it (``You are a helpful assistant.''), with the coding layer adding only ``You are an expert coding assistant operating inside pi, a coding agent harness.''
\oh{} adds a role: ``You are OpenHands agent, a helpful AI assistant that can interact with a computer to solve tasks''---now wrapped in a \texttt{<SOUL>} block loaded from a user-overridable \texttt{SOUL.md}.
\cdx{}'s GPT-5.2 prompt is the most explicit about model coupling: ``You are GPT-5.2 running in the Codex CLI, a terminal-based coding assistant''---an opening the GPT-5.5 generation drops in favor of ``You are Codex, a coding agent based on GPT-5\ldots'' (the 5.6 family shortens further to ``an agent based on GPT-5'') with an elaborate Personality section.
\ocd{} varies persona \emph{by model}: Claude, Codex, and Meta models are told ``You are OpenCode, the best coding agent on the planet,'' while the default lineage gets the sober ``an interactive CLI tool that helps users with software engineering tasks.''
\hrm{} layers a general identity (``You are Hermes Agent\ldots created by Nous Research\ldots prioritize being genuinely useful over being verbose'') with a coding posture injected only when auto-detected: ``Operate like a careful senior engineer.''
\vibe{}, which in its April version opened with a complaint about prior behavior (``\texttt{CRITICAL:} Users complain you are too verbose''), now opens with a formal \emph{instruction-hierarchy contract}: seven precedence levels (critical $>$ user $>$ repo AGENTS.md $>$ user AGENTS.md $>$ prompt defaults $>$ skills/MCP $>$ external-data-as-data)---prompt-injection defense expressed as a ranking rather than prohibitions.
\adr{} eschews persona entirely in favor of a role assertion: ``Act as an expert software developer. Always use best practices when coding.''
\clcode{} sits between these poles, with a short opening followed by extensive elaboration in later sections.

\paragraph{Verbosity control.}
Nine of the eleven prompts contain explicit verbosity directives, but the implementations differ sharply.
\ocd{}'s is now the most aggressive prompt in the corpus: ``You MUST answer concisely with fewer than 4 lines\ldots One word answers are best,'' complete with \texttt{<example>} blocks (``user: what is 2+2? assistant: 4'')---though its Claude-specific prompt, notably, \emph{drops} the quantified rule that the fallback lineage retains.
\vibe{} keeps its 150-word budget (``Most tasks need under 150 words of prose'') and structure-first rule (``Structure first. Prose after, if at all'').
\clcode{}'s response-length rules are quantified ($\le 25$ words for inter-tool updates, $\le 100$ words for end-of-turn summaries)---though, our re-audit finds, that numeric section is gated to Anthropic-internal builds as an A/B experiment; external users receive qualitative guidance.
\cdx{} is qualitative in the 5.2 generation (``concise, direct, and friendly''); the 5.6 prompts fold verbosity into a Personality section, while 5.5 keeps its own length rules.
\hrm{} is qualitative (``Be concise: lead with the change or answer, not a preamble''), \pib{} minimal (a single bullet: ``Be concise in your responses''), \adr{} requests ``a few short sentences'' before any patch.
\mswe{} reaches the same target structurally rather than verbally: exactly one \texttt{THOUGHT} block followed by one bash command per turn.
\oh{} is alone in leaving verbosity largely to the model: its base prompt says nothing on length, though its GPT-5-family model-specific blocks do instruct concise responses and short preambles.

\paragraph{Banned phrases and forbidden surface forms.}
\vibe{} pioneered the forbidden-phrase list; its rewritten prompt merges two April lists into one: ``No filler words: `robust', `elegant', `seamless', `powerful', `Great!', `Absolutely!', `Of course!', `Happy to help!'\,''---a direct attempt to suppress LLM register tics.
\ocd{} now runs the corpus's second banned-phrase list, model-dependent: its GPT prompt bans openers like ``Done ---'' and ``Got it,'' along with emoji \emph{and em dashes}.
\vibe{} also retains the strictest emoji prohibition in the corpus: ``No emoji of any kind. No smiley faces, icons, flags, or Unicode symbols\ldots This applies to prose, code comments, and commit messages.''
\clcode{} is conditional (``Only use emojis if the user explicitly requests it''), and \ocd{} carries the identical conditional rule in its Claude, default, Meta, and Trinity prompts while its GPT prompt bans emoji outright and its GPT-4-era \texttt{beast} prompt instead \emph{instructs} emoji status markers; \gemini{}, \mswe{}, \adr{}, \hrm{}, and \pib{} say nothing about emoji.
\cdx{}'s 5.2 prompt forbids a CLI-specific surface form---inline citations of the shape \texttt{[F:README.md L5-L14]} (with bracketed citation glyphs)---because the Codex terminal cannot render them; the rule is absent from the 5.5/5.6 prompts.

\paragraph{Anti-gold-plating directives.}
A near-universal pattern: scaffolds tell the model not to enlarge the request.
\clcode{}: ``Don't add features, refactor code, or make `improvements' beyond what was asked,'' with a separate bullet against speculative abstractions.
\vibe{} (rephrased since April but preserved in substance): ``Change minimally\ldots Don't touch what wasn't asked. Unused imports may have side effects. Redundant-looking code may be load-bearing. When fixing X, leave Y alone.''
\cdx{} (5.2 generation): ``Do not attempt to fix unrelated bugs. \dots\ Fix the problem at the root cause rather than applying surface-level patches''---absent from the 5.4 prompt and dropped entirely by 5.6.
\oh{}: ``NEVER create multiple versions of the same file with different suffixes (e.g., \texttt{file\_test.py}, \texttt{file\_fix.py}, \texttt{file\_simple.py}).''
\ocd{}: ``NEVER create files unless they're absolutely necessary.''
\hrm{}: no drive-by refactors, renames, or reformatting beyond the request.
The convergence is striking---six independently developed scaffolds anticipate the same failure mode (LLMs over-deliver on under-specified requests) and respond with phrasing that, while not lifted verbatim, is rhetorically isomorphic.
\pib{} is the deliberate exception: its prompt contains no such directive, policy being delegated wholesale to user-supplied AGENTS.md and extensions.

\paragraph{Read-before-edit and verify-before-claim.}
Three prompts make code-reading a precondition for editing.
\clcode{}: ``In general, do not propose changes to code you haven't read.''
\vibe{} strengthened its version between snapshots: ``Never edit a file you have not read in this session. Do not edit a file in the same turn you first read it---read, then act on the next turn.''
\ocd{}'s edit-tool description claims enforcement (``This tool will error if you attempt an edit without reading the file'')---but no runtime read-tracking exists in the tool's source; the description overclaims, a reminder that prompt text is a behavioral wish, not a mechanism.
\clcode{} additionally forbids false success claims---``Never claim `all tests pass' when output shows failures, never suppress or simplify failing checks (tests, lints, type errors) to manufacture a green result''---in a section that, like its word budgets, ships to internal builds first; and \hrm{} is the only system to \emph{mechanize} the same demand, via the verify-on-stop guard of Section~\ref{sec:agentloop} plus a universal anti-fabrication directive (``NEVER substitute plausible-looking fabricated output\ldots'').
The other systems treat this as implicit.

\paragraph{Git commit policy: a convergence that dissolved.}
In April~2026 this was the corpus's tightest rhetorical convergence: every prompt that mentioned \texttt{git} forbade autonomous commits.
By July the picture has split three ways.
The prohibition survives in \clcode{} (``NEVER commit changes unless the user explicitly asks you to''), \ocd{} (``NEVER commit changes unless the user explicitly asks''), \hrm{} (``don't commit, push, or rewrite history unless asked, and never read, print, or commit secrets''), and \cdx{}'s 5.2-generation prompt.
\vibe{} reversed course: it deleted its labeled ``Never Commit'' hard rule (the changelog reads ``Loosened the no-git-commit constraint'') and now actively instructs the model how to commit, with a mandated \texttt{Co-Authored-By} signature trailer. \oh{} never carried a commit prohibition to reverse---its \texttt{<VERSION\_CONTROL>} section has taught commit mechanics with a co-author trailer since before the window---while gating \emph{push} and pull-request creation on explicit user request in a separate \texttt{<PULL\_REQUESTS>} section.
And \cdx{}'s newest prompts drop the rule entirely: the GPT-5.6 instructions contain no occurrence of ``commit.'' A \texttt{codex\_git\_commit} flag briefly governed commit-attribution behavior but is retired (unused) by the July snapshot---the directive simply left the prompt.
What remains universal is the outer boundary: no prompt in the corpus permits autonomous \emph{push}, force-push, or history rewriting.
The commit convergence of April, in other words, was not a stable engineering conclusion but a snapshot of trust calibration---and trust moved measurably in one quarter.

\paragraph{Emphasis markers.}
The corpus shows three distinct conventions for in-prompt emphasis.
\clcode{}, \cdx{}, \oh{}, and \ocd{} use uppercase tags inline: \texttt{IMPORTANT:}, \texttt{CRITICAL:}, \texttt{NEVER} (\ocd{} adds \texttt{<system-reminder>} XML injections at runtime).
\mswe{} uses XML-like tags (\texttt{<important>\dots</important>}), \pib{} uses XML only as data delimiters, and \hrm{} combines \texttt{MUST}/\texttt{NEVER} capitals with checkmark/cross exemplar pairs---shipping its XML-tagged enforcement blocks only to GPT/Codex and Grok models.
\adr{} uses almost none---its enforcement comes from the patch-format examples rather than from emphatic prose (a sparse \texttt{IMPORTANT:} survives in the patch-format instructions).
\vibe{} has moved furthest: no \texttt{CRITICAL:} tags and no Hard Rules section remain; emphasis is now fully \emph{structural}, an explicit overridability contract with named ``Critical instructions---not overridable'' and ``Overridable defaults'' sections.
The contrast suggests three schools: \emph{rhetorical} emphasis (uppercase tags), \emph{structural} emphasis (named rule blocks and precedence contracts), and \emph{example-driven} emphasis (exemplar pairs and format samples), with most systems combining at least two.

\paragraph{Tool-use philosophy.}
Tool guidance varies in specificity.
\clcode{} insists on dedicated tools over shell: ``Do NOT use the Bash tool to run commands when a relevant dedicated tool is provided. \dots\ Using dedicated tools allows the user to better understand and review your work.''
\cdx{} optimizes for parallelism: ``Parallelize tool calls whenever possible---especially file reads.''
\adr{}'s instructions are format-mechanical: ``Your entire response containing the patch MUST start with \texttt{*** Begin Patch} on a line by itself. \dots\ Each file MUST appear only once in the patch.''
\mswe{} is the most restrictive: exactly one bash block per turn, with directory and environment changes prefixed inline (\texttt{MY\_VAR=val cd /path \&\&\,\dots}) because each call runs in a new subshell.

\paragraph{What is conspicuously absent.}
None of the eleven prompts contains explicit harmful-request refusal language.
There are no instructions of the form ``if the user asks you to do X, refuse and explain why''---the only refusal-shaped text concerns operational risks (destructive git operations, secrets hygiene, dangerous shell patterns), not policy-level harms.
The nearest thing in the corpus is refusal \emph{etiquette}: \ocd{}'s default prompt inherits early \clcode{} phrasing instructing the model, \emph{if} it refuses, not to preach about why (``keep your response to 1-2 sentences'')---style guidance for a refusal whose grounds are never stated.
Safety against misuse is fully delegated to the underlying model's pretraining and any provider-side policy layer.
This is a notable design choice: production scaffolds, even those bundled with their own provider, do not duplicate alignment work in the system prompt---and the finding now holds at 11/11 across three more vendors and community lineages.

\observation{\label{obs:rhetoric}Prompt rhetoric converges where engineering experience converges, then \emph{thins} as trust calibrates. Anti-gold-plating language appears in nearly isomorphic phrasings across six independently developed scaffolds, and the no-autonomous-commit rule was universal in April~2026. By July, the corpus shows the rules' life cycle: \vibe{} reversed its commit prohibition into commit \emph{instruction} with a mandated co-author trailer (\oh{} had taught commit mechanics all along), and \cdx{}'s newest model generation dropped both the commit rule and the anti-gold-plating directives from its prompts---behavioral policy thinning as models internalize the norms. Three rhetorical strategies now coexist: harness-enforced policy prose (\clcode{}, \ocd{}, \hrm{}), structural contracts (\vibe{}'s seven-level instruction hierarchy and overridability sections), and near-total delegation to user context (\pib{}'s 30-line prompt). Model-conditional rhetoric has emerged as its own axis: \hrm{} gates enforcement blocks by model family with comments citing observed per-model failures, and \ocd{} maintains nine per-family prompts whose strictness varies by recipient. The stable invariant across all eleven: no policy-level refusal language anywhere---alignment work is never duplicated in the harness prompt.}

\subsection{Streaming and Response Processing}

Response processing strategies span a wide range:

\begin{itemize}[itemsep=1pt]
  \item \textbf{\clcode{}}: SSE with fine-grained events (\texttt{content\_block\_start/delta/stop}, \texttt{message\_delta}). Supports extended thinking via \texttt{thinking} blocks with configurable \texttt{budget\_tokens}.
  \item \textbf{\cdx{}}: WebSocket + SSE fallback. \texttt{ResponseItem} events parsed from JSON-RPC. Connection prewarm with \texttt{generate=false}. Turn state maintained via \texttt{x-codex-turn-state} header.
  \item \textbf{\gemini{}}: \texttt{@google/genai} SDK \texttt{chat.sendMessageStream()} with a four-attempt mid-stream retry loop (\texttt{MidStream\-RetryOptions}) recovering from content, network, and invalid-stream errors. Wraps the SDK in a custom \texttt{GeminiChat} class to work around a function-response validation bug. Keeps extended-thinking parts out of persisted history at response-recording time so they don't pollute future cache lookups.
  \item \textbf{\vibe{}}: Mistral SDK \texttt{stream\_async()} aggregates \texttt{LLMChunk} blocks; \texttt{ThinkChunk} blocks are routed into a separate \texttt{reasoning\_content} field. Aggregated usage stats are produced across the stream.
  \item \textbf{\adr{}}: Token-level streaming with \texttt{mdstream} for rich terminal rendering. Retry loop with exponential backoff; handles \texttt{FinishReasonLength} by resuming with assistant prefill.
  \item \textbf{\mswe{}}: Full response (no streaming). Simplest approach; relies on model's native completion.
  \item \textbf{\oh{}}: SSE via LiteLLM wrapper, now with first-class streaming (an \texttt{LLM.stream} field and token deltas forwarded through \texttt{on\_token} callbacks) plus an OpenAI Responses path. Supports prompt caching markers for Anthropic models.
  \item \textbf{\hrm{}}: five owned transport implementations; a $\sim$23-value \texttt{FailoverReason} enum drives classified recovery (compress vs.\ rotate credential vs.\ fallback chain vs.\ abort).
  \item \textbf{\pib{}}: nine wire-protocol implementations with a \texttt{partial-json} salvage parser for streamed tool arguments---paired with the truncation guard of Section~\ref{sec:agentloop}, since salvaged arguments can validate while incomplete.
  \item \textbf{\ocd{}}: Vercel AI SDK \texttt{streamText} owns dispatch; token-granular part deltas are persisted and re-served over SSE by the embedded server, so every client (TUI, web, IDE) replays the same stream.
  \item \textbf{\ocl{}}: ACP delta events for text, thought, and tool calls; provider-specific transport streams handle the underlying SSE/WebSocket layer.
\end{itemize}

\subsection{Advanced API Features}

The single-provider systems exploit features unavailable to multi-provider abstractions:

\clcode{}'s extended-thinking integration exposes a \texttt{budget\_tokens} parameter that controls how much internal reasoning the model performs before responding; while LiteLLM does pass through Anthropic's \texttt{thinking} block, surfacing \texttt{budget\_tokens} as a first-class scaffold knob still requires Anthropic-specific conditional code in any multi-provider design.
\cdx{}'s reasoning-effort enum spans \texttt{minimal}/\texttt{low}/\texttt{medium}/\texttt{high}/\texttt{xhigh}, extended in mid-2026 with \texttt{max} and \texttt{ultra}---and \texttt{ultra} is not just more reasoning: the model catalog describes it as maximum reasoning \emph{with automatic sub-agent task delegation}, an effort level that changes orchestration behavior, unique in the corpus.
\vibe{}'s per-model \texttt{thinking} field now spans five levels (\texttt{off}/\texttt{low}/\texttt{medium}/\texttt{high}/\texttt{max}) mapped onto Mistral's \texttt{reasoning\_effort} enum, and \pib{} normalizes the whole landscape behind a unified seven-level scale translated through ten per-API \texttt{thinkingFormat} dialects.
\cdx{} implements prompt caching via a \texttt{prompt\_cache\_key} derived from the session's \texttt{ThreadId} (with an explicit override hook for delegate and guardian sessions); this is a lighter-weight strategy than \clcode{}'s Blake2b-hashed static/dynamic boundary.
\adr{}'s vision support warrants a note: when the active model's \texttt{supports\_vision} attribute is set (via LiteLLM metadata), \adr{} accepts image files in the chat, handles MIME types, and passes them to the model.
The support is model-gated rather than unconditional, hence marked ``Opt.'' in Table~\ref{tab:api_features}; \vibe{} joined the vision column in v2.14--2.18 (TUI @-mentions, clipboard paste, ACP inline image blocks), and \mswe{} retains an opt-in, default-off multimodal path (a \texttt{multimodal\_regex} expanding tagged content into image blocks), so no corpus system is unconditionally text-only.
\gemini{} exposes extended thinking (with thought-part stripping to keep caches clean) and then routes the choice of model itself through \texttt{ModelRouterService}; that treatment of per-request model selection as classifier-driven runtime scheduling remains unique in the corpus (\oh{}'s RouterLLM profiles route by configuration, not per-query classification), now extending to a managed \emph{on-device} runtime (a \texttt{gemini gemma} command group provisions a local LiteRT-LM server, and the routing chain can use the local Gemma model as its query classifier).
On the observability side, \gemini{} emits OpenTelemetry traces alongside per-token cost accounting.
\vibe{} also instruments its agent loop, tool execution, and hooks with OpenTelemetry (\texttt{agent\_span}/\texttt{tool\_span}/\texttt{hook\_span} async context managers with an OTLP exporter and GenAI semantic-convention attributes), and \pib{} contributes the corpus's most unusual observability metric: a per-turn \emph{cache-miss dollar-waste audit} that prices every avoidable cache invalidation.

\section{Tool and Action Systems}
\label{sec:tools}

The tool system defines what the agent can actually do: which actions it can invoke and how they are executed.
This subsystem absorbs a large share of the engineering effort in every mature agent we studied.

\subsection{Tool Count Spectrum}

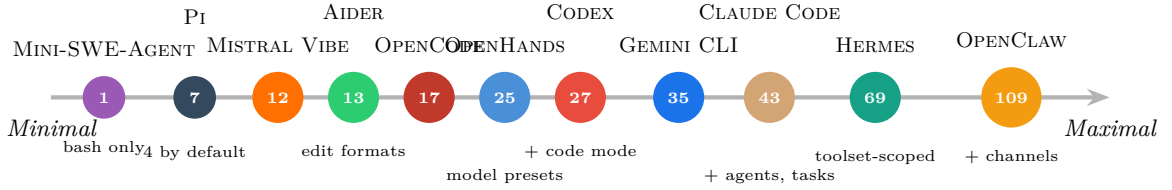
\begin{figure}[H]
\centering
\begin{tikzpicture}
  \draw[-{Stealth[length=3mm]}, ultra thick, gray!60] (0,0) -- (14,0);
  \node[font=\footnotesize\itshape] at (0,-0.4) {Minimal};
  \node[font=\footnotesize\itshape] at (14,-0.4) {Maximal};

  \node[circle, fill=miniswe, minimum size=0.55cm, font=\tiny\bfseries, text=white] (ms) at (0.7,0) {1};
  \node[font=\scriptsize, above=0.12cm of ms] {\mswe{}};

  \node[circle, fill=pi, minimum size=0.55cm, font=\tiny\bfseries, text=white] (pp) at (1.9,0) {7};
  \node[font=\scriptsize, above=0.55cm of pp] {\pib{}};

  \node[circle, fill=vibe, minimum size=0.55cm, font=\tiny\bfseries, text=white] (vb) at (3.0,0) {12};
  \node[font=\scriptsize, above=0.12cm of vb] {\vibe{}};

  \node[circle, fill=aider, minimum size=0.55cm, font=\tiny\bfseries, text=white] (ai) at (4.0,0) {13};
  \node[font=\scriptsize, above=0.55cm of ai] {\adr{}};

  \node[circle, fill=opencode, minimum size=0.55cm, font=\tiny\bfseries, text=white] (od) at (5.0,0) {17};
  \node[font=\scriptsize, above=0.12cm of od] {\ocd{}};

  \node[circle, fill=openhands, minimum size=0.55cm, font=\tiny\bfseries, text=white] (oh) at (6.0,0) {25};
  \node[font=\scriptsize, above=0.12cm of oh] {\oh{}};

  \node[circle, fill=codex, minimum size=0.55cm, font=\tiny\bfseries, text=white] (cx) at (7.0,0) {27};
  \node[font=\scriptsize, above=0.55cm of cx] {\cdx{}};

  \node[circle, fill=gemini, minimum size=0.55cm, font=\tiny\bfseries, text=white] (gm) at (8.3,0) {35};
  \node[font=\scriptsize, above=0.12cm of gm] {\gemini{}};

  \node[circle, fill=claudecode, minimum size=0.55cm, font=\tiny\bfseries, text=white] (cc) at (9.5,0) {43};
  \node[font=\scriptsize, above=0.55cm of cc] {\clcode{}};

  \node[circle, fill=hermes, minimum size=0.55cm, font=\tiny\bfseries, text=white] (hm) at (10.9,0) {69};
  \node[font=\scriptsize, above=0.12cm of hm] {\hrm{}};

  \node[circle, fill=openclaw, minimum size=0.55cm, font=\tiny\bfseries, text=white] (oc) at (12.7,0) {109};
  \node[font=\scriptsize, above=0.12cm of oc] {\ocl{}};

  \node[font=\tiny, below=0.15cm of ms] {bash only};
  \node[font=\tiny, below=0.15cm of pp, yshift=-2pt] {4 by default};
  \node[font=\tiny, below=0.15cm of ai] {edit formats};
  \node[font=\tiny, below=0.15cm of oh, yshift=-9pt] {model presets};
  \node[font=\tiny, below=0.15cm of cx] {+ code mode};
  \node[font=\tiny, below=0.15cm of cc, yshift=-9pt] {+ agents, tasks};
  \node[font=\tiny, below=0.15cm of hm, yshift=-2pt] {toolset-scoped};
  \node[font=\tiny, below=0.15cm of oc] {+ channels};
\end{tikzpicture}
\caption{Tool count spectrum (July~2026). Numbers indicate distinct built-in tools or edit formats; counts for \oh{} and \cdx{} are default surfaces (their full catalogs reach 30+). \pib{} builds 7 and exposes 4 by default; \hrm{} registers 69 but scopes visibility per platform through a toolset layer.}
\label{fig:tool_spectrum}
\end{figure}

Figure~\ref{fig:tool_spectrum} illustrates the spectrum from minimalist to maximal tool designs.
The systems span from a single bash tool to 109+ tools including media processing and channel integrations---with \pib{} making the sparse end a philosophy (seven tools built, four exposed by default, everything else an extension) and \hrm{} pushing the general-purpose end to 69 built-ins behind a toolset layer that scopes what each platform's model actually sees.

\subsection{Tool Interface Architectures}

Tool interfaces range from almost no abstraction---\mswe{} sends everything through a single shell call (\texttt{subprocess.Popen(shell=True)}, wrapped only to kill the process group on timeout)---to rich typed contracts.
\clcode{}'s 43 tools each declare validation, permission checking, concurrency safety, and UI rendering as separate interface concerns.
\cdx{} uses Rust trait objects for runtime polymorphism---the registry stores \texttt{Arc<dyn CoreToolRuntime>} where \texttt{CoreToolRuntime: ToolExecutor}, the abstraction now extracted into a standalone \texttt{codex-tools} crate shared with code mode and extensions.
\gemini{} routes calls through a state-machine \texttt{Scheduler} that walks each invocation through deterministic \emph{Validating~$\to$ Executing~$\to$ Completed/Errored} phases.
\vibe{} adds \emph{tree-sitter-parsed bash}---an \texttt{asyncio.create\_subprocess\_shell}-based execution path that uses \texttt{tree\_sitter\_bash} to validate command structure before spawning---and fine-grained permission scopes (command pattern, file pattern, URL allowlist, directory containment).
\hrm{}'s singleton registry separates \emph{registration} from \emph{exposure}: 69 tools self-register at import, a toolset layer scopes what each platform's model sees, per-tool availability probes (30-second TTL with a last-good grace window) keep one flaky \texttt{docker version} from stripping a toolset, and a coercion layer repairs schema drift (\texttt{"42"}$\to$42) motivated by observed open-model output.
\pib{} gives each of its seven tools a pluggable \texttt{Operations} interface (\texttt{BashOperations}, \texttt{EditOperations}, \ldots)---the single remoting seam through which SSH, container, and micro-VM extensions relocate execution without touching the tools.
\ocd{} registers 17 first-party tools with Effect schemas and swaps the \emph{tool surface} per model: GPT-family models receive a port of \cdx{}'s \texttt{apply\_patch} DSL and lose \texttt{edit}/\texttt{write} entirely (Section~\ref{sec:editing}).
The spectrum reflects the simplicity-versus-capability axis: more interface surface means more safety guarantees but more engineering cost.

\subsection{Deferred Tool Loading}
\label{sec:deferred}

\clcode{} introduced \emph{deferred tool loading}: tools marked with \texttt{shouldDefer = true} are excluded from the initial system prompt.
The LLM discovers them on-demand via \texttt{ToolSearchTool}, which supports keyword search and direct selection (\texttt{select:<tool\_name>}).
This reduces prompt size significantly: of 43 tools, only a core subset is loaded initially.
The deferred tool set is cached via \texttt{getDeferredToolsCacheKey()}, invalidated when the set changes.

The April edition presented this as a \clcode{} exclusive; by July it is a spreading pattern with independent implementations.
\cdx{} marks tools (by default, MCP tools) with a \texttt{defer\_loading} flag and exposes a \texttt{tool\_search} tool backed by a \emph{BM25 lexical index} over tool specs---the same prompt economics with ranked search instead of substring matching.
\hrm{} generalizes the idea with a threshold: when MCP and plugin schemas would exceed 10\% of the context window, they collapse into three bridge tools (\texttt{tool\_search}/\texttt{tool\_describe}/\texttt{tool\_call}) searched by an inlined BM25 engine.
\ocd{} defers \emph{skills} (catalog lists name and description only; a native \texttt{skill} tool fetches bodies) and treats its experimental code mode as MCP-catalog deferral: instead of exposing every MCP tool, one \texttt{execute} tool runs model-written scripts against the catalog.
\vibe{}'s \texttt{defer\_mcp} flag, by contrast, defers only MCP \emph{connection} at startup for latency, not prompt visibility---once integrated, every MCP tool is model-visible, so we no longer group it with the prompt-level deferral pattern.

\subsection{File Editing Strategies}
\label{sec:editing}

File editing is the core action for SWE agents.
The eleven systems implement eight fundamentally different strategies, summarized in Table~\ref{tab:editing}.

\begin{table}[H]
\centering
\caption{File editing strategies across the eleven systems (July~2026).}
\label{tab:editing}
\scriptsize
\renewcommand{\arraystretch}{1.25}
\begin{tabularx}{\textwidth}{@{}l L{2.9cm} L{3.1cm} L{2.5cm} L{3.4cm}@{}}
\toprule
\textbf{System} & \textbf{Format} & \textbf{Matching} & \textbf{Fallback} & \textbf{Innovation} \\
\midrule
\clcode{} & Exact string replacement & Unique substring in file & Error + context & Zod schema validation \\
\cdx{}    & \texttt{*** Begin/End Patch} & Unified diff hunks & Hunk-level retry & Custom patch DSL; Lark-grammar-constrained variant \\
\adr{}    & 13 formats (SEARCH/ REPLACE, udiff, diff-fenced, \ldots) & Exact $\to$ fuzzy (dmp) & Similar line suggestions & RelativeIndenter \\
\oh{}     & str\_replace\_editor (5 cmds) + \texttt{apply\_patch} + Gemini-ported \texttt{edit} & Exact unique match & Error message & undo\_edit; three coexisting edit dialects \\
\gemini{} & \texttt{edit} (old/new string) + \texttt{write\_file} & Cascading: exact $\to$ flexible $\to$ regex $\to$ fuzzy & LLM ``edit fixer'' subcall repairs the pair & LLM-assisted edit repair \\
\vibe{}   & Exact \texttt{edit} (old/new string, \texttt{replace\_all}) + create-only \texttt{write\_file} & Exact unique substring & Tool error on ambiguity & Migrated \emph{off} fuzzy SEARCH/REPLACE (Apr$\to$Jul) \\
\mswe{}   & sed/awk via bash & Regex-based & Shell error & None (model-driven) \\
\hrm{}    & One \texttt{patch} tool: SEARCH/REPLACE + V4A DSL (+ \texttt{write\_file}) & 9-strategy fuzzy chain (exact $\to$ \ldots $\to$ block-anchor 0.50/0.70 $\to$ context-aware 0.80) & Closest-line suggestions; escalate to \texttt{write\_file} after 3 failures & Two-phase validate-then-apply; backend-transparent shell file ops \\
\pib{}    & Multi-edit exact replacement (\texttt{edits[]}, overlap rejection) & Exact $\to$ Unicode/whitespace canonicalization (no similarity threshold) & Uniqueness guidance error & Byte-fidelity overlay; async diff preview before execution \\
\ocd{}    & Model-conditional: \texttt{apply\_patch} DSL (GPT family) or SEARCH/REPLACE \texttt{edit} (others) & 9-stage replacer cascade (Levenshtein 0.65) / 4-pass patch relaxation & Distinct not-found vs.\ ambiguous errors & Editing dialect swapped per model in the registry; LSP diagnostics appended to results \\
\ocl{}    & N/A & N/A & N/A & Not a code editor \\
\bottomrule
\end{tabularx}
\end{table}

\paragraph{The editing landscape reorganized between snapshots.}
The April edition described \vibe{} and \gemini{} as a ``polymorphic-by-degree'' pair---coarse whole-file tool plus fuzzy fine-grained patch tool.
Neither half survived the quarter.
\vibe{} deleted its SEARCH/REPLACE tool outright and converged on \clcode{}'s exact unique-substring contract (with a \texttt{replace\_all} escape hatch and a now create-only \texttt{write\_file}): a rare observable case of a production harness migrating between editing-strategy clusters, and evidence that stronger models shift the optimum from tool-side drift tolerance toward strict contracts.\footnote{The April text also overstated the old mechanism: re-audit shows v2.7.5's $\geq$0.90 fuzzy matcher only generated ``closest match'' diagnostics for error messages; it never applied fuzzy edits.}
\gemini{}'s \texttt{edit}, on closer inspection, is an old/new-string tool with a cascading matcher (exact $\to$ whitespace-flexible $\to$ regex $\to$ fuzzy) that ends in something no other system had: an \emph{LLM edit-fixer subcall} that repairs the failed old/new pair against a model-supplied \texttt{instruction}---the corpus's first in-tool LLM-assisted edit repair.

\paragraph{The fuzzy-cascade family and its visible lineage.}
Where \vibe{} left, two newcomers arrived with the corpus's most elaborate drift tolerance, and their source code documents its ancestry.
\ocd{}'s nine-stage replacer cascade (exact $\to$ line-trimmed $\to$ block-anchor $\to$ whitespace/indentation/escape-normalized $\to$ context-aware, under Levenshtein~0.65 with a disproportionate-match guard) carries header comments crediting Cline and \gemini{} evals; \hrm{}'s nine-strategy chain is annotated ``inspired by OpenCode.''
Editing machinery now has a visible cross-harness genealogy---convergence by inheritance, not just rediscovery.
\pib{} stakes out a third position: no similarity threshold at all, only Unicode/whitespace \emph{canonicalization} (NFKC, smart quotes, trailing whitespace), with replacements computed in normalized space overlaid line-wise onto the original bytes so untouched lines keep their exact original whitespace---plus an async diff preview rendered before the tool executes.

\adr{}'s \texttt{RelativeIndenter} still deserves special attention.
It converts absolute indentation to relative changes using Unicode markers (default: left arrow $\leftarrow$, selected to avoid collisions with file content).
The \texttt{make\_relative()} method tracks indent differences between consecutive lines, and \texttt{make\_absolute()} reconstructs original indentation.
This enables edit blocks to work across different indentation levels, a common failure mode for other systems.
Combined with fuzzy matching via \texttt{diff\_match\_patch} (threshold 0.95, distance 500), this creates a robust editing pipeline.

\observation{\label{obs:editing}The file-editing strategy is one of the most important determinants of code-modification accuracy in an SWE agent, and model-aware polymorphism is no longer \adr{}'s alone. \adr{} selects the edit \emph{format} per model through a prompt-class factory; \ocd{} reaches the same insight through the tool registry, swapping the \emph{toolset} per model (GPT-family models get a \cdx{}-style patch DSL and lose the string-replacement tools entirely). The corpus's editing machinery now shows explicit cross-harness lineage (\hrm{}'s matcher ``inspired by OpenCode''; \ocd{}'s cascade crediting Cline and \gemini{}), and it is in visible motion: \vibe{} migrated from fuzzy SEARCH/REPLACE to \clcode{}-style exact matching within one quarter, while \gemini{} added LLM-assisted edit repair. \adr{}'s \texttt{RelativeIndenter} remains the most sophisticated solution to indentation-sensitive matching.}

\subsection{Execution and Sandboxing}

Table~\ref{tab:sandbox} compares sandboxing approaches in detail.

\begin{table}[H]
\centering
\caption{Execution sandboxing mechanisms (July~2026).}
\label{tab:sandbox}
\small
\renewcommand{\arraystretch}{1.2}
\begin{tabularx}{\textwidth}{@{}l L{4.4cm} L{3.2cm} L{2.4cm}@{}}
\toprule
\textbf{System} & \textbf{Mechanism} & \textbf{Isolation Level} & \textbf{Investment} \\
\midrule
\oh{}     & \texttt{Workspace} abstraction: Docker, Apptainer, remote-API, cloud workspaces (KVM passthrough app-side) & Process + filesystem + network & Substantial \\
\cdx{}    & Bubblewrap (vendored, compiled in-tree) + Landlock legacy fallback (Linux), Seatbelt (macOS), restricted tokens (Windows) & OS-level namespaces + filesystem + network & Substantial \\
\gemini{} & Docker / Bubblewrap / gVisor / LXC (Linux), Seatbelt via \texttt{sandbox-exec} (macOS), restricted tokens (Windows); TOML policies & OS-level + per-mode policy + env scrubbing & Moderate \\
\clcode{} & Opt-in OS sandbox via Anthropic's \texttt{sandbox-runtime} (Bubblewrap/Seatbelt) with network restrictions; git worktrees for branch isolation & OS-level (opt-in) + filesystem & Moderate \\
\mswe{}   & Docker, Singularity, Bubblewrap (pluggable) & Configurable & Light \\
\hrm{}    & Six pluggable execution backends (local, Docker, SSH, Singularity, Modal, Daytona); local backend = policy floor, no OS primitives & Per-backend (container/VM); none locally & Delegated \\
\vibe{}   & None OS-level; opt-in \texttt{--worktree} branch isolation & Filesystem (branch-level, opt-in) & None \\
\pib{}    & None built-in, with documented rationale; opt-in extensions (Anthropic \texttt{sandbox-runtime}, QEMU micro-VM) via per-tool \texttt{Operations} seam & None by default & None (extension-space) \\
\ocd{}    & None (permission rules + tree-sitter command parsing; shadow-git checkpoints; ``sandbox'' in-tree means worktrees) & None & None (policy-only) \\
\adr{}    & None (direct local execution) & None & None \\
\ocl{}    & None (local gateway execution) & None & None \\
\bottomrule
\end{tabularx}
\end{table}

\cdx{} is the most serious sandboxing investment in the corpus, now packaged as a dedicated \texttt{sandboxing} crate.
On Linux it runs Bubblewrap---since mid-2026 \emph{vendored and compiled in-tree}, invoked through C FFI---for namespace isolation (\texttt{--ro-bind}, \texttt{--unshare-net}, \texttt{--unshare-user}, \texttt{--unshare-pid}), with Landlock demoted to a legacy fallback behind a feature flag.
On macOS it drives Seatbelt profiles through \texttt{/usr/bin/sandbox-exec}.
On Windows it uses restricted-token processes.
Protected paths (\texttt{.git}, \texttt{.agents}, \texttt{.codex}) are re-marked read-only even when they sit inside a writable root.
The result is OS-level isolation without the overhead of a container runtime.

\paragraph{Gemini CLI joins Codex as the second cross-platform sandboxer.}
\gemini{}'s \texttt{SandboxManager} implements the same three-platform stack as \cdx{}: Docker or Bubblewrap on Linux, Seatbelt via \texttt{/usr/bin/sandbox-exec} on macOS, restricted-token processes on Windows.
Where \cdx{} bakes its policy into Starlark \texttt{execpolicy} rules (Section~\ref{sec:safety}), \gemini{} layers a TOML-based per-mode sandbox policy (\texttt{plan} / \texttt{default} / \texttt{accepting\_edits}) plus environment scrubbing of provider credentials (name-pattern regexes over \texttt{TOKEN}/\texttt{SECRET}/\texttt{KEY}/\texttt{AUTH}/\texttt{CREDENTIAL}, a denylist, and GitHub-token value regexes) before each command runs.
\gemini{}'s sandbox investment is meaningfully lighter than \cdx{}'s because \gemini{} reuses Node's child-process API and OS-provided binaries rather than implementing low-level namespace plumbing itself; \cdx{} pays for its larger footprint in vendored-Bubblewrap integration and Rust-native namespace control.
The April edition observed that the two largest systems in the corpus were also the two with native cross-platform sandboxing and asked whether the correlation was structural; the expanded corpus answers \emph{no}: \hrm{} ($\sim$642K lines) ships zero OS-level isolation primitives, delegating containment to six pluggable execution backends while investing its safety budget in content-borne threats, and \ocd{} ($\sim$578K lines) is policy-only.
Size predicts \emph{some} large safety investment, not specifically a sandbox; Section~\ref{sec:safety} maps where each system spends instead.

\section{Memory and Context Management}
\label{sec:memory}

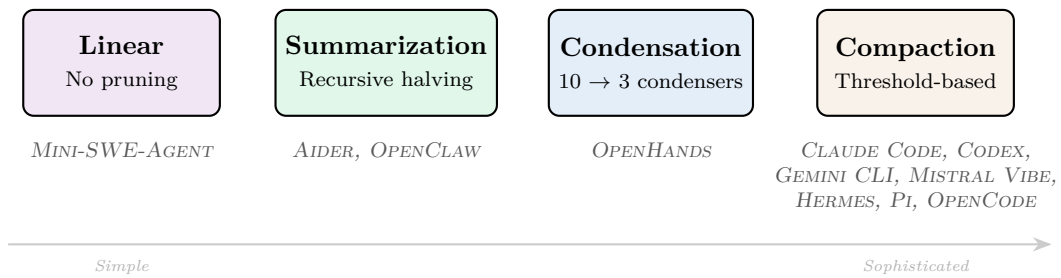
\begin{figure}[H]
\centering
\begin{tikzpicture}[
  strat/.style={rectangle, draw, rounded corners=4pt, minimum width=2.6cm, minimum height=1.4cm, font=\small, align=center, thick},
  label/.style={font=\scriptsize\itshape, text=gray!70!black},
]
  \node[strat, fill=miniswe!15] (linear) at (0,0) {\textbf{Linear}\\\scriptsize No pruning};
  \node[strat, fill=aider!15] (summary) at (3.5,0) {\textbf{Summarization}\\\scriptsize Recursive halving};
  \node[strat, fill=openhands!15] (condense) at (7.0,0) {\textbf{Condensation}\\\scriptsize 10 $\to$ 3 condensers};
  \node[strat, fill=claudecode!15] (compact) at (10.5,0) {\textbf{Compaction}\\\scriptsize Threshold-based};

  \node[label, below=0.2cm of linear] {\mswe{}};
  \node[label, below=0.2cm of summary] {\adr{}, \ocl{}};
  \node[label, below=0.2cm of condense] {\oh{}};
  \node[label, below=0.2cm of compact, align=center, text width=3.8cm] {\clcode{}, \cdx{}, \gemini{}, \vibe{}, \hrm{}, \pib{}, \ocd{}};

  \draw[-{Stealth[length=2.5mm]}, thick, gray!40] (-1.5,-2.4) -- (12.3,-2.4);
  \node[font=\tiny\itshape, gray!60] at (0,-2.7) {Simple};
  \node[font=\tiny\itshape, gray!60] at (10.5,-2.7) {Sophisticated};
\end{tikzpicture}
\caption{Memory management strategies ordered by sophistication. Seven of the eleven systems---the four provider-native harnesses and all three systems added in this revision---converge on threshold-triggered LLM compaction: a de facto standard for production-grade context management, though each newcomer hybridizes it with traits of the other strategies (iterative summary merging, pluggable engines, session-tree substrates).}
\label{fig:memory}
\end{figure}

Every agent we looked at runs into the same problem eventually: the conversation outgrows the context window.
The literature offers two broad families of solutions: tiered memory architectures (such as MemGPT~\cite{packer2023memgpt}, which shuffles data between an in-context working set and an external store) and prompt-compression methods (such as LLMLingua~\cite{jiang2023llmlingua}, which shortens text before it reaches the model).
The eleven systems in the corpus implement four practical strategies (Figure~\ref{fig:memory}), which we present in roughly increasing order of sophistication.

\subsection{Strategy Taxonomy}

\subsection{Linear History (Mini-SWE-Agent)}

\mswe{} makes no attempt to manage context.
The full message list grows unboundedly, relying on the LLM's native context window (up to 1M tokens with the Claude~4 family).
This simplicity enables perfect reproducibility and straightforward trajectory analysis, a deliberate choice for a research baseline system.

\subsection{Recursive Summarization (Aider)}

\adr{}'s \texttt{ChatSummary} class implements a recursive halving algorithm.
When the message history exceeds \texttt{max\_tokens} (default: 1024), it splits messages into two halves by token count, keeping the tail (most recent 50\%) intact.
The head is summarized via an LLM call (\texttt{summarize\_all()}).
If the combined summary~+~tail still exceeds the limit, the process recurses up to 3 levels.
A separate ``weak model'' (e.g., GPT-3.5-turbo) can be configured for cost-efficient summarization.

\subsection{Pluggable Condensation (OpenHands)}

\oh{} pioneered pluggable memory management through an abstract \texttt{Condenser} base class.
The V0 application shipped ten implementations; the V1 SDK consolidated the zoo to three (\texttt{NoOpCondenser}, \texttt{LLMSummarizingCondenser}, \texttt{PipelineCondenser}) over a \texttt{RollingCondenser} base---a telling simplification, most of the specialized variants having proven unnecessary.
Two upgrades came with the consolidation: condensation is itself \emph{event-sourced} (\texttt{CondensationRequest}/\texttt{Condensation} events in the log, so compression history is replayable), and the condenser doubles as \emph{error recovery}---context-window overflows and malformed-history errors trigger a condensation rather than a crash.
The summarizing condenser also gained a token-count trigger (\texttt{max\_tokens}) alongside its event-count threshold, aligning \oh{} with the threshold-compaction family below.

\subsection{Threshold Compaction (Seven Systems)}

\clcode{}'s compaction triggers when the running token estimate enters a buffer below the model's effective context window (\texttt{AUTOCOMPACT\_BUFFER\_TOKENS = 13\,000}, with an env-overridable percent threshold for testing).
The process: (1)~strip images from messages (replace with \texttt{[image]} markers), (2)~group messages by API round, (3)~invoke the LLM to generate a conversation summary, (4)~create a \texttt{SystemCompactBoundaryMessage} separating pre-compaction history from the summary.
Post-compaction cleanup restores up to 5 files (\texttt{POST\_COMPACT\_TOKEN\_BUDGET = 50\,000}) and truncates skills to a per-skill 25\,000-token budget.

\cdx{} supports both local in-process and cloud-based summarization via a dedicated \texttt{/responses/compact} endpoint, with \texttt{model\_auto\_compact\_token\_limit} controlling when summarization triggers (the separate \texttt{/memories/\allowbreak trace\_summarize} endpoint serves the memories pipeline of Section~\ref{sec:memory_pipelines}); a remote-compaction v2 path and per-thread rollout token budgets (with remaining-budget reminders and turn aborts on exhaustion) were added in the spring 2026 releases.

\paragraph{Gemini CLI: Multi-Stage Context Pipeline.}
\gemini{}'s context management is more sophisticated than simple compression.
At the top level, a \texttt{ChatCompressionService} fires when history exceeds 50\% of the model's token limit, calls Gemini to generate a state snapshot, and \emph{preserves the most recent 30\%} of history verbatim (a token-budget truncation pre-pass and a \texttt{PreCompress} hook now fire before compression).
Beneath this, a dedicated context-processors module implements a multi-stage context distillation pipeline with pluggable processors (rolling summary, node distillation, node truncation, blob degradation, tool masking, state snapshot), consolidated in v0.43--0.45 into a \texttt{ContextManager} that maintains a \emph{pristine-versus-active} context graph in a working buffer, runs processors through an orchestrator with hysteresis-based trigger damping, and replaces static chars-per-token estimates with an adaptive token calculator calibrated against observed API token counts.
A separate \texttt{ContextCompressionService} tracks per-file compression levels (FULL, PARTIAL, SUMMARY, EXCLUDED), providing finer granularity than \clcode{}'s binary compacted/uncompacted split, and a pluggable \texttt{AgentHistoryProvider} interface lets extensions substitute their own compression logic.

\paragraph{Mistral Vibe: Middleware-Triggered, Now Two-Tier and Reactive.}
\vibe{}'s \texttt{AutoCompactMiddleware} checks \texttt{context\_tokens >= auto\_compact\_threshold} before each turn; the compaction routine itself, extracted into a \texttt{CompactionManager}, has grown three refinements since April: the post-compaction state is a context \emph{envelope} that re-injects prior user messages alongside the summary (so original task goals survive resets), a cache-friendly primary summarization falls back to a dedicated tool-less summarizer, and the same routine fires \emph{reactively} when a turn overflows the context window mid-flight.
The integration through the middleware pipeline remains architecturally cleaner than in-loop checks: compaction is just another turn policy.
\vibe{}'s complementary \texttt{RewindManager} creates filesystem-level checkpoints per user message and supports rewinding with optional file restoration---now forkable to a new session and exposed \emph{over ACP} so IDE clients can drive the rewind.
Once unique, checkpointing now has a finer-grained sibling: \ocd{} maintains a \emph{shadow git repository} per project (separate \texttt{--git-dir} over the real worktree) snapshotting at pre-stream, step start, and step finish, yielding per-step patch parts and full conversation-plus-filesystem revert.

\paragraph{Hermes: Lineage Compaction.}
\hrm{} lands in the threshold family (trigger at 50\% of context minus max-tokens, 64K floor; protect the first three messages plus a token-budgeted tail; iterative auxiliary-model summary capped at min(5\% of context, 12K tokens)) but adds a construct no other system has: compaction does not rewrite the transcript---it \emph{ends the session}.
Each compaction closes the current SQLite session (\texttt{end\_reason='compression'}) and rotates to a child chained by \texttt{parent\_session\_id}; lineage helpers walk the ancestry, and session search deduplicates across it.
Context management and session persistence become one mechanism, and no history is ever destroyed.
The engine is pluggable (a \texttt{ContextEngine} ABC), echoing \oh{}'s condensers.

\paragraph{Pi: Compaction over a Session Tree.}
\pib{} triggers at \texttt{contextWindow} $-$ 16{,}384 tokens (keeping 20K recent), with two traits borrowed from the other strategies: summaries are \emph{iteratively merged}---the update prompt re-ingests the previous summary rather than re-summarizing from scratch (a recursive-summarization trait)---and a \texttt{session\_before\_compact} hook lets extensions veto or replace the result entirely (a pluggable-condensation trait).
The distinctive substrate is the session itself: an append-only JSONL \emph{tree} where every entry carries \texttt{id}/\texttt{parentId} and a movable leaf pointer defines the active branch.
\texttt{/tree} navigation, \texttt{/fork}, and \texttt{/clone} unify what other harnesses implement as three separate features (checkpointing, rewind, alternative exploration), and abandoning a branch optionally generates an LLM \emph{branch summary} spliced into the new position, so exploration is never lost---with the explicit trade-off that filesystem state is not restored.
Cumulative \texttt{<read-files>}/\texttt{<modified-files>} lists persist as structured memory across compactions.

\paragraph{OpenCode: Anchored Incremental Summaries.}
\ocd{} triggers at \texttt{limit.input} minus a reserved-output margin, using a chars/4 estimate (no tokenizer exists anywhere in the tree), and produces the corpus's most explicitly \emph{incremental} summary: the previous summary is passed back in a \texttt{<previous-summary>} block and merged (``preserve still-true details, remove stale details'') by a hidden, all-tools-denied \texttt{compaction} agent, into mandated Markdown sections (Objective / Important Details / Work State / Next Move / Relevant Files).
A verbatim tail of two user turns is preserved; opt-in pruning soft-deletes old tool outputs, protecting the newest 40K tokens and firing only if more than 20K are reclaimable.

\subsection{Persistent Memory Pipelines}
\label{sec:memory_pipelines}

The April edition of this study could describe cross-session memory in one sentence per system: a Markdown context file here, a transcript store there.
Three months later, persistent memory is where the provider-native systems are differentiating most aggressively, and the designs diverge instructively on a single axis: \emph{who writes the memory, and who reviews it}.

\cdx{} runs the most autonomous design.
A background two-phase pipeline extracts structured memories from recent session rollouts (per-rollout LLM extraction into a SQLite state database, with leased jobs, retry backoff, and secret redaction) and then \emph{consolidates} them into filesystem artifacts under a git-baselined \texttt{\textasciitilde/.codex/memories/} root---the consolidation performed by a sandboxed internal sub-agent (no approvals, no network, local-write-only) reviewing a git-style workspace diff.
Memories are injected into new sessions with citation tracking and usage-based ranking.
\cdx{} is thus the first system in the corpus whose long-term memory is maintained \emph{by an agent} rather than by code; an experimental ``chronicle'' sidecar extends the same machinery toward passive screen context.

\gemini{} moved in the opposite governance direction.
Its April-era callable \texttt{save\_memory} tool is gone---the system prompt now states flatly that no such tool exists---replaced by (a)~the model editing \texttt{GEMINI.md} or a private per-project \texttt{MEMORY.md} index directly with ordinary edit tools, and (b)~an asynchronous \emph{skill-extraction sub-agent} that mines completed session transcripts and emits canonical patch files into a per-project \texttt{.inbox/}, which the user reviews and applies via \texttt{/memory}.
Where \cdx{} lets an agent consolidate memory autonomously, \gemini{} inserts a human-gated review inbox between extraction and persistence---the only such design in the corpus.

\hrm{} keeps persistent memory deliberately tiny and cache-friendly: two bounded Markdown files (\texttt{MEMORY.md}, 2{,}200 characters; \texttt{USER.md}, 1{,}375) injected as a \emph{frozen snapshot}, so mid-session writes hit disk without invalidating the prompt cache.
Conversation recall is deterministic---a \texttt{session\_search} tool over trigger-maintained SQLite FTS5 tables (BM25, plus trigram indexing for CJK), ``no LLM calls anywhere''---and embeddings exist only in opt-in memory plugins.
\ocl{} ships an \emph{Active Memory} plugin (present, we note for the record, since the very April release our earlier edition audited) that runs a dedicated memory sub-agent immediately \emph{before} the main reply, plus its older opt-in LanceDB conversation-memory extension.
\oh{} folds memory into its context-file convention: a \texttt{<MEMORY>} prompt section instructs the model to persist durable facts into \texttt{AGENTS.md} itself.
\clcode{} persists per-project memory directories of Markdown facts; \pib{} and \mswe{} deliberately persist nothing beyond the session substrate.

\observation{\label{obs:memory}Persistent memory has replaced compaction as the frontier of context engineering. Compaction designs have converged (seven of eleven systems use threshold-triggered LLM summarization, increasingly with incremental merging); what now differentiates the field is the memory write path. Four governance models coexist: agent-maintained (\cdx{}: a sandboxed sub-agent consolidates memories over a git-baselined store), human-gated (\gemini{}: extraction sub-agent writes to a patch inbox the user reviews via \texttt{/memory}), model-direct-but-bounded (\hrm{}: character-capped Markdown snapshots, frozen per session for cache hygiene; \oh{}, \clcode{}: context-file conventions), and pre-turn agentic recall (\ocl{}'s Active Memory sub-agent). Notably, none of the eleven uses embedding-based retrieval as its primary memory substrate---the deterministic-retrieval finding of Observation~\ref{obs:absences} extends from code to memory, with SQLite full-text search (\hrm{}) as the production ceiling.}

\subsection{Repository Context}

Beyond conversation history, agents must manage code context:

\begin{itemize}[itemsep=1pt]
  \item \textbf{\adr{} RepoMap}: Tree-sitter-based symbol indexing with diskcache (SQLite v4)---still the only ranked repository map in the corpus.
  Symbols are ranked by relevance to the current conversation.
  Token budget is adaptive (\texttt{map\_tokens}, default 1024) with a \texttt{map\_mul\_no\_files} multiplier for large repositories.

  \item \textbf{\clcode{} Memory System}: Auto-discovers \texttt{CLAUDE.md} files via a dedicated \texttt{claudemd} subsystem (managed, user, and project scopes traversed from the working directory), injected as nested memory attachments with deduplication; the separate \texttt{memdir} machinery manages the per-project persistent auto-memory directory.

  \item \textbf{\cdx{} Instructions}: Hierarchical \texttt{AGENTS.md} instructions---precisely, a \emph{concatenation} from the project root down to the working directory (nested content appended last, so ``precedence'' holds only by position), plus an \texttt{AGENTS.override.md} escape hatch and configurable root markers and fallback filenames.

  \item \textbf{\gemini{} Context Files}: Three-scope \texttt{GEMINI.md} merge (global, extension, project) gated by folder trust, plus JIT injection of subdirectory context files through tool output and the private \texttt{MEMORY.md} project index.

  \item \textbf{\vibe{} JIT Discovery}: Top-level \texttt{AGENTS.md} in the prompt (user and project scopes), and---since v2.19---nested \texttt{AGENTS.md} files surfaced just-in-time when \texttt{read\_file} touches files beneath them.

  \item \textbf{\hrm{} Hierarchy with Threat Scanning}: First-found-wins walk (\texttt{.hermes.md}/\texttt{HERMES.md} to the git root, then \texttt{AGENTS.md}, \texttt{CLAUDE.md}, \texttt{.cursorrules}); every context file is scanned for prompt-injection patterns and \emph{blocked wholesale} on a match; a subdirectory hint tracker appends per-directory context files to tool results, keeping the cached prompt untouched.

  \item \textbf{\pib{} Ancestor Walk}: \texttt{AGENTS.md}/\texttt{CLAUDE.md} collected from cwd to root into \texttt{<project\_context>} tags; loading of repo-controlled \emph{configuration} (extensions, skills) is trust-gated, but context files load regardless---a documented, deliberate injection-surface acceptance.

  \item \textbf{\ocd{} Pull-Based Attach}: \texttt{AGENTS.md}/\texttt{CLAUDE.md}/\texttt{CONTEXT.md} walk-up (first filename wins) plus remote-URL instruction files; nested \texttt{AGENTS.md} lazily attach when the \texttt{read} tool touches files beneath them.

  \item \textbf{\oh{} Third-Party Ingestion}: the skills subsystem ingests \texttt{AGENTS.md}, \texttt{.cursorrules}, and nested per-directory \texttt{AGENTS.md} as scoped rules---one system reading three other ecosystems' conventions.

  \item \textbf{\ocl{} Transcripts}: JSONL transcript files persisted per agent/session at \linebreak\texttt{\textasciitilde/.openclaw/agents/<agentId>/sessions/<sessionKey>.jsonl}.
\end{itemize}

\section{Safety and Permission Models}
\label{sec:safety}

Safety mechanisms become important in direct proportion to the autonomy the agent is given.
A recent line of academic work has formalized the threat model: indirect prompt injection through file or web content~\cite{greshake2023indirect}, AgentDojo~\cite{debenedetti2024agentdojo} as a dynamic evaluation environment for injection attacks and defenses, risk-awareness benchmarks like R-Judge~\cite{yuan2024rjudge}, and ToolSword's~\cite{ye2024toolsword} catalog of tool-use safety failure modes.
The corpus covers a wide range, from systems with essentially no safety controls to multi-layered architectures.

\subsection{Claude Code: Three-Layered Permission System}

\clcode{} implements the most architecturally elegant safety system, with three layers (Figure~\ref{fig:safety}):

\begin{figure}[H]
\centering
\begin{tikzpicture}[
  level/.style={rectangle, draw, rounded corners=3pt, minimum width=10cm, minimum height=0.7cm, font=\small, thick},
]
  \node[level, fill=red!8] (l1) at (0,0) {\textbf{Layer 3:} Interactive User Dialog (approve / reject / edit)};
  \node[level, fill=orange!10] (l2) at (0,1) {\textbf{Layer 2:} LLM-based Permission Classifier (allow / ask / deny)};
  \node[level, fill=green!8] (l3) at (0,2) {\textbf{Layer 1:} Static Hook Rules (\texttt{PreToolUse} pattern matching)};

  \draw[decorate, decoration={brace, amplitude=6pt, mirror}, thick] (5.5,-0.35) -- (5.5,2.35) node[midway, right=8pt, font=\small\itshape] {\clcode{}};

  \node[font=\scriptsize, gray, anchor=east] at (-5.3,0) {Slowest, most reliable};
  \node[font=\scriptsize, gray, anchor=east] at (-5.3,2) {Fastest, least context};
\end{tikzpicture}
\caption{\clcode{}'s three-layered permission system.}
\label{fig:safety}
\end{figure}
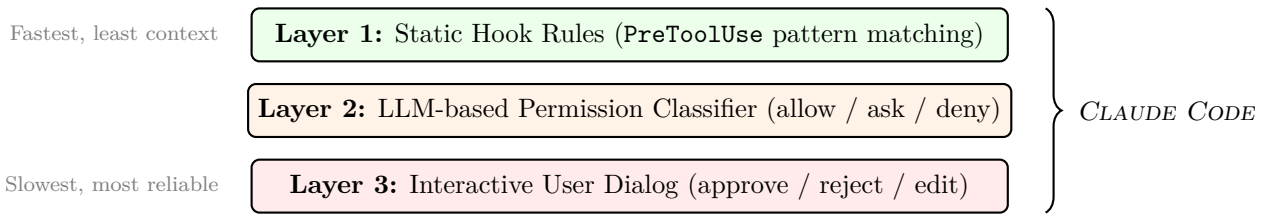

For \emph{background} and forked sub-agents, only layers~1--2 are available---their would-be prompts resolve to denials---while default foreground sub-agents can still surface interactive dialogs.
Unresolvable cases escalate to the parent coordinator.
A \textbf{denial tracking} mechanism counts permission refusals per session; after $N$ denials, the system falls back to prompting.
Each async agent maintains local denial state to prevent cross-agent pollution.

\subsection{Codex: A Four-Layer Permission Stack with Native Sandboxing}

The April edition described \cdx{}'s safety as two mechanisms---policy rules plus OS sandbox---and characterized its approval gate as ``near-invisible.''
The July snapshot shows four layers, two of them converging on \clcode{}'s design:

\paragraph{Layer 1: Execution Policy (Starlark, not TOML).}
Executable rules (\texttt{execpolicy/}) are written in \emph{Starlark}: \texttt{prefix\_rule(pattern=[\ldots], decision="allow|prompt|forbidden", match=[\ldots], not\_match=[\ldots])} and \texttt{network\_rule(\ldots)} functions, loaded from \texttt{CODEX\_HOME/rules/} and per-config-layer rules directories.\footnote{The April edition rendered these rules as TOML; that was an error in our earlier audit---the policy language has been Starlark throughout.}
A distinctive detail: rules may carry inline \texttt{match}/\texttt{not\_match} examples that are \emph{validated at parse time}---executable test cases inside the policy file.

\begin{lstlisting}[caption={Codex execution policy example (Starlark).},label={lst:policy}]
prefix_rule(
    pattern = ["git", "status"],
    decision = "allow",
    match = [["git", "status"], ["git", "status", "--short"]],
    not_match = [["git", "push"]],
)
network_rule(
    protocol = "https",
    host = "*.github.com",
    decision = "allow",
)
\end{lstlisting}

\paragraph{Layer 2: Lifecycle Hooks.}
A hooks crate exposes external hooks whose event vocabulary is near-verbatim \clcode{}'s: \texttt{PreToolUse}, \texttt{PermissionRequest}, \texttt{PostToolUse}, \texttt{PreCompact}, \texttt{SessionStart}, \texttt{UserPromptSubmit}, \texttt{SubagentStart}, \texttt{SubagentStop}, \texttt{Stop}---plus a \texttt{PostCompact} event with no \clcode{} counterpart---with block/allow decision outputs.
Section~\ref{sec:evolution_window} discusses this as the corpus's clearest case of one provider-native harness adopting another's extension interface as a de facto standard.

\paragraph{Layer 3: Guardian, an LLM Approval Reviewer.}
When a command would require user approval, a dedicated \emph{Guardian} session re-assesses the exact planned action against a policy prompt (compact transcript reconstruction, strict-JSON verdict), fails \emph{closed} on timeout or malformed output, and carries a per-turn rejection circuit breaker.
The ``no LLM classifier'' contrast the April edition drew against \clcode{} no longer holds: both provider-native flagships now implement the static-rules~$\to$~LLM-classifier~$\to$~interactive/OS backstop pattern.
A user-facing collaboration-mode selector (Plan/Default presets that mask model, reasoning effort, and developer instructions) completes the convergence: all four provider-native systems now ship a read-only planning mode.

\paragraph{Layer 4: Native OS Sandboxing.}
The sandbox implementation spans three platforms:
\begin{itemize}[itemsep=1pt]
  \item \textbf{Linux}: vendored, in-tree Bubblewrap (invoked via C FFI) with read-only filesystem (\texttt{--ro-bind / /}), writable roots via \texttt{--bind}, network namespace isolation (\texttt{--unshare-net}), user/PID namespace isolation; Landlock demoted to a legacy fallback.
  \item \textbf{macOS}: Seatbelt profiles via \texttt{/usr/bin/sandbox-exec} with configurable allow/deny rules.
  \item \textbf{Windows}: Restricted-token processes with ACL-based read/write restrictions.
\end{itemize}

\observation{\label{obs:sandboxcost}OS-level sandboxing is one of the most code-expensive capabilities in the corpus, and it is a \emph{choice} rather than a consequence of scale. The April corpus suggested a size correlation (its two largest systems were its two cross-platform sandboxers); the expanded corpus breaks it. \hrm{}, among the largest systems here, ships zero OS-level isolation primitives: it delegates containment to six pluggable execution backends and spends its safety budget on content-borne threats instead (promptware scanning---pattern-matching for prompt-injection payloads---a skills supply chain, a command-policy floor that survives \texttt{--yolo}). \ocd{}, comparably large, is policy-only, investing in syntax-aware command permissioning rather than isolation. \pib{} documents the refusal as a security argument: a partial in-process sandbox ``would be easy to misunderstand as a security boundary.'' What remains true is the cost side: where native sandboxing is built (\cdx{}, \gemini{}; \clcode{} wraps Anthropic's reusable \texttt{sandbox-runtime} as an opt-in), it is a substantial multi-thousand-line investment---and Section~\ref{sec:omnigent} shows the meta-harness paying the same bill a second time, one layer up.}

\subsection{OpenHands: Ensemble Defense-in-Depth}

\oh{}'s pluggable \texttt{SecurityAnalyzer} framework matured substantially in the V1 SDK.
The Invariant analyzer of the V0 era is gone; the current set pairs an \texttt{LLMSecurityAnalyzer} (the agent self-rates each action's \texttt{security\_risk}, a parameter the system prompt instructs it to fill) and a \texttt{GraySwanAnalyzer} (external adversarial API) with two new \emph{deterministic} analyzers---\texttt{PatternSecurityAnalyzer} and \texttt{PolicyRailSecurityAnalyzer}, the latter shipping named rails such as \texttt{fetch-to-exec}, \texttt{raw-disk-op}, and \texttt{catastrophic-delete} over shell-AST parsing.
An \texttt{EnsembleSecurityAnalyzer} fuses verdicts worst-case-wins, with broken child analyzers failing closed to HIGH.
The LOW/MEDIUM/HIGH/UNKNOWN risk enum survives; confirmation is now a policy object (\texttt{AlwaysConfirm}/\texttt{NeverConfirm}/\texttt{ConfirmRisky}).
The prompt side completes a two-sided injection defense: repository-derived context is wrapped in \texttt{<UNTRUSTED\_CONTENT>} markers, and a dedicated prompt section teaches the risk-assessment protocol.

\subsection{OpenClaw: Scope-Based Authorization}

\ocl{} implements scope-based authorization with \texttt{operator} and \texttt{node} connection roles over namespaced scopes (\texttt{operator.read}/\texttt{write}/\texttt{admin}/\texttt{pairing}/\ldots).
The gateway enforces per-method scope requirements.
Additional safety mechanisms include: fixed-window rate limiting per IP/token, SSRF policy enforcement, DM allowlist with pairing approval workflow, exec approval gates for browser actions and plugin actions, and chat sanitization.
The 2MB \texttt{MAX\_PROMPT\_BYTES} limit provides DoS protection (CWE-400).

\subsection{Gemini CLI: Four Approval Modes Plus Cross-Platform Sandbox}

\gemini{} layers two complementary safety mechanisms.
First, an \texttt{ApprovalMode} enum gates every tool call into one of four modes:
\begin{itemize}[itemsep=1pt]
  \item \textbf{PLAN}: Read-only; the model sees the full tool list but can only invoke read tools (Glob, Read, Grep, etc.). Its response is presented as a plan, and the user must explicitly approve before exiting plan mode.
  \item \textbf{DEFAULT}: Interactive approval dialog per tool call; users can approve once, approve always, or reject.
  \item \textbf{AUTO\_EDIT}: Edit operations are auto-approved while shell and other side-effecting tools still prompt; intended for fast iteration on local code.
  \item \textbf{YOLO}: All actions auto-approved. Intended for CI/scripted use only.
\end{itemize}
Second, a \texttt{SandboxPolicyManager} enforces TOML-based per-mode policies: which paths are readable/writable, which network hosts are reachable, which environment variables are scrubbed.
The combination is structurally similar to \cdx{}'s policy-as-code + sandbox stack but layers an explicit user-facing mode selector on top---an ergonomic that \cdx{} itself has since adopted with its collaboration-mode presets.
A separate \texttt{isTrustedFolder()} check further widens or narrows context-discovery scope based on whether the directory is in the user's trust list, providing a fourth layer of opt-in granularity.
The release cadence between our snapshots was dominated by policy-engine and supply-chain hardening---trust-gated \texttt{.env} loading in headless mode, shell-command validation with a core-tools allowlist, skill-install path-traversal checks---supporting Observation~\ref{obs:sandboxcost}'s claim that safety absorbs a disproportionate share of engineering in the largest systems.

\subsection{Mistral Vibe: Permission-Scope Patterns Plus Agent-Profile Gates}

\vibe{} has no OS sandbox (an opt-in \texttt{--worktree} flag now provides branch-level isolation) but compensates with a finer-grained permission model than the other no-sandbox systems.
Each tool call is checked against a hierarchy:
\begin{enumerate}[itemsep=1pt]
  \item Tool-wide \texttt{permission} setting (\texttt{ALWAYS}, \texttt{ASK}, \texttt{NEVER}) at the agent-profile level.
  \item Tool-specific validators (e.g., \texttt{resolve\_file\_tool\_permission} pattern-matches paths), plus a default read-only command allowlist that auto-approves \texttt{ls}/\texttt{cat}/\texttt{grep}-class commands.
  \item User-declared shell hooks (\texttt{hooks.toml}, \texttt{before\_tool}) that can deny a call or rewrite its inputs before permission evaluation.
  \item Session rules added at runtime---now persisted across sessions for ``always allow'' grants.
  \item An interactive approval callback presented to the user unless \texttt{bypass\_tool\_permissions} (the renamed \texttt{auto\_approve}) is set.
\end{enumerate}
Agent profiles bake in safety classifications: \texttt{plan} and \texttt{chat} profiles are \texttt{SAFE} (read-only); \texttt{default} is \texttt{NEUTRAL}; \texttt{accept-edits} is \texttt{DESTRUCTIVE}; \texttt{auto-approve} is YOLO.
Tools can also override the agent default: \texttt{write\_file} forces \texttt{ASK} for sensitive patterns (\texttt{.env} files) regardless of profile.
There is no LLM-based risk classifier and no OS-level isolation; safety reduces to explicit per-pattern matching, hooks, and user prompts.

\subsection{Hermes: A Policy Floor That Survives YOLO}

\hrm{} inverts the sandbox question: the largest Python system in the corpus ships \emph{zero} OS-level isolation primitives---no Seatbelt, Landlock, seccomp, or Bubblewrap anywhere in its tools tree---delegating isolation to its six pluggable execution backends.
On the local backend, safety is \emph{policy-as-code at scale}: a 3{,}200-line approval module whose documentation states ``config.yaml IS the security policy.''
Its layers: user deny-globs; a twelve-pattern \textbf{hardline floor} (\texttt{rm -rf /}, \texttt{mkfs}, \texttt{dd} to block devices, fork bombs, shutdown) that survives \texttt{--yolo}---the YOLO environment variable is frozen at module import precisely so a prompt-injected skill cannot flip it at runtime; 47 dangerous-command patterns matched on \emph{deobfuscated} variants (quote-splice removal, command-substitution folding, command-position anchoring); an external Rust content scanner (Tirith, cosign-verified install); and an optional auxiliary-LLM gate (\texttt{\_smart\_approve}: temperature~0, 16 max tokens, APPROVE/DENY/ESCALATE---its code credits \cdx{}'s Smart Approvals).
The distinctive threat model is \emph{content-borne}: promptware scanning of context files, memory writes, MCP tool descriptions, and skill installs (with builtin/trusted/community trust tiers and quarantine); SSRF guards that always block cloud metadata endpoints; untrusted tool results wrapped in \texttt{<untrusted\_tool\_result>} delimiters with lookalike defanging.
Container backends skip the dangerous-command layer entirely---until a host bind-mount re-enters it.
Notable absences: no per-tool allow/ask/deny matrix, no append-only audit log.

\subsection{Pi: Documented Absence as a Security Argument}

\pib{} is the corpus's only system that documents the \emph{absence} of safety infrastructure as a design principle: its security documentation argues that a partial in-process sandbox ``would be easy to misunderstand as a security boundary,'' declares prompt injection via context files unprotectable, and lets built-in tools execute unconditionally with the invoking user's permissions.
The one built-in gate is \emph{project trust}: repo-controlled configuration (extensions, skills, prompts, themes) loads only from trusted paths, decided per canonical path with nearest-ancestor lookup---explicitly ``not a sandbox,'' and context files load regardless.
Permissioning is policy-as-code taken literally: the \texttt{tool\_call} extension event exposes mutable tool arguments and a \texttt{\{block, reason\}} veto, and the reference \texttt{permission-gate} example implements the familiar regex-plus-confirm flow (\texttt{rm -rf}, \texttt{sudo}, \texttt{chmod 777}) in $\sim$80 lines of user space; isolation, when wanted, arrives as extensions (Anthropic's \texttt{sandbox-runtime}, a QEMU micro-VM) through the per-tool \texttt{Operations} seam.
The append-only session tree doubles as a complete audit trail.

\subsection{OpenCode: Syntax-Aware Permissioning, Permissive by Default}

\ocd{} is, after \hrm{}, the second-largest system shipping no OS-level isolation (``sandbox'' in its code means git worktrees), and its default posture is the inverse of \cdx{}'s: \texttt{"*": allow} \emph{inside} the project, with asks reserved for \texttt{external\_directory} access, \texttt{.env} reads, and doom loops.
Its distinctive investment is \emph{syntax-aware command permissioning}: every bash command is parsed with web-tree-sitter (bash and PowerShell grammars); the exact command text becomes the ask pattern, while ``always allow'' grants are scoped by a $\sim$130-entry, explicitly LLM-generated command-arity dictionary (\texttt{git}$\to$2, \texttt{npm run}$\to$3), producing grants like \texttt{git commit *} rather than blanket approval; file-manipulating commands get their argument paths resolved, and out-of-project paths raise \texttt{external\_directory} asks.
Permission rules (\texttt{\{permission, pattern, action\}}, last-match-wins over glob wildcards) are layered defaults~$\to$~built-in agent~$\to$~user config~$\to$~per-agent~$\to$~per-session, and denying \texttt{"*"} for a tool removes it from the LLM's view entirely---tool availability itself is permission-derived.
A parallel v2 engine inverts to deny-by-default and persists approvals to SQLite; headless runs auto-reject all asks.
There is no dangerous-command denylist and no risk classifier.

\subsection{Minimal Safety (Aider, Mini-SWE-Agent)}

\adr{} offers an interactive confirmation mode but no automated risk assessment.
However, its reflection loop with three-stage Python linting (syntax $\to$ compile $\to$ flake8) provides implicit safety by catching errors before they propagate.

\mswe{} implements only resource limits as safety mechanisms---\texttt{step\_limit}, \texttt{cost\_limit}, and (added in the v2.4 line) a wall-clock limit and a consecutive-format-error cap, plus an opt-in interactive confirm mode---sufficient for benchmark evaluation but inadequate for production use.

\section{Multi-Agent Orchestration}
\label{sec:multiagent}

Nine of the eleven systems support multi-agent execution, and their architectures diverge substantially.
This is the dimension on which the corpus is most spread out, so we go into more detail here than in the other sections.

\subsection{Multi-Agent Taxonomy}

We identify six distinct multi-agent patterns:

\begin{enumerate}[itemsep=1pt]
  \item \textbf{Single-agent (no multi-agent)}: \mswe{}, \adr{}---and \pib{}'s \emph{core}, which ships no spawn or task tool; its sub-agents live in extension space (below).
  \item \textbf{Sequential delegation}: \vibe{} (parent spawns a sub-agent via the \texttt{task} tool and awaits completion).
  \item \textbf{Parallel child sessions}: \oh{} (task and delegate tools run children as separate conversations, concurrently in threads), \ocd{} (\texttt{task} creates child sessions running the same loop; multiple calls in one message execute concurrently).
  \item \textbf{Hierarchical thread tree with fan-out}: \cdx{} (dedicated threads, fork-mode-controlled inheritance, per-model tool generations, map-reduce fan-out, persisted parent/child topology).
  \item \textbf{Recursive composition}: \clcode{} (agents spawn sub-agents composably, with prompt-cache-sharing forks).
  \item \textbf{Registry + cross-process protocol}: \gemini{} (named agent definitions behind a symmetric local/remote session protocol, remote agents via A2A), \ocl{} (ACP session spawn), \hrm{}'s Kanban swarm (subprocesses over a SQLite blackboard).
\end{enumerate}

\begin{figure}[H]
\centering
\tikzset{
  ma/.style={rectangle, draw, rounded corners=3pt, minimum width=1.35cm, minimum height=0.55cm, font=\tiny, align=center, thick},
  maarr/.style={-{Stealth[length=1.6mm]}, thick},
  maret/.style={-{Stealth[length=1.6mm]}, thick, dashed},
}
\begin{minipage}[t]{0.32\textwidth}
\centering
{\scriptsize\textbf{(1) Single-agent}}\\{\tiny \mswe{}, \adr{}, \pib{} core}\\[5pt]
\begin{tikzpicture}
  \node[ma, fill=blue!10, minimum width=1.7cm] (a) {Agent};
  \draw[maarr] (a.20) .. controls +(0.9,0.5) and +(0.9,-0.5) .. (a.-20) node[midway, right=14pt, font=\tiny, align=left] {act--\\observe};
  \node[font=\tiny\itshape, below=0.45cm of a] {no spawn tool};
\end{tikzpicture}
\end{minipage}\hfill
\begin{minipage}[t]{0.32\textwidth}
\centering
{\scriptsize\textbf{(2) Sequential delegation}}\\{\tiny \vibe{}}\\[5pt]
\begin{tikzpicture}
  \node[ma, fill=red!15] (p) {Parent};
  \node[ma, fill=blue!10, below=0.8cm of p] (c) {Sub-agent};
  \draw[maarr] ([xshift=-3.5mm]p.south) -- node[font=\tiny, left] {task} ([xshift=-3.5mm]c.north);
  \draw[maret] ([xshift=3.5mm]c.north) -- node[font=\tiny, right] {summary} ([xshift=3.5mm]p.south);
  \node[font=\tiny\itshape, below=0.15cm of c, align=center] {parent blocks;\\one child at a time};
\end{tikzpicture}
\end{minipage}\hfill
\begin{minipage}[t]{0.32\textwidth}
\centering
{\scriptsize\textbf{(3) Parallel child sessions}}\\{\tiny \oh{}, \ocd{}}\\[5pt]
\begin{tikzpicture}
  \node[ma, fill=red!15] (p) {Parent};
  \node[ma, fill=blue!10, below left=0.8cm and 0.15cm of p] (c1) {Session A};
  \node[ma, fill=blue!10, below right=0.8cm and 0.15cm of p] (c2) {Session B};
  \draw[maarr] (p.250) -- (c1.70);
  \draw[maret] (c1.20) -- (p.212);
  \draw[maarr] (p.290) -- (c2.110);
  \draw[maret] (c2.160) -- (p.328);
  \node[font=\tiny\itshape, below=1.98cm of p, align=center] {concurrent; separate histories};
\end{tikzpicture}
\end{minipage}

\vspace{1.3em}
\begin{minipage}[t]{0.32\textwidth}
\centering
{\scriptsize\textbf{(4) Hierarchical thread tree}}\\{\tiny \cdx{}}\\[5pt]
\begin{tikzpicture}
  \node[ma, fill=red!15] (r) {Root thread};
  \node[ma, fill=blue!10, below left=0.55cm and 0.05cm of r] (t1) {Thread};
  \node[ma, fill=blue!10, below right=0.55cm and 0.05cm of r] (t2) {Thread};
  \node[ma, fill=blue!10, below left=0.5cm and -0.35cm of t2, minimum width=0.9cm] (g1) {\ldots};
  \node[ma, fill=blue!10, below right=0.5cm and -0.35cm of t2, minimum width=0.9cm] (g2) {\ldots};
  \draw[maarr] (r) -- (t1);
  \draw[maarr] (r) -- (t2);
  \draw[maarr] (t2) -- node[font=\tiny, left, pos=0.7] {fork mode} (g1);
  \draw[maarr] (t2) -- (g2);
  \node[font=\tiny\itshape, below=0.12cm of g1, xshift=0.5cm, align=center] {depth-tracked; CSV fan-out};
\end{tikzpicture}
\end{minipage}\hfill
\begin{minipage}[t]{0.32\textwidth}
\centering
{\scriptsize\textbf{(5) Recursive composition}}\\{\tiny \clcode{}}\\[5pt]
\begin{tikzpicture}
  \node[ma, fill=red!15] (a) {Agent};
  \node[ma, fill=blue!10, below=0.5cm of a] (b) {Sub-agent};
  \node[ma, fill=blue!10, below=0.5cm of b] (c) {Sub-sub-agent};
  \draw[maarr] (a) -- node[font=\tiny, right] {fork context} (b);
  \draw[maarr] (b) -- node[font=\tiny, right] {fork context} (c);
  \node[font=\tiny, below=0.05cm of c] {$\vdots$};
  \node[font=\tiny\itshape, below=0.4cm of c, align=center] {any agent spawns agents;\\shared prompt cache};
\end{tikzpicture}
\end{minipage}\hfill
\begin{minipage}[t]{0.32\textwidth}
\centering
{\scriptsize\textbf{(6) Registry + protocol}}\\{\tiny \gemini{}, \ocl{}, \hrm{} swarm}\\[5pt]
\begin{tikzpicture}
  \node[ma, fill=red!15] (p) {Parent};
  \node[ma, fill=green!10, below=0.7cm of p, minimum width=1.6cm] (reg) {Registry\\(agent defs)};
  \node[ma, fill=blue!10, right=0.75cm of reg] (rem) {Remote\\agent};
  \draw[maarr] (p) -- node[font=\tiny, left] {invoke by name} (reg);
  \draw[maarr] (reg) -- (rem);
  \draw[dashed, gray, thick] ($(reg.east)!0.5!(rem.west)+(0,0.55)$) -- ($(reg.east)!0.5!(rem.west)+(0,-1.15)$);
  \node[font=\tiny, gray, align=center, anchor=south] at ($(reg.east)!0.5!(rem.west)+(0,0.6)$) {process\\boundary};
  \node[font=\tiny\itshape, below=0.15cm of reg, align=center, xshift=-2mm] {A2A / ACP /\\SQLite blackboard};
\end{tikzpicture}
\end{minipage}
\caption{The six multi-agent orchestration patterns of the taxonomy above, schematically. Red nodes spawn, blue nodes are spawned, green is the definition registry; solid arrows spawn or invoke, dashed arrows return results. Systems beneath each panel are the pattern's corpus representatives (Table~\ref{tab:multiagent}).}
\label{fig:ma_patterns}
\end{figure}
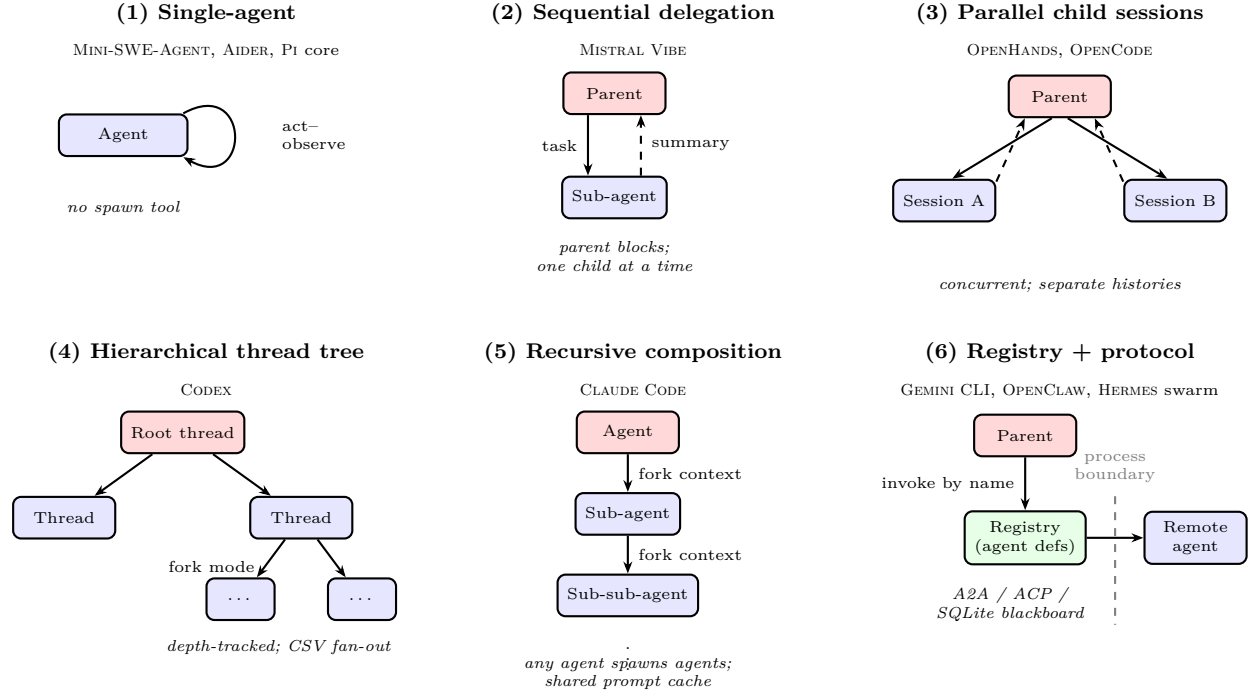

Table~\ref{tab:multiagent} provides a detailed comparison; Figure~\ref{fig:ma_patterns} gives each pattern's shape at a glance.

\begin{table}[H]
\centering
\caption{Multi-agent orchestration across the nine multi-agent systems (\pib{} via its reference extension).}
\label{tab:multiagent}
\scriptsize
\renewcommand{\arraystretch}{1.3}
\setlength{\tabcolsep}{3.5pt}
\begin{tabularx}{\textwidth}{@{}l L{2.6cm} L{2.6cm} L{2.5cm} L{2.4cm} L{2.6cm}@{}}
\toprule
\textbf{System} & \textbf{Mechanism} & \textbf{Isolation / inherit.} & \textbf{Communication} & \textbf{Depth / parallelism} & \textbf{Tool filtering} \\
\midrule
\oh{}     & \texttt{task} + \texttt{delegate} tools $\to$ separate conversations & Fresh event log per child; metrics synced back & Summary return; conversation tree navigation & Nesting supported / concurrent threads & Markdown frontmatter (tools, skills, permission mode) \\
\clcode{} & AgentTool + context fork & Forked context (6 dimensions), shared rendered prompt & Return + XML notifications & Recursive / true parallel & 16-tool whitelist \\
\cdx{}    & \texttt{spawn\_agent} v1/v2 (per-model) & Thread tree; \texttt{SpawnAgentForkMode} & Session input queue with mailbox phases; typed \texttt{InterAgentCommunication} records & Depth-tracked / parallel + CSV fan-out & Inherited + filtered; TOML roles (explorer, awaiter) \\
\gemini{} & \texttt{invoke\_agent} tool; registry defs & Isolated registries per instance & Symmetric local/remote session protocol; A2A RPC & Flat + remote / sequential in turn & Per-definition \texttt{toolConfig} \\
\vibe{}   & \texttt{task} tool $\to$ new AgentLoop & Fresh config + session dir; inherits permission store, scratchpad, hooks & Event forwarding & 1 hop / sequential & Subagent-tagged profiles only (explore) \\
\hrm{}    & \texttt{delegate\_task} in-process forks; Kanban swarm subprocesses & Fresh context, own task id; provider/creds inherited, no history & Spawn-time goal; streamed events; summary return; SQLite blackboard (swarm) & 1 by default, orchestrator role unlocks nesting / 3 concurrent (pool) & Intersection with parent + 6-tool blocklist \\
\pib{} (ext.) & Extension spawns \texttt{pi --mode json -p} OS processes & Hard process isolation; fresh context & Child JSONL stdout parsed for progress/cost & Unbounded (no counter) / pool of 4, max 8; sequential \texttt{chain} & Child \texttt{--tools} flag from agent frontmatter \\
\ocd{}    & \texttt{task} tool $\to$ child Session (same loop) & \texttt{parentID} sessions; child inherits parent's deny + external-directory rules & \texttt{<task\_result>} XML; background results as synthetic user messages & Recursion off by default, opt-in unbounded / parallel & Permission rulesets (deny \texttt{"*"} hides tool) \\
\ocl{}    & ACP session spawn & Child processes; fresh session & ACP delta events & Recursive / parallel (RPC) & Session-scoped \\
\bottomrule
\end{tabularx}
\end{table}

\subsection{Deep Dive: Claude Code Recursive Composition}
\label{sec:cc_multiagent}

\clcode{}'s multi-agent architecture is the most sophisticated among the systems studied.
It supports three levels of coordination: ad-hoc sub-agents, typed agent definitions, and full coordinator mode.

\subsubsection{Agent Definitions System}

Agents are defined via a JSON/Zod schema specifying a description, prompt, optional tool whitelist/blacklist, model selection, turn limits, permission mode, MCP server bindings, isolation level, background execution flag, and skills.

Agent definitions are loaded from a priority hierarchy: built-in agents $\to$ plugin agents $\to$ user settings $\to$ project settings $\to$ flag settings $\to$ policy/managed agents.
Later sources override earlier ones.
Built-in agents include Explore (fast codebase search), Plan (architecture design), and VerificationAgent (test execution).

\subsubsection{Context Forking}

When a sub-agent is spawned via \texttt{AgentTool}, the parent's context is forked along six dimensions:

\begin{enumerate}[itemsep=1pt]
  \item \textbf{AbortController}: New controller linked to parent (parent abort $\to$ child abort, but not vice versa).
  \item \textbf{File State Cache}: LRU cache cloned to prevent concurrent mutation.
  \item \textbf{Permission Prompts}: Suppressed for async agents (\texttt{shouldAvoidPermissionPrompts = true}).
  \item \textbf{App State}: Set to no-op for async agents to prevent dead session mutation.
  \item \textbf{Denial Tracking}: Fresh local state (refusals don't accumulate globally).
  \item \textbf{Tool Decisions}: Fresh Map per agent (no cross-agent pollution).
\end{enumerate}

A critical optimization is \textbf{prompt cache sharing}: the parent's \texttt{renderedSystemPrompt} is frozen at fork time and passed verbatim to the child.
This ensures prompt cache hits are maintained across fork/resume boundaries, preventing GrowthBook feature gate divergence from invalidating caches.

\subsubsection{Coordinator Mode}

When activated (via environment variable or feature gate), a Lead agent orchestrates Workers (Figure~\ref{fig:cc_coordinator}):

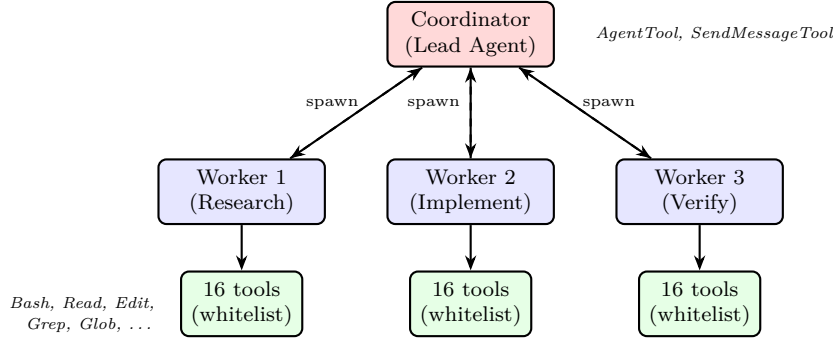
\begin{figure}[H]
\centering
\begin{tikzpicture}[
  node distance=0.6cm,
  agent/.style={rectangle, draw, rounded corners=3pt, minimum width=2.2cm, minimum height=0.8cm, font=\scriptsize, align=center, thick},
  arr/.style={-{Stealth[length=2mm]}, thick},
  darr/.style={{Stealth[length=2mm]}-{Stealth[length=2mm]}, thick},
]
  \node[agent, fill=red!15] (coord) {Coordinator\\(Lead Agent)};

  \node[agent, fill=blue!10, below left=1.2cm and 0.8cm of coord] (w1) {Worker 1\\(Research)};
  \node[agent, fill=blue!10, below=1.2cm of coord] (w2) {Worker 2\\(Implement)};
  \node[agent, fill=blue!10, below right=1.2cm and 0.8cm of coord] (w3) {Worker 3\\(Verify)};

  \node[agent, fill=green!10, below=0.6cm of w1, minimum width=1.6cm] (t1) {16 tools\\(whitelist)};
  \node[agent, fill=green!10, below=0.6cm of w2, minimum width=1.6cm] (t2) {16 tools\\(whitelist)};
  \node[agent, fill=green!10, below=0.6cm of w3, minimum width=1.6cm] (t3) {16 tools\\(whitelist)};

  \draw[arr] (coord) -- node[font=\tiny, left, pos=0.4] {spawn} (w1);
  \draw[arr] (coord) -- node[font=\tiny, left, pos=0.4] {spawn} (w2);
  \draw[arr] (coord) -- node[font=\tiny, right, pos=0.4] {spawn} (w3);
  \draw[arr, dashed] (w1) -- (coord);
  \draw[arr, dashed] (w2) -- (coord);
  \draw[arr, dashed] (w3) -- (coord);
  \draw[arr] (w1) -- (t1);
  \draw[arr] (w2) -- (t2);
  \draw[arr] (w3) -- (t3);

  \node[font=\tiny\itshape, right=0.4cm of coord] {AgentTool, SendMessageTool};
  \node[font=\tiny\itshape, left=0.2cm of t1] {Bash, Read, Edit,};
  \node[font=\tiny\itshape, left=0.2cm of t1, yshift=-8pt] {Grep, Glob, \ldots};
\end{tikzpicture}
\caption{\clcode{} coordinator mode architecture. Workers receive a whitelist of 16 tools and report via XML notifications.}
\label{fig:cc_coordinator}
\end{figure}

Workers are restricted to a 16-tool whitelist (\texttt{ASYNC\_AGENT\_ALLOWED\_TOOLS}): FileRead, WebSearch, TodoWrite, Grep, WebFetch, Glob, the two shell tools (Bash and PowerShell), FileEdit, FileWrite, NotebookEdit, Skill, SyntheticOutput, ToolSearch, EnterWorktree, and ExitWorktree.
The coordinator synthesizes worker findings before delegating the next phase, following a structured workflow: Research $\to$ Synthesis $\to$ Implementation $\to$ Verification.

\subsection{Deep Dive: Codex Thread Tree Model}
\label{sec:codex_multiagent}

\cdx{}'s multi-agent architecture is built around three core abstractions:

\paragraph{AgentControl} is the central control plane for spawning and managing sub-agents.

\paragraph{AgentRegistry} tracks all live agents in a session tree, maintaining \texttt{LiveAgent} metadata structs.

\paragraph{Spawn Flow.}
The \texttt{spawn\_agent\_internal()} function reserves a spawn slot, inherits the parent's sandbox and execution policy, resolves fork mode and metadata from the source, creates a thread with the appropriate history (full, truncated, or fresh), and emits a session-started notification.

\paragraph{SpawnAgentForkMode.}
A distinctive feature is \texttt{SpawnAgentForkMode}, which controls how much conversation history is inherited by sub-agents:
\begin{itemize}[itemsep=1pt]
  \item \texttt{FullHistory}: Inherit all messages (full context).
  \item \texttt{LastNTurns(N)}: Inherit only the most recent $N$ turns (reduced context).
\end{itemize}

Message filtering (\texttt{keep\_forked\_rollout\_item()}) preserves only system/developer/user messages and final assistant answers, filtering out intermediate tool calls and reasoning.
This selective inheritance prevents context bloat in sub-agents.

\paragraph{Inter-Agent Communication.}
The dedicated mailbox module of early versions has been folded into the session's input queue: inter-agent messages flow through \texttt{InputQueue} activities with mailbox delivery phases, carried as typed \texttt{InterAgentCommunication} records (\texttt{Spawn}/\texttt{Message}/\texttt{Followup}/\texttt{Result}).
Two tool generations coexist---\texttt{multi\_agents} (spawn/wait/send\_message/interrupt/list) and \texttt{multi\_agents\_v2} (spawn/wait/send\_input/resume\_agent/close\_agent)---selected \emph{per model} via a \texttt{multi\_agent\_version} field in the model metadata, and parent/child topology is persisted by a dedicated \texttt{agent-graph-store} crate.
The parent monitors child status via an \texttt{AgentStatus} enum.

\paragraph{Fan-Out and Declarative Roles.}
Beyond spawn and mailbox, \cdx{} has grown map-reduce-style orchestration: \texttt{spawn\_agents\_on\_csv} launches one sub-agent per CSV row under a shared JSON-schema result contract with concurrency normalization and runtime caps, folding results back via \texttt{report\_agent\_job\_result}.
Sub-agents are typed by declarative TOML \emph{roles} (built-in \texttt{explorer} and \texttt{worker} roles ship active in-tree; an \texttt{awaiter} role file remains but is currently unregistered) applied as high-precedence configuration layers---\cdx{}'s answer to \clcode{}'s agent definitions---and delegation itself is model-governed: the \texttt{ultra} reasoning tier couples maximum reasoning with \emph{automatic} task delegation, and configurable delegation modes (disabled / explicit-request-only / proactive) apply at thread and turn level.

\subsection{OpenHands: Parallel Delegation over Conversation Trees}

The V0-era \texttt{AgentDelegateAction} (sequential, one delegate at a time, shared event stream) is gone.
The V1 SDK delegates through tools: a \texttt{TaskTool} whose shape is conspicuously \clcode{}-like (\texttt{prompt}, \texttt{subagent\_type}, \texttt{description}, \texttt{resume}), with a \texttt{TaskManager} creating a \emph{separate} conversation per task (own event log, metrics synced back), and a \texttt{DelegateTool} with \texttt{spawn}/\texttt{delegate} commands that runs delegated tasks \emph{concurrently in threads}.
Sub-agents are declared as Markdown files with YAML frontmatter (name, description, model, tools, skills, hooks, MCP config, permission mode, condenser) at project/user/builtin/plugin levels---built-ins include \texttt{code-explorer}, \texttt{bash-runner}, and \texttt{web-researcher} (plus a \texttt{general-purpose} default)---and nesting is explicitly supported.
Above the delegation layer sits a scaffold-level outer loop: a \texttt{/goal} endpoint runs an LLM \emph{judge} over the transcript after each run, either injecting a follow-up prompt or stopping with a completion status, complemented by pluggable critics with configurable refinement-iteration caps.

\subsection{Mistral Vibe Sequential Subagent Delegation}

\vibe{}'s multi-agent support is intentionally simple: a parent agent invokes the \texttt{task} tool, which constructs a fresh \texttt{AgentLoop} instance with the requested agent profile (e.g., \texttt{explore}), runs it to completion in-process, and returns the sub-agent's \texttt{AssistantEvent} outputs and \texttt{ToolResultEvent} summaries to the parent.
A guard enforces that only profiles tagged as ``subagent'' (currently just \texttt{explore}) can be delegated to, preventing accidental recursive escalation to write-capable profiles.
Each sub-agent gets a fresh \texttt{VibeConfig}, its own session directory, and an independent message history, providing isolation comparable to \cdx{}'s thread tree but at the cost of true parallelism: the parent blocks on \texttt{await} until the sub-agent exits.
The architectural simplicity (one tool, sequential, in-process) means \vibe{} sits on the \emph{simplicity end} of the multi-agent spectrum---the corpus's simplest delegation short of none, reminiscent of \oh{}'s retired V0 \texttt{AgentDelegateAction}.

\subsection{Gemini CLI: Registry, Symmetric Session Protocol, and A2A}

\gemini{}'s multi-agent model splits across two layers.
Locally, an \texttt{AgentRegistry} holds named agent definitions---Markdown files with mandatory YAML frontmatter---loaded from user, project, and extension scopes.
The built-in \texttt{generalist} agent, for example, grants access to all registered tools with a 10-minute, 20-turn run budget.
Invocation flows through the \texttt{/agent} slash command, \texttt{@agent-name} prompt notation, or---since the sub-agent unification of v0.39---a model-invocable \texttt{invoke\_agent} tool, each run receiving an isolated tool registry.
The invocation path has been abstracted behind an \emph{agent session protocol} with symmetric local and remote implementations, so a registry entry backed by an in-process executor and one backed by a remote A2A agent are driven through the same interface.

Where \gemini{} diverges from every other system in our corpus is the \texttt{packages/a2a-server} module, an experimental implementation of Google's \emph{Agent-to-Agent} (A2A) protocol.
A2A is to inter-agent communication what MCP is to tool integration: a JSON-RPC standard for one agent process to discover, authenticate to, and exchange messages with another, possibly on a different machine and built by a different vendor.
\gemini{}'s A2A server lets a remote orchestrator drive a local \gemini{} instance as if it were a registered local agent---now with usage metadata reported over the protocol for cross-vendor resource accounting---opening the door to cross-vendor multi-agent topologies that no other corpus system supports natively (the meta-harness of Section~\ref{sec:omnigent} builds exactly such topologies from outside).

\subsection{Hermes: Guarded Delegation, Swarms, and Mixture-of-Agents}

\hrm{} is flat-by-default delegation with two escape hatches upward.
\texttt{delegate\_task} builds child agents \emph{in-process} on a thread pool: fresh conversation, focused system prompt, summary-only return, toolsets intersected with the parent's (``subagent must not gain tools the parent lacks''), and a six-tool blocklist (no recursion, no user interaction, no memory writes, no scheduling).
Defaults are conservative---three concurrent children, 50-iteration budgets, depth 1---but a \texttt{role="orchestrator"} configuration plus a spawn-depth setting unlocks nested trees, with the orchestrator prompt block explicitly ``modeled on OpenClaw's \texttt{buildSubagentSystemPrompt}.''
Background delegation returns a handle and re-enters as a \emph{fresh turn} through a completion queue rather than splicing mid-turn---a prompt-cache-preserving choice.
Above this sits a process-level tier: the \emph{Kanban swarm} runs planning-root~$\to$~parallel workers~$\to$~verifier~$\to$~synthesizer as separate \texttt{hermes -p <profile>} subprocesses coordinated through a SQLite board and a blackboard of structured JSON comments---coordinator-worker via shared database rather than protocol or channel.
A third overlay, \texttt{/moa}, fans out advisory reference models (up to eight concurrently) whose outputs season the main model's next iteration.

\subsection{Pi: Sub-Agents as an Extension, Fleets as a Package}

Core \pib{} has no spawn or task tool---sub-agents are the flagship demonstration of its extension thesis, shipped as a $\sim$1{,}000-line example.
Each invocation spawns a separate \texttt{pi --mode json -p --no-session} OS process, giving hard context isolation by construction; the parent parses the child's JSONL stdout for progress and cost aggregation.
One tool offers three shapes: \texttt{single}, \texttt{parallel} (worker pool of four, max eight tasks), and \texttt{chain} with \texttt{\{previous\}} substitution.
Agent definitions are Markdown-plus-frontmatter with tool filtering via the child's \texttt{--tools} flag; there is no depth counter and no mid-run parent-child messaging.
Above the single process, an experimental \texttt{orchestrator} package supervises fleets of \pib{} instances in RPC mode and registers presence with a hosted coordinator---multi-\emph{instance} management rather than an in-loop coordinator.
Notably, when \pib{} does do multi-agent, it lands on \emph{cross-process JSONL} rather than in-process primitives---a data point Section~\ref{sec:protocol_placement} returns to.

\subsection{OpenCode: Child Sessions with Permission-Derived Capabilities}

\ocd{}'s sub-agents are child sessions, not a separate engine: the \texttt{task} tool creates a Session with a \texttt{parentID} running the same loop, resumable via \texttt{task\_id}, returning only the final assistant text wrapped in \texttt{<task\_result>} XML; background results are injected as synthetic user messages.
Agents and modes are unified in one schema (\texttt{mode: primary|subagent|all}): built-ins are \texttt{build}, \texttt{plan} (edit denied except plan files), \texttt{general}, read-only \texttt{explore}, plus hidden \texttt{compaction}/\texttt{title}/\texttt{summary} utility agents---plan mode is a permission ruleset plus a \texttt{<system-reminder>} injection, not an architectural overlay.
Tool availability is everywhere \emph{permission-derived} (denying \texttt{"*"} for a tool removes it from the LLM's view), and containment is asymmetric: children inherit only the parent's deny rules.
Recursion is off by default (\texttt{task} auto-denied to children) but opt-in per agent with no numeric depth cap; custom agents come from config, Markdown files, or---uniquely---\emph{LLM generation} (\texttt{generateObject} over \texttt{\{identifier, whenToUse, systemPrompt\}}).

\subsection{OpenClaw Session-Based Orchestration}

\ocl{}'s multi-agent model is fundamentally different: it operates through an ACP (Agent Client Protocol, Zed/Google's editor-oriented JSON-RPC standard at agentclientprotocol.com, not IBM's older homonymous Agent Communication Protocol) translator that bridges the Gateway wire protocol to agent processes.
Sub-agents are spawned as child processes with RPC binding.
Each agent gets its own session with JSONL transcript persistence.
Communication occurs via ACP delta events (text, thought, tool calls) rather than shared memory.
A rate limiter (\texttt{MAX\_PROMPT\_BYTES = 2MB}) provides DoS protection.

\observation{\label{obs:coordinator}Coordinator-worker emerges independently across the corpus: in all four provider-native systems (\clcode{}, \cdx{}, \gemini{}, \vibe{}), in \oh{} (parallel delegation over conversation trees), in \hrm{} (configuration-gated orchestrator role plus a subprocess swarm over a SQLite blackboard---a novel coordination substrate), in \ocd{} (concurrent child sessions), and in \ocl{} at the protocol layer. We use coordinator-worker broadly here, for any parent-spawns-workers delegation; Section~\ref{sec:agentloop}'s narrower prescriptive overlay applies to \clcode{}, \cdx{}, and (gated) \hrm{} only. So many independently developed codebases converging on the same hierarchical shape is strong evidence of convergent evolution---though the convergence should be scoped to systems that build multi-agent into core: \pib{} is a production SWE harness that deliberately ships single-agent, relegating sub-agents to extension space. The implementations diverge instructively in what they optimize: \clcode{} forks context and shares the prompt cache across children (cost); \cdx{} builds a depth-tracked thread tree with typed inter-agent records, per-model tool generations, and CSV fan-out (isolation and scale); \gemini{} abstracts invocation behind a symmetric local/remote session protocol (portability); \vibe{} runs sub-agents in-process and sequentially (simplicity); \hrm{} intersects toolsets and blocklists recursion by default (containment). Hadfield et~al.~\cite{hadfield2025multiagent} describe orchestrator-worker as the natural pattern for ``valuable tasks that involve heavy parallelization, information that exceeds single context windows, and interfacing with numerous complex tools,'' which is a fair description of a long-horizon coding session.}

\subsection{Orchestration Pipelines: Claude Code vs.\ Codex}
\label{sec:orchestration_pipelines}

\clcode{} and \cdx{} are the two dominant commercial coding agents, each representing the state of the art in its provider's ecosystem.
Their orchestration pipelines, end-to-end from user request to completed task, take very different design positions.
Table~\ref{tab:pipeline_detail} highlights the key divergences.

The table reveals two coherent philosophies whose gap narrowed measurably between our snapshots.
\clcode{} is \emph{compositional-prescriptive}: it enforces structured phases (Research $\to$ Synthesis $\to$ Implementation $\to$ Verification), restricts sub-agent capabilities via tool whitelists, composes agents recursively, and optimizes aggressively around prompt-cache economics.
\cdx{} was \emph{sandboxed-emergent}---safety guaranteed at the OS level, workflow organization left to the model---but has grown prescriptive structure of its own: a user-facing collaboration-mode selector with a Plan preset, declarative sub-agent roles, depth-tracked spawn control, and a Guardian classifier that mirrors \clcode{}'s middle permission layer, while \clcode{}'s hook vocabulary and plugin format have become \cdx{} interfaces (Section~\ref{sec:evolution_window}).
Both produce strong results on real-world benchmarks; the residual philosophical difference is where safety is \emph{guaranteed} (OS enforcement vs.\ layered review), not how workflows are organized.

\section{Extensibility Mechanisms}
\label{sec:extensibility}

Every system in the corpus provides extension points, though through quite different mechanisms.

\subsection{Plugin Architectures}

\ocl{} provides the most mature plugin architecture in the corpus---and since the spring 2026 releases, an \emph{externalized} one: official provider and channel plugins were cut over to standalone first-class npm packages with manifest-first metadata, moving model catalogs and routing tables out of core.
Plugins are manifest-driven (\texttt{openclaw.plugin.json}), discovered in the workspace directory, loaded dynamically via jiti, and registered through typed SDK contracts:

\begin{itemize}[itemsep=1pt]
  \item \textbf{Channel plugins}: Implement \texttt{ChannelEntry} (receive, send, status, configure).
  \item \textbf{Provider plugins}: Implement \texttt{ProviderEntry} (inference stream).
  \item \textbf{Skill plugins}: Register via capability system.
\end{itemize}

Strict import boundaries are enforced: extension production code may only import the public plugin-SDK surface and local API barrels.
Direct imports of \texttt{src/**} are prohibited.

\ocl{} is, however, no longer alone in the manifest-driven-plugin camp.
\cdx{}'s plugin architecture expanded substantially between our snapshots: a marketplace-add pipeline (GitHub, git, local paths, URLs), a model-callable in-conversation approval tool (\texttt{request\_plugin\_install}), \texttt{.claude-plugin} format compatibility, and workspace-level sharing joined an already-present plugin manager; plugins contribute tools, skills, and lifecycle hooks.
\oh{}'s plugin system makes the convergence explicit: its plugins bundle skills, hooks, MCP config, agents, and commands, and its manifest loader accepts the directories \texttt{[".plugin", ".claude-plugin"]} with \texttt{plugin.json}---\emph{it deliberately reads \clcode{}'s plugin format}, the clearest interoperability-by-adoption case in the corpus.
\clcode{} itself supports MCP servers (for external tools), skills (for specialized workflows), and agent definitions (for custom agent types) via a JSON/Zod schema with built-in, custom, and plugin sources (Section~\ref{sec:cc_multiagent}).
\pib{} relocates the entire question: extensions are runtime-loaded TypeScript modules against an API of $\sim$33 typed events spanning session lifecycle, tool interception (blocking and in-place argument mutation), whole-context rewriting, and---unusually---raw provider I/O surgery (\texttt{before\_provider\_request}, \texttt{after\_provider\_response}); the repo ships 78 examples that are, feature for feature, other harnesses' built-ins, most at 1--16\,KB each, distributed through a package manager (\texttt{pi install npm:\ldots|git:\ldots}).
\ocd{}'s plugins are async functions returning $\sim$20 typed hooks (\texttt{chat.params}, \texttt{tool.execute.before/after}, \texttt{tool.definition} rewriting, multi-step OAuth \texttt{auth} flows), executed sequentially for determinism; custom tools need no plugin at all---any \texttt{.opencode/tool/*.ts} export becomes a tool.

\paragraph{Gemini CLI: Extensions + Slash Commands + Early MCP Adoption.}
\gemini{} ships an explicit \texttt{ExtensionLoader} interface backed by the \texttt{GeminiCLIExtension} type.
Extensions can contribute custom tools (typically via MCP), slash commands, hooks, skills, and \texttt{GEMINI.md} snippets, and are discovered from the \texttt{\textasciitilde/.gemini/extensions/} directory.
\gemini{} was an early adopter of MCP, with first-class implementations for \texttt{stdio}-spawned local servers, SSE connections, and Streamable HTTP transports, plus built-in auth providers (\texttt{google-credentials}, OAuth).
MCP support extends beyond tools to \emph{prompts} and \emph{resources} (the broader MCP spec), which most other systems' adapters omit.
Slash commands (\texttt{/agent}, \texttt{/mcp}, \texttt{/skill}, \texttt{/memory}, \texttt{/help}) are themselves extensible via skill metadata, allowing extensions to register new commands.

\paragraph{Mistral Vibe: Skills, MCP, Custom Tools, Agents, and Hooks.}
\vibe{}'s extensibility surface has grown to five orthogonal axes.
Skills follow the \texttt{agentskills.io} specification~\cite{agentskills2025}: each is a directory containing a \texttt{SKILL.md} with YAML frontmatter, discovered from \texttt{.agents/skills/}, \texttt{.vibe/skills/}, \texttt{\textasciitilde/.vibe/skills/}, \texttt{\textasciitilde/.agents/skills/}, and any user-configured paths---now supplemented by built-in skills (including a ``self-awareness'' skill that documents the harness to itself) and a \emph{remote registry client} that installs versioned skills from a hosted catalog. See Section~\ref{sec:skills} for full cross-system comparison.
MCP servers are declared as \texttt{[[mcp\_servers]]} blocks supporting \texttt{http}, \texttt{streamable-http}, and \texttt{stdio} transports, with tools namespaced as \texttt{\{server\_name\}\_\{tool\_name\}} and permissioned identically to built-ins; the layer matured between snapshots into a managed integration surface---OAuth login flows driven from the TUI (\texttt{/mcp add|login|logout|status}), a curated registry of hosted Mistral ``connectors,'' and an \emph{MCP sampling} handler that lets connected servers borrow \vibe{}'s LLM backend for their own completions, a reverse dependency no other corpus system implements.
Custom tools can be added via path-based discovery (\texttt{tool\_paths}); custom system prompts as \texttt{\textasciitilde/.vibe/prompts/\{prompt\_id\}.md}; and custom agents as \texttt{\textasciitilde/.vibe/agents/\{agent\_name\}.toml} with a deep-merge override mechanism on the base config---itself now part of an explicit configuration-layer stack (default, user TOML, trusted-project \texttt{.vibe/config.toml}, \texttt{VIBE\_*} environment, runtime overrides, agent profile).
The fifth axis is user-facing lifecycle hooks: \texttt{before\_tool}/\texttt{after\_tool}/\texttt{post\_agent\_turn} shell hooks declared in \texttt{hooks.toml} can deny a tool call, rewrite its inputs, or append context---a direct structural sibling of \clcode{}'s \texttt{PreToolUse}/\texttt{PostToolUse}, layered on top of (not replacing) the internal middleware pipeline.

\subsection{Protocol-Based Interfaces}

\mswe{} uses Python Protocols for structural subtyping: any class implementing the \texttt{Model}, \texttt{Agent}, or \texttt{Environment} protocol can be substituted without inheritance.
This is the lightest-weight extensibility mechanism, requiring no registration, no manifest, and no SDK, just interface conformance.

\subsection{Configuration-Driven Extensibility}

\adr{}'s model registry contains per-model metadata for 350+ models, including edit format, weak model name, cache control settings, extra API parameters, reasoning tag handling, and editor model configuration.
This data-driven approach enables support for new models without code changes.

\cdx{}'s layered TOML configuration now extends \emph{above} the user: mobile-device-management (MDM), host-wide system, and enterprise cloud-delivered layers enter as base defaults beneath user (\texttt{config.toml} plus named profiles), project (\texttt{.codex} directory), and session-flag scopes, while a separately composed requirements/constraint engine (fed from the same MDM, system, and cloud sources) validates and can hard-restrict what the lower layers may set. It is the most enterprise-governance-oriented configuration model in the corpus.
\vibe{} now mirrors the layered approach (six layers merged by a \texttt{ConfigBuilder} with per-field merge strategies), making explicit config-layer stacks a two-system pattern.

\ocl{}'s config system is the most complex in the corpus: a Zod-validated \texttt{openclaw.json} with hundreds of nested structures, plus a large auto-generated TypeScript file of validation rules for all bundled channels alongside a hand-maintained schema-help surface.

\subsection{Model Context Protocol (MCP)}

MCP remains a cross-cutting extension mechanism, supported by \clcode{}, \cdx{}, \gemini{}, \vibe{}, \oh{}, \hrm{}, \ocd{}, and \ocl{}---8 of 11 systems (7 of 10 coding-first).
\oh{} implements MCP via \texttt{MCPToolAction}/\texttt{MCPToolObservation} event types, with a single \texttt{MCPServer} model validating four transports (\texttt{stdio}, \texttt{http}, \texttt{streamable-http}, \texttt{sse}) and first-class authentication up to full OAuth flows with token state.
\hrm{}'s support is near-complete (stdio/Streamable HTTP/SSE, resources, prompts, sampling, elicitation, OAuth; only prompt-list-change notifications ignored), and \hrm{} is \emph{itself} an MCP server, matching \ocl{}'s channel bridge.
\ocd{} covers tools, prompts, resources, and resource templates with OAuth dynamic client registration (sampling and elicitation explicitly disabled with tracking-issue comments).
The corpus's one principled holdout is \pib{}, whose documentation rejects MCP outright---``build CLI tools with READMEs''---making skills-plus-shell its articulated alternative to a wire protocol.

\subsection{Skills as an Emerging Capability-Bundle Standard}
\label{sec:skills}

Alongside MCP, a second cross-cutting extensibility mechanism has shown up in the corpus: \emph{skills}.
MCP standardizes how an agent talks to an external tool process; skills standardize how an agent packages a unit of expertise into a discoverable directory (instructions, optional scripts, optional tool whitelists, metadata) that can be dropped into a project or user home and picked up by the agent at discovery time.
Mitra et~al.~\cite{mitra2024agentinstruct} frame the same idea from the training side: skills as teachable units in agentic flows.
Nine of the eleven systems implement some form of skills system; only \adr{} and \mswe{} omit it (Table~\ref{tab:skills}).

\begin{table}[H]
\centering
\caption{Skills support across the eleven systems. ``Spec'' indicates conformance with the \texttt{agentskills.io} specification (\texttt{SKILL.md} + YAML frontmatter, \texttt{name}/\texttt{description} required).}
\label{tab:skills}
\scriptsize
\renewcommand{\arraystretch}{1.25}
\setlength{\tabcolsep}{3.5pt}
\begin{tabularx}{\textwidth}{@{}l c L{4.2cm} L{3.0cm} L{4.2cm}@{}}
\toprule
\textbf{System} & \textbf{Skills?} & \textbf{Discovery paths} & \textbf{Invocation} & \textbf{Spec / loading} \\
\midrule
\clcode{} & \cmark & \texttt{.claude/skills/}, \texttt{\textasciitilde/.claude/skills/}, plus dynamic discovery on file edits gated by \texttt{paths} frontmatter & \texttt{SkillTool} (deferred); MCP prompts dedup'd via \texttt{uniqBy} on name & Custom (path-gated extension); hybrid eager + on-demand \\
\cdx{}    & \cmark & \texttt{core-skills/} + \texttt{skills/} roots; plugin-provided skill roots & \texttt{skills/list} app-server RPC + TUI \texttt{\$}-mention; implicit-invocation detection & Custom Rust; metadata eager under a context-scaled token budget, body injected on \$-mention or detected invocation; skills declare MCP/env dependencies that \cdx{} auto-installs \\
\gemini{} & \cmark & \texttt{\textasciitilde/.gemini/skills/}, \texttt{.gemini/skills/}, \texttt{.agents/skills/}, plus built-ins (\texttt{skill-creator}, \texttt{antigravity-support}); precedence workspace$>$user$>$built-in & \texttt{ActivateSkill} tool wraps skill in XML with \texttt{available\_resources} tree; permission-confirmation UI & agentskills.io-aligned with \texttt{isBuiltin} extension; eager startup walk; path-traversal-hardened installs \\
\vibe{}   & \cmark & \texttt{.agents/skills/}, \texttt{.vibe/skills/}, \texttt{\textasciitilde/.vibe/skills/}, \texttt{\textasciitilde/.agents/skills/}, user \texttt{skill\_paths}; trust-gated project discovery & \texttt{skill} tool (deferred); user \texttt{/skill-name} invocations materialized as synthetic tool calls & Full \texttt{agentskills.io} compliance; remote registry client (versioned catalog); builtin self-awareness skill \\
\oh{}     & \cmark & \texttt{\{workdir,git-root\}/.agents/skills/}, \texttt{.openhands/skills/} (+legacy microagents); user equivalents + managed installs & Metadata in \texttt{<available\_skills>}; body on demand via builtin \texttt{invoke\_skill} tool & agentskills.io flag; \emph{progressive disclosure}; \texttt{PathTrigger} rules on file-touch; server CRUD/sync/marketplace API; ingests AGENTS.md/.cursorrules as scoped rules \\
\hrm{}    & \cmark & \texttt{\textasciitilde/.hermes/skills/} (seeded from 72 bundled; 102 more official skills installable via the Skills Hub) + read-only \texttt{external\_dirs} & Three-tier progressive disclosure: index in prompt $\to$ \texttt{skill\_view} $\to$ linked assets; skills double as slash commands & agentskills.io-compatible; trust-tiered Skills Hub (builtin/trusted/community) with pre-install scanning and quarantine; self-authoring via \texttt{skill\_manage} \\
\pib{}    & \cmark & \texttt{\textasciitilde/.pi/agent/skills/}, \texttt{\textasciitilde/.agents/skills/}, \texttt{.pi/skills/}, \texttt{.agents/skills/} (cwd to git root), packages, \texttt{--skill}; project paths trust-gated & No dedicated tool: \texttt{<available\_skills>} XML index; body loaded via the ordinary \texttt{read} tool; \texttt{/skill:name} expansion & agentskills.io, deliberately lenient (warns, still loads); \texttt{disable-model-invocation} gating \\
\ocd{}    & \cmark & \texttt{\{skill,skills\}/**/SKILL.md} in config dirs, \texttt{\textasciitilde/.claude/skills}, \texttt{.claude/}, \texttt{.agents/} walk-up, extra paths, remote URL registries (versioned \texttt{index.json} cache) & Native \texttt{skill} tool (deferred: catalog lists name+description only) & agentskills.io field set; per-skill permission gating \\
\ocl{}    & \cmark & \texttt{\textasciitilde/.openclaw/skills/}, \texttt{.agents/skills/}, plugin-provided, \texttt{skills.load.extraDirs} config & Injected as XML in system prompt; \texttt{/skills} slash command lists eligible & agentskills.io-compatible; gating via \texttt{metadata.openclaw.requires} (\texttt{bins}, \texttt{env}, OS); governed installs (Skill Workshop approval flow, provenance-verified ClawHub) \\
\adr{}    & \xmark & ---  & --- & --- \\
\mswe{}   & \xmark & ---  & --- & --- \\
\bottomrule
\end{tabularx}
\end{table}

\paragraph{Convergence on the SKILL.md format and the \texttt{.agents/skills/} path.}
The format is essentially universal among the nine adopters: a directory named after the skill, containing a \texttt{SKILL.md} file with YAML frontmatter (at minimum a \texttt{name} and \texttt{description}, often with optional fields like \texttt{paths}, \texttt{requires}, or \texttt{user-invocable}).
The \texttt{.agents/skills/} discovery path, the canonical agentskills.io location, is accepted by six systems (\vibe{}, \gemini{}, \ocl{}, \oh{}, \pib{}, \ocd{}), and the vendor-specific homes (\texttt{\textasciitilde/.claude/skills/}, \texttt{\textasciitilde/.gemini/skills/}, \texttt{\textasciitilde/.vibe/skills/}, \texttt{\textasciitilde/.hermes/skills/}, \texttt{\textasciitilde/.pi/agent/skills/}, \texttt{\textasciitilde/.openclaw/skills/}) coexist with project-local equivalents.
The sharpest interoperability datum: \ocd{} \emph{deliberately searches a competitor's home}---its discovery list includes \texttt{\textasciitilde/.claude/skills} and \texttt{.claude/} directories, so skills installed for \clcode{} work in \ocd{} unmodified.

\paragraph{Skills versus MCP: the tie has broken.}
MCP and skills address different layers.
MCP is a \emph{wire protocol} for an agent to talk to a separately running process that exposes tools, prompts, and resources; skills are a \emph{file-system convention} for packaging an instruction-plus-script bundle the agent reads directly.
The two compose: \clcode{}'s skill registry deduplicates MCP-supplied prompts that look like skills, and \vibe{} composes in the other direction, letting MCP servers borrow its LLM via sampling while skills arrive from a hosted registry.
In April the two standards were tied at 6/8 adoption; the expanded July corpus breaks the tie in skills' favor---\textbf{skills 9/11, MCP 8/11}---because \pib{} implements agentskills.io while rejecting MCP outright.
Skills-as-MCP-replacement is now an articulated position, not a coexistence.

\paragraph{Deferred loading as the dominant pattern.}
Echoing \clcode{}'s deferred tool loading (Section~\ref{sec:deferred}), eight of the nine adopters now load only skill \emph{metadata} eagerly and fetch bodies on demand---via a dedicated tool (\texttt{SkillTool}, \texttt{ActivateSkill}, \texttt{skill}, \oh{}'s \texttt{invoke\_skill}), via \$-mention and implicit-invocation detection (\cdx{}), via a three-tier disclosure chain (\hrm{}), or, most minimally, via the ordinary \texttt{read} tool against an XML index (\pib{}).
\oh{}, an eager outlier in April, moved to progressive disclosure with the V1 SDK; \ocl{} remains the one eager loader, compensating with eligibility gating so ineligible skills never reach the prompt.

\paragraph{Gating and conditional activation.}
The most architecturally interesting variation is conditional activation.
\ocl{}'s skills declare runtime requirements (binaries on \texttt{PATH}, environment variables, OS family) and are filtered at load time, preventing ``how to use \texttt{kubectl}'' from polluting the prompt on a machine without it.
\clcode{}'s \texttt{paths} frontmatter activates a skill when the model touches matching files (via the Read, Edit, or Write tools), and \oh{} now implements the same idea as \texttt{PathTrigger} rules injected on file-touch---marrying skills to the JIT context philosophy of~\cite{rajasekaran2025context}.
This is a structural advance over MCP, which treats every connected server symmetrically regardless of whether its tools are relevant to the current task.

\paragraph{Skills grow a supply chain---and authors.}
Distribution has professionalized in one quarter.
Remote registries now exist in four systems (\vibe{}'s hosted catalog client, \ocd{}'s URL registries with versioned index caches, \hrm{}'s Skills Hub, \oh{}'s marketplace API), and with distribution comes supply-chain security: \hrm{} applies trust tiers, pre-install static scanning, quarantine, and dependency checks; \ocl{} routes installs through an approval workflow with provenance-verified sources; \gemini{} hardened skill installs against path traversal; \cdx{} auto-installs a skill's declared MCP-server dependencies (optionally after prompting).
More striking still, skills have acquired \emph{non-human authors}: \hrm{}'s background-review agent creates and patches skills from completed tasks (with a curator maintaining the collection), and \gemini{}'s extraction sub-agent mines sessions into skill patches for inbox review.
The capability-package layer is acquiring package-manager economics---registries, provenance, and now generated packages---at remarkable speed.

\observation{\label{obs:skillsmcp}Skills have overtaken MCP as the corpus's most-adopted extensibility standard: 9/11 systems implement SKILL.md bundles (only \adr{} and \mswe{} abstain) versus 8/11 for MCP, the tie of April broken by \pib{}'s explicit skills-yes-MCP-no position. Three second-order developments mark the layer's maturation: deferred loading is now the near-universal strategy (8 of 9 adopters); conditional activation (\clcode{} \texttt{paths}, \oh{} \texttt{PathTrigger}, \ocl{} \texttt{requires}) pushes JIT context engineering into the extensibility layer; and a supply chain has emerged---hosted registries in four systems, trust tiers and quarantine in \hrm{}, provenance verification in \ocl{}---together with the first agent-\emph{authored} skills (\hrm{}'s self-improving loop, \gemini{}'s extraction inbox).}

\section{Cross-Cutting Observations and Implications}
\label{sec:synthesis}

\subsection{Architectural Pattern Catalog}

Tables~\ref{tab:patterns} and~\ref{tab:patterns2} catalog the 29 recurring design patterns identified across the eleven systems: the seventeen of the April edition (with membership updated) plus twelve contributed or crystallized by the expanded corpus.

\begin{table}[H]
\centering
\caption{Catalog of 29 recurring architectural patterns (1/2): the April-edition seventeen, membership updated to July~2026.}
\label{tab:patterns}
\scriptsize
\renewcommand{\arraystretch}{1.2}
\begin{tabularx}{\textwidth}{@{}L{3cm} L{4.6cm} L{4.9cm}@{}}
\toprule
\textbf{Pattern} & \textbf{Description} & \textbf{Used By} \\
\midrule
Event Sourcing            & Actions/observations appended to a persistent event log & \oh{}, \pib{} (session log-as-tree), \ocd{} (log-as-queue; v2 tables) \\
Policy-as-Code            & Safety rules as executable configurations & \cdx{} (Starlark, with inline validated examples), \clcode{} (hooks), \gemini{} (TOML modes), \hrm{} (config-as-policy + hardline floor), \ocd{} (rulesets), \pib{} (extension hooks) \\
Recursive Composition     & Agents spawn sub-agents with forked/linked context & \clcode{}, \cdx{}, \oh{}, \ocl{}; config-gated in \hrm{}, opt-in in \ocd{} \\
Polymorphic Edits         & Model-aware edit format/toolset selection & \adr{} (prompt factory), \ocd{} (tool registry) \\
Deferred Loading          & Tools/skills hidden from prompt, discovered on demand & \clcode{}, \cdx{} (BM25 \texttt{tool\_search}), \hrm{} (BM25 bridge tools), + 8 skills implementations \\
Template Method           & Base class defines flow; subclasses override parse/format & \adr{} (Coder), \oh{} (Agent) \\
Protocol Interfaces       & Structural subtyping / interface seams for pluggable components & \mswe{}, \pib{} (per-tool \texttt{Operations} remoting seam) \\
LLM Summarization         & LLM compresses conversation history & 9 systems (all but \mswe{}, \adr{}-partial) \\
Stuck Detection           & Automated detection of repetitive agent behavior & \oh{} (5 scenarios), \gemini{} (hybrid hash+LLM), \hrm{} (warn-first signatures), \ocd{} (doom-loop ask), \mswe{} (format-error cap) \\
Reflection Loop           & Inner self-correction cycle with lint/test feedback & \adr{}; cousins: \gemini{} (edit fixer), \ocd{} (LSP feedback) \\
Prompt Caching            & Cache-boundary prompt structure or session-key reuse & \clcode{}, \cdx{}, \oh{} (cache tiers), \hrm{} (prefix normalization), \pib{} (breakpoints+TTL), \ocd{} (dialect fanout) \\
Context Forking           & Cloning parent state for sub-agent isolation & \clcode{}, \cdx{}, \hrm{} (cache-sharing background fork) \\
Middleware Pipeline       & Composable turn-level policies orthogonal to loop body & \vibe{} \\
JIT Repo Context          & Hierarchical Markdown context files auto-discovered + on-demand injection & \clcode{}, \cdx{}, \gemini{}, \vibe{}, \hrm{}, \pib{}, \ocd{}, \oh{} \\
Skills (capability bundles) & SKILL.md directories with YAML frontmatter & 9 systems (all but \adr{}, \mswe{}) \\
Conditional Activation    & Skills/tools gated by environment or file paths & \clcode{} (\texttt{paths}), \oh{} (\texttt{PathTrigger}), \ocl{} (\texttt{requires}) \\
Turn-Level Checkpoint     & Filesystem snapshots with rewind and file restoration & \vibe{} (per user message), \ocd{} (shadow-git, per step), \hrm{} (shadow-git store), \pib{} (conversation-only) \\
\bottomrule
\end{tabularx}
\end{table}

\begin{table}[H]
\centering
\caption{Catalog of 29 recurring architectural patterns (2/2): the twelve patterns new to the July corpus.}
\label{tab:patterns2}
\scriptsize
\renewcommand{\arraystretch}{1.2}
\begin{tabularx}{\textwidth}{@{}L{3cm} L{4.6cm} L{4.9cm}@{}}
\toprule
\textbf{Pattern} & \textbf{Description} & \textbf{Used By} \\
\midrule
Agent-Maintained Memory   & Background sub-agent extracts and consolidates cross-session memory & \cdx{} (two-phase, git-baselined); human-gated variant in \gemini{} (patch inbox) \\
Outer Verification Loop   & Scaffold-level judge/guard validating completion, outside the turn loop & \oh{} (\texttt{/goal} judge + critics), \hrm{} (verify-on-stop) \\
Self-Improving Skill Loop & Agent authors, patches, and curates its own capability bundles & \hrm{}; partial: \gemini{} (skill-extraction inbox) \\
Lineage Compaction        & Compaction as session rotation with searchable ancestry chain & \hrm{} \\
Session-Tree Version Control & Append-only entry tree with movable head; fork/rewind/branch summaries & \pib{}; \oh{} (conversation tree) \\
Minimal-Core / Extension-Host & Safety, sandbox, sub-agents, plan mode relocated to a runtime event bus & \pib{} \\
Client/Server Harness     & Embedded API server; every UI (TUI, web, IDE, CI) is a client & \ocd{}; \oh{} (agent-server) \\
Model-Family Prompt Matrix & Distinct base prompts dispatched per model family/generation & \cdx{} (server catalog), \ocd{} (9 prompts), \hrm{} (gated blocks) \\
Cache-Dialect Fanout      & All providers' cache-control dialects emitted simultaneously & \ocd{} \\
Syntax-Aware Command Permissioning & Commands parsed (tree-sitter) and grants scoped by argument arity & \ocd{}; \vibe{} (parse-validation), \hrm{} (deobfuscated matching) \\
Untrusted-Content Delimiting & Tool/web results wrapped in taint markers with delimiter defanging & \hrm{}, \oh{} (\texttt{<UNTRUSTED\_CONTENT>}) \\
Harness Mimicry           & Client presents a first-party harness's identity (headers, prompt opening, tool-name casing) to ride its subscription OAuth backend & \pib{} (Claude Code identity on Anthropic OAuth; ChatGPT-plan Codex backend) \\
\bottomrule
\end{tabularx}
\end{table}

\subsection{The Twin Absences: No Agentic Frameworks, No Code RAG}
\label{sec:twin_absences}

Two technologies that the broader LLM-application literature treats as central to any production agent turn out to be missing from all eleven systems.\footnote{We expected to find at least some use of LangChain or another framework in the open-source projects. The uniformity surprised us; the search for counterexamples (vendored dependencies, dynamic imports, transpiled TypeScript builds) ran for several weeks before we accepted the result. The July re-audit repeated the manifest and import sweep across all twelve trees, including the three systems new to the corpus and the meta-harness.}
For this analysis, the absences are at least as informative as the presences.

\paragraph{Absence 1: agentic frameworks.}
Every dependency manifest was inspected, and every source tree grepped for imports of the widely deployed agentic frameworks: LangChain, LangGraph, LlamaIndex, AutoGen, CrewAI, Pydantic AI, Genkit, Haystack agents, Semantic Kernel, Google's ADK, and several smaller libraries (Smolagents, Swarm, Agno).
Across roughly 4\,M lines of Python, TypeScript, and Rust, no production agent code path imports any of them---a result worth savoring in two directions: \gemini{} uses neither of Google's own frameworks (Genkit, ADK), and \hrm{}'s only ``LangChain'' strings sit inside bundled skill \emph{documentation} teaching the agent how users might use vector databases.
Two boundary cases deserve precision.
\adr{} ships one optional framework-adjacent extra: its \texttt{/help} command can install \texttt{llama-index} to run doc-RAG over \emph{aider's own documentation}---an opt-in feature dependency, not agent-loop orchestration.
And \ocd{} delegates its inner LLM/tool plumbing to Vercel's AI SDK---a provider-abstraction layer, not an agent-orchestration framework, but the first corpus system whose inner loop plumbing is a third-party SDK at all (an in-house replacement client sits behind a flag, suggesting the dependency is transitional).

Every loop is hand-rolled in the host language's native primitives: \texttt{asyncio} (\oh{}, \vibe{}, \hrm{}), blocking synchronous Python (\adr{}, \mswe{}), \texttt{Promise}/async-iterator (\clcode{}, \gemini{}, \pib{}, \ocd{}, \ocl{}), Tokio (\cdx{}).
Every tool registry is custom-built around Pydantic, Zod, TypeBox, Effect Schema, or Rust enums.
Every prompt template is plain Markdown, Jinja2, or string concatenation.

\paragraph{Absence 2: retrieval-augmented generation over code.}
A parallel search covered vector-store dependencies (Chroma, Pinecone, Weaviate, Qdrant, Milvus, FAISS, LanceDB, sqlite-vec, Elasticsearch in vector mode), embedding libraries, and folders or files matching \texttt{embedding}, \texttt{vector\_store}, \texttt{vectordb}, \texttt{rag}, or \texttt{retrieval}.
For \emph{code} retrieval the result is unchanged and now spans eleven systems: zero.
The exceptions all concern \emph{conversation memory}, and here the July picture did shift: \ocl{}'s default memory plugin (\texttt{memory-core}) now runs a hybrid sqlite-vec KNN + FTS5/BM25 search with embeddings on by default (the provider defaults to OpenAI; a local GGUF model is opt-in; the April-era LanceDB extension survives as an optional plugin), making \ocl{} the one system where embeddings are on by default---for chat recall, never for reading the source tree.
\hrm{} demonstrates the opposite choice at scale: its core past-conversation search is deliberately lexical (trigger-maintained SQLite FTS5, BM25 plus trigram tokenization for CJK, ``no LLM calls anywhere''), with embeddings confined to opt-in memory plugins.

What the eleven systems use \emph{instead} of RAG for code retrieval is summarized in Table~\ref{tab:retrieval}.

\begin{table}[H]
\centering
\caption{Code-retrieval mechanisms used by each system, in lieu of vector-based RAG (July~2026).}
\label{tab:retrieval}
\small
\renewcommand{\arraystretch}{1.2}
\begin{tabularx}{\textwidth}{@{}l L{3.2cm} L{7.8cm}@{}}
\toprule
\textbf{System} & \textbf{Embeddings used?} & \textbf{Code-retrieval mechanism} \\
\midrule
\oh{}     & No & \texttt{GrepTool}/\texttt{GlobTool} + terminal; context-file ingestion (AGENTS.md, .cursorrules); tree-sitter no longer present in the SDK \\
\adr{}    & Optional \texttt{/help} extra only (doc-RAG over aider's own docs) & \texttt{RepoMap}: tree-sitter symbol extraction with token-budget-constrained PageRank-style ranking \\
\clcode{} & No & ripgrep keyword search + BashTool + on-demand file Read; CLAUDE.md auto-discovery \\
\cdx{}    & No & Rust-native file search; AGENTS.md root-to-cwd concatenation \\
\gemini{} & No & Bundled ripgrep + glob tools; GEMINI.md auto-discovery; \texttt{MEMORY.md} project index; JIT subdirectory context \\
\vibe{}   & No & ripgrep + tree-sitter (bash parsing) + filesystem search; git status injection; JIT nested AGENTS.md \\
\mswe{}   & No & Grep-based search via shell tools \\
\hrm{}    & Opt-in memory plugins only & ripgrep-backed \texttt{search\_files}; SQLite FTS5 (BM25 + trigram) over \emph{session history}; LSP for post-write diagnostics only; no repo map \\
\pib{}    & No & Auto-downloaded ripgrep + fd binaries (latest release; system binaries preferred) behind \texttt{grep}/\texttt{find} tools; ancestor-walk context files; session search is a deterministic linear scan \\
\ocd{}    & No & Bundled ripgrep (\texttt{grep}/\texttt{glob}, 100-result cap); $\sim$25 auto-downloaded LSP servers feed diagnostics (no persistent index); lazy nested AGENTS.md attach \\
\ocl{}    & \textbf{Default} hybrid memory search (sqlite-vec KNN + FTS5/BM25; OpenAI embedder by default, local GGUF opt-in) for \emph{conversation memory only}, never code & N/A for code retrieval; standard filesystem APIs \\
\bottomrule
\end{tabularx}
\end{table}

\paragraph{Why the twin absences matter.}
The first absence corroborates advice from the agent designers themselves.
Schluntz \& Zhang's \emph{Building Effective Agents}~\cite{schluntz2024agents} cautions that frameworks ``often create extra layers of abstraction that can obscure the underlying prompts and responses, making them harder to debug,'' and recommends starting from raw SDK calls.
That advice appeared in December~2024; the source-code audit we run here, in 2026, shows that every commercial provider has followed it.
This is worth noting because the broader Python LLM-application community has kept investing in framework abstractions over the same period.
Production SWE agents appear to operate on a different complexity budget from generic LLM applications.
Hand-rolled debuggable code wins out over reusable abstractions once the agent is mutating real source code, because the failure modes (silent prompt corruption, opaque caching, version-incompatible tool schemas) become too expensive to shrug off.

The second absence is consistent with recent guidance from Rajasekaran et~al.~\cite{rajasekaran2025context} (September~2025), who favor ``just-in-time'' (JIT) retrieval while recommending a hybrid approach: ``We don't memorize entire corpuses of information, but rather introduce external organization and indexing systems like file systems, inboxes, and bookmarks to retrieve relevant information on demand.''
The blog post does not flatly reject pre-indexed retrieval, but its practical recommendations all point toward JIT methods.
Every system in the corpus follows this approach, even though repository-level retrieval over code is a well-studied area: RepoCoder~\cite{zhang2023repocoder}, RepoBench~\cite{liu2024repobench}, and CrossCodeEval~\cite{ding2023crosscodeeval} all build retrieval pipelines for code, and Long Code Arena~\cite{bogomolov2024longcodearena} provides matching long-context evaluation.
Wang et~al.'s CodeRAG-Bench~\cite{wang2025coderagbench} asks the very question this section raises, ``can retrieval augment code generation?'', and finds gains that are highly variable across tasks: substantial for documentation-lookup and library-use scenarios but marginal or absent for tasks where the model already has sufficient parametric knowledge. The inconsistency of these gains may help explain why production SWE agents skip the RAG layer entirely rather than investing in task-dependent retrieval routing.
The reasons are domain-specific.
Code carries dense deterministic structural metadata---file paths, language servers, tree-sitter parses, type information---that semantic-similarity retrieval cannot replicate (and that structure-aware tools such as RepoAgent~\cite{luo2024repoagent} and \adr{}'s \texttt{RepoMap} exploit directly).
Code changes minute to minute, so pre-indexed embeddings are stale almost by construction.
Every coding environment already ships a near-optimal retrieval system in the form of \texttt{ripgrep}, \texttt{find}, and \texttt{glob}.
On the typical SWE-agent task, RAG adds operational cost (embedding compute, vector-store maintenance, drift management) without offering marginal value.

\observation{\label{obs:absences}Two technologies that the broader LLM-application literature treats as central are missing from all eleven systems---a finding that survived a threefold corpus expansion and a three-month re-audit. No system uses a general-purpose agentic framework in its agent runtime (LangChain, LangGraph, AutoGen, CrewAI, or any of a dozen others checked; \gemini{} uses neither of Google's own); every loop is hand-rolled in the host language's async primitives, with \ocd{}'s use of a provider-abstraction SDK for inner plumbing as the closest boundary case. No system uses vector-embedding RAG for code retrieval; all rely on \texttt{ripgrep}, tree-sitter, glob, and auto-discovered Markdown context files, and where conversation-scale recall is needed, the production answer is lexical search (\hrm{}'s SQLite FTS5) or, in exactly one default configuration (\ocl{}), hybrid embeddings over chat history---never over the source tree. Production harnesses operate on a different complexity budget from generic LLM applications: debuggability and prompt transparency outweigh framework reuse when the failure mode is mutating real code. Section~\ref{sec:merger} gives the absence its historical resolution.}

\observation{\label{obs:guidance}The architectural patterns described in Anthropic's \emph{Effective Agents} engineering series (December~2024 through September~2025) line up closely with the architectures observed across the four provider-native systems, which were built independently. Whether this reflects shared empirical reality, public-guidance influence, or both is an open question.}

\subsection{Inter-Agent Protocols: Outward Adoption, Inward In-Process}
\label{sec:protocol_placement}

The twin absences are about two technologies that simply do not appear in the corpus.
Inter-agent protocols (ACP, A2A) follow a different pattern again---and this is the dimension that moved most between our snapshots.
ACP now ships as a first-class production dependency or implementation in \emph{six} of the eleven systems: \vibe{} (\texttt{agent-client-protocol==0.10.1}, with session fork, workspace trust, and rewind exposed over the protocol), \ocl{} (gateway translator), \ocd{} (\texttt{opencode acp} serves the agent side over stdio), \hrm{} (ACP adapter and registry modules), \oh{} (an \texttt{ACPAgent} class, discussed below), and \gemini{} (whose A2A module is a separate protocol for the mesh role).
Outside the corpus, xAI's Grok Build launched with a documented ACP server mode on day one (Section~\ref{sec:landscape}; vendor-documented, not source-verified).
None of the corpus adoptions is experimental.

What is interesting is the placement---and ACP now occupies \emph{three distinct architectural roles}:

\begin{enumerate}[itemsep=2pt]
  \item \textbf{Editor $\leftrightarrow$ agent (outward server).} The LSP role for ACP~\cite{acp2025}: Zed, JetBrains, and other IDEs drive a local agent as they would a language server. \vibe{}, \ocd{}, \hrm{}, \ocl{}, and Grok Build all serve this boundary.
  \item \textbf{Agent-as-backend (inward host).} The role nobody held in April: \oh{}'s \texttt{ACPAgent} delegates its \texttt{step()} to an external ACP server, with provider metadata for pinned \texttt{claude-agent-acp}, \texttt{codex-acp}, and \texttt{gemini --acp} binaries---\emph{rival harnesses become interchangeable brains inside an \oh{} conversation}. The meta-harness of Section~\ref{sec:omnigent} drives its Goose and Qwen adapters the same way (and taps Kiro's ACP permission stream while driving its TUI), and \hrm{} consumes an ACP agent as a \emph{model transport} (GitHub Copilot's CLI as a chat backend). The protocol built for editors turned out to be the interface for hosting.
  \item \textbf{Cross-vendor mesh (A2A~\cite{a2a2025}).} Still \gemini{}'s alone in the corpus: a remote orchestrator drives a local \gemini{} as one node in a multi-vendor topology, now with usage metadata over the wire.
\end{enumerate}

Table~\ref{tab:inter_agent_layers} lists the communication mechanism each multi-agent system uses between a main agent and its own sub-agents, alongside its protocol roles.

\begin{table}[H]
\centering
\caption{Main-agent $\leftrightarrow$ sub-agent communication and protocol placement (July~2026).}
\label{tab:inter_agent_layers}
\scriptsize
\renewcommand{\arraystretch}{1.25}
\begin{tabularx}{\textwidth}{@{}l L{4.0cm} L{4.6cm} L{2.2cm}@{}}
\toprule
\textbf{System} & \textbf{Main $\leftrightarrow$ sub-agent} & \textbf{Protocol roles} & \textbf{Sub-agents cross-proc.?} \\
\midrule
\clcode{}  & \texttt{AgentTool} function call + context fork + XML notifications & None at agent layer (ACP served by a separate adapter binary) & No \\
\cdx{}     & Session input queue with mailbox phases; typed records & None at agent layer (\texttt{codex-acp} adapter external) & No \\
\gemini{}  & \texttt{invoke\_agent} behind local/remote session protocol & \textbf{A2A} server (mesh role) & No (internal) \\
\vibe{}    & \texttt{task} tool, in-process \texttt{asyncio} & \textbf{ACP} server (editor role; rewind over protocol) & No \\
\oh{}      & Task/delegate tools, concurrent threads & \textbf{ACP} \emph{client/host}: rival harnesses as step() backends & No (own sub-agents) \\
\hrm{}     & In-process thread forks; swarm via SQLite blackboard subprocesses & \textbf{ACP} server (editors) \emph{and} client (Copilot CLI as model backend) & Swarm: yes (DB, not protocol) \\
\pib{}     & Extension spawns OS processes, JSONL over stdio & Proprietary JSONL RPC ($\sim$30 commands) for embedding; ACP/A2A/MCP absent by design & \textbf{Yes} (ext.; JSONL, not ACP) \\
\ocd{}     & \texttt{task} tool, in-process child sessions & \textbf{ACP} server (agent side, stdio ndjson) & No \\
\ocl{}     & \textbf{ACP session spawn} over RPC (child process) & Same ACP inward and outward & \textbf{Yes} \\
\bottomrule
\end{tabularx}
\end{table}

The inward/outward split of the April edition survives in refined form.
For coordinating \emph{their own} sub-agents, eight of the nine multi-agent systems still use in-process primitives or, where they cross processes, something other than the standard protocols: \pib{}'s extension spawns OS processes speaking plain JSONL, and \hrm{}'s swarm coordinates through a SQLite blackboard.
What changed is the \emph{inward consumption of whole harnesses}: hosting a rival agent as a swappable backend---\oh{}'s ACP hosting, \hrm{}'s ACP-as-model-transport, the meta-harness's adapter fleet---is a production pattern that did not exist in April, and ACP is its lingua franca.
\pib{} adds a final nuance: it rejects the standard protocols yet \emph{invents} a proprietary outward RPC for the same embedding role---evidence that the outward-facing protocol pressure is real even where the standards are declined.

\observation{\label{obs:protocols}Inter-agent protocol placement has evolved from a two-role story (outward: editor integration and mesh; inward: nothing) to a three-role story. ACP ships in six of eleven systems and now serves (1)~its designed editor$\leftrightarrow$agent boundary, (2)~a role outside the protocol's design brief---\emph{harness hosting}, where \oh{} runs \clcode{}, \cdx{}, or \gemini{} as interchangeable step-backends and \hrm{} consumes an ACP agent as a model transport---and (3)~via A2A, still only in \gemini{}, the cross-vendor mesh. For a harness's \emph{own} sub-agents, eight of nine multi-agent systems still use in-process primitives or, where they cross processes, something other than the standard protocols (\pib{}'s JSONL extension, \hrm{}'s SQLite-blackboard swarm); \ocl{}'s ACP spawn is the lone case that routes sub-agent coordination over a standard protocol. The practitioner guidance sharpens accordingly: build an ACP \emph{server} (it now buys you editors, hosts, and meta-orchestrators at once); keep your sub-agents in-process; and treat A2A as a bet on a cross-vendor mesh whose at-scale demand remains unproven.}

\subsection{Trade-off Framework}

We formalize five fundamental trade-off axes:

\paragraph{Axis 1: Simplicity vs.\ Capability.}
\mswe{}'s minimal scaffold reportedly achieves 74\%+ on SWE-Bench Verified, while \cdx{}'s far larger codebase reports 69.1\%.
These figures are not directly comparable (different underlying models, different evaluation runs, different deployment configurations: \mswe{} is unsandboxed and \cdx{} is sandboxed) and we do not draw a head-to-head conclusion from them.
What the gap does illustrate, qualitatively, is that the additional code in \cdx{} is not invested in raw task-completion logic: a substantial share goes into safety (cross-platform sandboxing), user experience (TUI, streaming), extensibility (MCP, plugins), and robustness.
Whether the production scaffolding produces measurable improvements on the same benchmark under matched conditions is an open question that this study does not answer (see Section~\ref{sec:discussion} on threats to validity).

\paragraph{Axis 2: Safety vs.\ Autonomy.}
Systems with more safety infrastructure (\cdx{}, \clcode{}) impose more friction on agent actions.
Systems with less safety (\mswe{}, \adr{}) enable faster execution but are unsuitable for enterprise deployment without additional safeguards.

\paragraph{Axis 3: Provider Coupling vs.\ Agnosticism.}
\clcode{} and \cdx{} exploit provider-specific features (prompt caching, extended thinking, model-specific prompts) at the cost of vendor lock-in.
LiteLLM-based systems sacrifice these optimizations for provider flexibility.

\paragraph{Axis 4: Monolithic vs.\ Modular.}
\cdx{}'s 126-crate, million-line Rust workspace optimizes for type safety and performance.
\mswe{}'s single-file agent maximizes comprehensibility.
\pib{} relocates the entire axis into runtime composition: a minimal core plus an extension event bus.
\ocl{}'s plugin SDK and \ocd{}'s client/server split enable extensibility and embeddability respectively.
Each point on this spectrum serves different user populations.

\paragraph{Axis 5: Scaffold Complexity vs.\ Model Capability.}
This is the most philosophically interesting axis.
\mswe{}'s competitive benchmark performance demonstrates that current models are capable enough to succeed with minimal scaffolding.
Production systems invest in scaffolding not because models need it for task completion, but because users need it for safety, reliability, and workflow integration.

\section{The Platform Turn}
\label{sec:platformturn}

The preceding sections dissected the harness as an artifact.
This section argues that the artifact has changed category: between 2025 and mid-2026, the coding-agent harness completed a turn from \emph{tool} to \emph{platform}---a runtime with its own extension ecosystem, package economics, governance layers, switching costs, and, since June~2026, its own meta-layer.
The April edition of this study advanced a cautious version of this claim as the ``CLI-as-framework hypothesis.''
The evidence accumulated since---in the corpus's own source trees, in the vendor SDKs, and in the market---lets us state it as a thesis.

\subsection{From Hypothesis to Thesis: The CLI as Framework}
\label{sec:cli_as_framework}

The standard reading of the absence of LangChain, AutoGen, and the rest (Observation~\ref{obs:absences}) is that production harnesses do not \emph{need} frameworks.
But there is a second reading: they do not need frameworks because the harness \emph{is itself the framework}---the default orchestration layer through which developers ship software, not merely a tool they query.

Consider what a framework traditionally provides: a loop, a tool registry, a memory layer, a configuration surface, an extension mechanism.
Every one of these is present in every production system studied here.
\clcode{} ships a streaming ReAct loop, 43 registered tools, a compaction-based memory manager, CLAUDE.md-driven configuration, MCP for integrations and Skills for capability bundles---and \cdx{}, \gemini{}, \vibe{}, \hrm{}, and \ocd{} do the same, each in its own idiom (\pib{} matches every element but MCP, which it pointedly rejects in favor of CLI tools plus its extension bus).
The difference is that the ``framework'' is not a library the developer imports into their code; it is a runtime the developer works \emph{inside of}.
The developer's ``program'' is a natural-language task plus a tree of Markdown files (CLAUDE.md, AGENTS.md, GEMINI.md, SKILL.md) that configure the agent's behavior, tool access, and domain knowledge.
The ``API'' is the filesystem, the terminal, and git.

Four convergent signals from the corpus support this reading.

\paragraph{Signal 1: Skills as declarative programs.}
Nine of the eleven systems implement Skills (Observation~\ref{obs:skillsmcp}): directories containing a \texttt{SKILL.md} with instructions, optional scripts, and metadata.
A skill is, functionally, a program written for an LLM runtime rather than for a CPU runtime.
It declares what the agent should do (\texttt{SKILL.md} body), when it should activate (\texttt{paths} frontmatter in \clcode{}, \texttt{PathTrigger} rules in \oh{}, \texttt{requires} in \ocl{}), and what tools it may use.
This is the structural analogue of a plugin in a traditional framework, except the programming language is English plus YAML frontmatter---and the layer now has registries, supply-chain security, and agent authors (Section~\ref{sec:skills}).

\paragraph{Signal 2: Hooks and event buses as the extension substrate.}
In April, user-facing lifecycle hooks were a \clcode{} distinctive.
By July they are the dominant extension substrate---nine of the eleven systems, with only \adr{} and \mswe{} abstaining: \cdx{} ships hooks whose event names are \emph{verbatim} \clcode{}'s; \oh{} wires \texttt{pre\_tool\_use}/\texttt{post\_tool\_use}/\texttt{stop} hooks into \texttt{Agent.step()}, with an \emph{agent} allowed as a hook handler; \vibe{} added \texttt{hooks.toml} shell hooks (deny, rewrite inputs, append context) alongside its internal middleware pipeline; \gemini{} exposes eleven loop-lifecycle events to extensions; \pib{} \emph{is} an event bus, with $\sim$33 typed events spanning tool interception to raw provider I/O; \ocd{}'s plugins are functions returning twenty typed hooks.
This is the same extension model that web frameworks and build systems have used for decades, applied to an LLM execution loop.
The developer is not writing \emph{for} the agent; they are writing \emph{inside} it.

\paragraph{Signal 3: The disappearing boundary between tool and workflow.}
In a traditional development workflow, the developer writes code, runs tests, reads logs, commits, and opens PRs as discrete steps.
In a harness workflow, these steps are delegated to the agent via natural-language instructions and executed through the agent's tool system.
The agent loop \emph{is} the workflow engine: \clcode{}'s coordinator phases, \cdx{}'s \texttt{update\_plan} tool and \texttt{/goal} workflows, \oh{}'s judge-audited goal loop.
When the developer types a task and the agent spawns sub-agents, dispatches tools, manages context, and produces the result, the CLI has subsumed the role of the build system, the task runner, and the IDE.

\paragraph{Signal 4: The harness as a service surface.}
The clearest post-April signal: harnesses now ship the interfaces of platforms.
\ocd{} embeds an HTTP server publishing an OpenAPI spec and a generated SDK, with every UI---TUI, desktop, web, IDE plugin, CI action---as a client.
\oh{} exposes conversations through an OpenAI-\emph{compatible} gateway, so any tooling that can call a chat-completions endpoint can drive an agent: the agent as a model.
\vibe{} teleports local sessions to a hosted runtime; \clcode{} pairs its SDK with a hosted Managed Agents API.
A runtime with clients, SDKs, and hosted tiers is not a tool with an ecosystem; it is a platform with distribution.

\subsection{The Harness--Framework Merger}
\label{sec:merger}

If the harness is the framework, the two artifact categories should merge---and in 2026 they are merging, visibly, from both directions (Table~\ref{tab:merger}).

\begin{table}[H]
\centering
\caption{The harness--framework merger, July~2026: named, versioned, installable artifacts.}
\label{tab:merger}
\small
\renewcommand{\arraystretch}{1.25}
\begin{tabularx}{\textwidth}{@{}l l L{7.6cm}@{}}
\toprule
\textbf{Direction} & \textbf{Artifact} & \textbf{What it is} \\
\midrule
Harness $\to$ framework & Claude Agent SDK~\cite{claudeagentsdk2026} & ``Claude Code as a library'': the same tools, loop, context management, hooks, and subagents, importable in Python/TypeScript; plus a hosted Managed Agents API \\
 & \texttt{openai-codex} SDK & \texttt{pip install openai-codex}: start \cdx{} threads, run turns, stream progress, control workspace access; Python/TypeScript, shipped in the \cdx{} repo \\
 & OpenHands agent SDK & The V1 restructuring \emph{is} an SDK-ization: the agent became \texttt{openhands-sdk}/\texttt{openhands-tools} packages plus an agent server \\
 & \pib{} packages / \ocd{} SDK & \texttt{pi-agent-core} importable + RPC mode; generated \texttt{@opencode-ai/sdk} over the embedded server \\
\midrule
Framework $\to$ harness & Deep Agents~\cite{deepagents2026} & LangChain's harness on LangGraph: todo planning, virtual filesystem with permissions, subagents, SKILL.md skills with progressive disclosure, AGENTS.md memory \\
 & Pydantic AI Harness~\cite{pydanticharness2026} & ``The batteries'' for pydantic-ai: composable capabilities (filesystem, code mode in a Rust-sandboxed interpreter) that turn a framework agent into a coding agent \\
 & Strands harness-sdk~\cite{strandsharness2026} & The merger in a repository name: a framework organization shipping ``the agent harness itself'' as an SDK \\
\bottomrule
\end{tabularx}
\end{table}

Both directions converge on the same artifact shape---loop + tools + skills + sub-agents + hooks + MCP + sessions---whether one starts from a terminal product or from an orchestration library.
The convergence is itself validation: when LangChain, the framework whose absence from every harness runtime Observation~\ref{obs:absences} documents, finally built a harness, it independently adopted the corpus's conventions (SKILL.md progressive disclosure, AGENTS.md memory, sub-agent spawning, todo planning).
And the meta-harness of the next subsection completes the picture from above: its baseline installation imports \texttt{claude-agent-sdk} and \texttt{openai-agents} as production dependencies---the ``frameworks'' a 2026 orchestration layer builds on \emph{are} the harness SDKs.
The twin absence of 2025-era frameworks thus gets its historical resolution: harnesses did not adopt the frameworks; they replaced them, became importable themselves, and turned the question ``which agentic framework should I use?'' into ``which harness do you already run?''

\subsection{Platform Economics: Marketplaces, Switching Costs, Governance}
\label{sec:platform_econ}

Three developments from the ninety-day window are textbook platform economics.

\paragraph{Marketplaces.}
\cdx{} grew a plugin marketplace (GitHub/git/URL sources, in-conversation install approval, workspace sharing); skills registries appeared in four systems with trust tiers, provenance verification, and quarantine (Section~\ref{sec:skills}); \ocl{} externalized its official plugins to npm.
Capability distribution now has app-store mechanics---including the app store's security pathologies, as SkillProbe's audit of published skills shows~\cite{guo2026skillprobe}.

\paragraph{Switching costs.}
\cdx{} ships a first-class \emph{importer for \clcode{}'s on-disk state}: it detects \texttt{\textasciitilde/.claude/projects} session JSONL files, converts them into \cdx{} rollout items under an import ledger, and offers to translate \texttt{\textasciitilde/.claude/settings.json} into \texttt{\textasciitilde/.codex/config.toml}.
Vendors writing importers for each other's session stores is the stage of platform competition where user data becomes the moat---and the countervailing force is equally visible: \ocd{} reads \clcode{}'s skills directory, \oh{} reads \clcode{}'s plugin manifests and hosts rival harnesses outright, and the meta-harness arbitrages every vendor's lock-in mechanisms into its adapter surface.

\paragraph{Enterprise governance.}
\cdx{}'s configuration stack now extends above the user---MDM-managed preferences, host-wide system files, and enterprise cloud-delivered bundles feed both base defaults and a constraint engine that can hard-restrict what user and project scopes may set.
Google, meanwhile, announced \gemini{}'s transition toward an Antigravity-branded CLI---closed-source, where \gemini{} is Apache-2.0 (Section~\ref{sec:landscape})---and withdrew its consumer free tier, and the SpaceX--xAI merger plus the announced \$60B Cursor acquisition (pending as of July~2026) priced the harness layer at acquisition scale.
Governance layers, gated tiers, and acquisition-scale pricing are the signatures of platforms, not of standalone tools.

\subsection{The Meta-Harness Layer}
\label{sec:omnigent}

\omni{}~\cite{omnigent2026} is not a twelfth harness; it is a bet that the harness has become a commodity component and that the durable value sits one layer up.
Databricks open-sourced it in June~2026 (Apache~2.0, $\sim$1M lines total, $\sim$312K of production Python): a four-tier process topology (server~$\rightleftarrows$~host daemon~$\rightleftarrows$~runner~$\rightleftarrows$~per-conversation harness subprocess) whose adapter boundary is, by explicit design, a recursive subset of \omni{}'s own public REST API.
Its registry ships 23 canonical harness adapters (plus 16 aliases and a community entry-point group) spanning \clcode{}, \cdx{}, Cursor, \ocd{}, \hrm{}, \pib{}, Goose, Qwen, Kimi, Kiro, Copilot, and Antigravity, formalized into five integration modes (\texttt{sdk-in-process}, \texttt{cli-subprocess}, \texttt{acp-subprocess}, \texttt{native-tui}, \texttt{native-server})---and the declared capabilities are reconciled against a \emph{conformance bench} (probes for basic turns, tool calling, streaming, interrupts, model override, policy denial), with live verification landed for the four flagship SDK adapters and best-effort declarations for the rest.
Harnesses are tested like hardware.

Above that boundary the meta-layer adds four things no single harness provides.
\emph{Composition}: any registry harness is addressable as a sub-agent session of any other, so a Claude-brained orchestrator can dispatch work to \cdx{} and have Cursor review it.
\emph{Cross-harness policy}: one three-level policy plane (session~$\to$~agent~$\to$~admin; six phases; CEL (Common Expression Language), Python, or LLM-classifier evaluators; cross-session per-user budgets) enforced on foreign harnesses \emph{through each vendor's own extension mechanism}---\clcode{} hooks, Cursor hooks, \hrm{} hooks, ACP permission requests---with fail-closed semantics.
The lock-in mechanisms of Observation~\ref{obs:coupling} become, from one layer up, an adapter surface.
\emph{A uniform sandbox and egress stack}: bubblewrap-plus-seccomp on Linux, generated Seatbelt profiles on macOS, Job Objects on Windows---the same architecture the corpus documents inside \cdx{}, re-implemented at the meta-layer, plus an L7 MITM egress proxy (own CA, per-host rules, default-deny) hosting a \emph{secretless credential proxy}, where real tokens never enter the sandbox and synthetic placeholders are swapped for credentials in flight---a design \cdx{} also ships (its \texttt{network-proxy} crate), here lifted to the meta-layer across all wrapped harnesses.
\emph{Shareable sessions}: server-durable transcripts with multi-device sync, ACL grants, review comments, fork, and mid-session harness switching.

Just as informative is what it does \emph{not} do: it implements no editing loop, no repository context, no edit-application strategy---the anatomy's D3/D4 core stays below the line---and it does not pretend the harnesses are equivalent: per-harness capability records, vendor-specific webhook endpoints, a Claude-specific \texttt{todos} field, and a Codex-only goal-mode extension leak through the ``common'' API by design.
Its flagship example, Polly, is a cross-vendor coordinator-worker: a \clcode{}-brained orchestrator that writes no code, fans work out to six vendor harnesses in per-task git worktrees, and mandates \emph{cross-vendor review} in its orchestrator prompt---the reviewer must be a different vendor than the implementer (the mechanism-layer guardrails gate fan-out and blast radius, not vendor identity).
Even its restraint is informative: sandboxing is applied inconsistently \emph{by design} (it exec-wraps the Claude CLI in its own sandbox but \emph{delegates} to \cdx{}'s native sandbox modes), evidence that OS-level isolation resists being factored out as a shared service.
And the twin absences hold at the meta-layer: no agentic framework, no RAG, in a million lines---unless one counts its two baseline dependencies, \texttt{claude-agent-sdk} and \texttt{openai-agents}, which is precisely the point of Section~\ref{sec:merger}.

\subsection{Ninety Days of Harness Evolution}
\label{sec:evolution_window}

Because the eight systems of the April edition were re-pinned rather than replaced, the corpus contains a controlled longitudinal sample: the same harnesses, source-diffed across one quarter.
Four movements characterize the window.

\paragraph{Convergence became imitation.}
April's convergences were mostly independent rediscovery; July's are traceable.
\cdx{} adopted \clcode{}'s hook event vocabulary verbatim and its plan-mode ergonomics; \oh{} adopted \clcode{}'s plugin manifest format, its task-tool signature, and its static/dynamic cache boundary; \ocd{} reads \clcode{}'s skills directory; \hrm{}'s source comments credit \ocd{} (edit matcher), \cdx{} (smart approvals), \ocl{} (orchestrator prompt), and Goose (context hints).
Cross-harness lineage is now written in the code itself.

\paragraph{Patterns diffused down the corpus.}
Deferred tool loading went from one system to three (plus seven skills variants); read-only plan modes from two to all four provider-native systems; LLM approval classifiers from one (\clcode{}) to two (\cdx{}'s Guardian); turn-level checkpointing from one (\vibe{}) to three; safety-aware scheduler partitioning from one to two.
The half-life of a competitive distinctive in this field is currently measurable in weeks.

\paragraph{Policy migrated out of prose.}
The clearest rhetorical trend of the quarter (Observation~\ref{obs:rhetoric}): \cdx{}'s newest model prompts dropped the no-commit and anti-gold-plating rules in favor of feature flags; \vibe{} deleted its ``Never Commit'' hard rule and re-founded its prompt on a seven-level precedence contract, A/B-tested server-side; \oh{} teaches commit mechanics.
As models internalize norms and harnesses grow governance surfaces, behavioral policy is moving from the prompt (where the model reads it) to configuration (where the platform enforces it).

\paragraph{The trees themselves moved at platform speed.}
\cdx{}'s workspace nearly doubled (621K $\to$ ${\sim}$1.12M lines of Rust, 89 $\to$ 126 crates) in one quarter---memories, Guardian, marketplace plugins, code mode, realtime voice; \vibe{} grew 77\% (35.6K~$\to$~63K lines); \oh{} re-architected into an SDK while its application repo became an automation control center; \gemini{} announced its brand pivot and gated its free tier; \adr{}, the field's pioneer, settled into community maintenance with 18 commits in the window.
Three of the April edition's observations required substantive revision in place---the coupling dichotomy (Observation~\ref{obs:coupling}), the size-implies-sandbox correlation (Observation~\ref{obs:sandboxcost}), and the protocol placement story (Observation~\ref{obs:protocols})---each corrected by evidence that did not exist, or that we could not see, ninety days earlier.
The methodological lesson generalizes: in this field, \emph{inventory} claims (tool counts, feature cells, version pins) decay in weeks, while \emph{structural} claims (loop taxonomy, subsystem anatomy, the absences) have so far proven durable.
We have separated the two accordingly throughout this edition.

\observation{\label{obs:platform}The coding-agent harness completed its platform turn in the first half of 2026, and every layer of the turn is now visible in source: extension substrates (hooks, skills, plugins) converged across all production systems; capability distribution acquired marketplaces, registries, trust tiers, and agent authors. Vendors shipped importers for each other's on-disk state and MDM-grade governance layers; the harnesses became importable SDKs while the framework vendors shipped harnesses (the merger of Section~\ref{sec:merger}); the agent itself became addressable as a model behind an OpenAI-compatible endpoint; and a meta-harness now orchestrates a fleet of eleven vendor harnesses---five of the systems studied here among them---behind one API, re-implementing the expensive parts (sandbox, policy) and arbitraging the proprietary parts (hooks, session stores) of each. The competitive unit of the field is no longer the agent loop; it is the ecosystem surface around it.}

\section{Discussion}
\label{sec:discussion}

\subsection{The Anthropic Effective-Agents Series as Empirical Validation}
\label{sec:anthropic_validation}

Between December~2024 and September~2025, Anthropic published four engineering articles offering prescriptive guidance on agent design: \emph{Building Effective Agents}~\cite{schluntz2024agents}, \emph{Effective Context Engineering for AI Agents}~\cite{rajasekaran2025context}, \emph{Writing Effective Tools for AI Agents}~\cite{aizawa2025tools}, and the case study \emph{How We Built Our Multi-Agent Research System}~\cite{hadfield2025multiagent}.
Across the four pieces, a coherent design philosophy emerges: hand-rolled loops in preference to framework abstractions, a small set of high-signal tools in preference to wrap-everything APIs, just-in-time structural retrieval in preference to pre-indexed RAG, orchestrator-worker patterns in preference to flat monoliths, transparency about agent planning, and explicit context budgeting as an engineering discipline.

Eleven independently developed systems (only one of which, \clcode{}, is from Anthropic) line up with these prescriptions:

\begin{itemize}[itemsep=2pt]
  \item ``Don't use frameworks unless necessary''~\cite{schluntz2024agents}: 0 of 11 systems use LangChain, LangGraph, AutoGen, CrewAI, ADK, LlamaIndex, Pydantic AI, Genkit, or Semantic Kernel in the agent runtime (Observation~\ref{obs:absences}).
  \item ``ACI matters as much as HCI''~\cite{schluntz2024agents} (the concept originates in this article; Aizawa et~al.~\cite{aizawa2025tools} extend it with practical tooling advice) and ``few thoughtful tools''~\cite{aizawa2025tools}: \adr{}'s 13 polymorphic edit formats, \clcode{}'s deferred tool loading, \cdx{}'s custom patch format, \hrm{}'s nine-strategy edit chain, and \ocd{}'s model-conditional tool surfaces are all heavy ACI investments (Observation~\ref{obs:editing}).
  \item ``Just-in-time over RAG''~\cite{rajasekaran2025context}: 0 of 11 systems use vector embeddings over code; all use \texttt{grep}, tree-sitter, glob, and the file system, plus auto-discovered Markdown context files (Observation~\ref{obs:absences}).
  \item ``Orchestrator-workers for parallelizable, context-spanning tasks''~\cite{hadfield2025multiagent}: emerges in \clcode{} (recursive composition), \cdx{} (thread tree with fan-out), \gemini{} (registry + session protocol), \vibe{} (\texttt{task} tool), \oh{} (parallel delegation), \hrm{} (orchestrator role + swarm), and \ocd{} (concurrent child sessions) (Observation~\ref{obs:coordinator}).
  \item ``Save plans to external memory before subagent spawn''~\cite{hadfield2025multiagent}: \clcode{}'s CLAUDE.md, \gemini{}'s plan files and \texttt{MEMORY.md} project index, \cdx{}'s AGENTS.md hierarchy and its agent-maintained memories root.
  \item ``Compaction + structured note-taking + sub-agents'' for long-horizon tasks~\cite{rajasekaran2025context}: all four provider-native systems---and all three newcomers---implement all three.
  \item ``Multi-agent systems use $\sim$15$\times$ more tokens than chat''~\cite{hadfield2025multiagent} (vs.\ a single chat baseline, not a single-agent pipeline): this is what motivates \clcode{}'s prompt-cache-sharing fork mechanism, \hrm{}'s cache-sharing background forks and summary-only returns, and \gemini{}'s thought-stripping (extended-thinking text is kept out of persisted history when recording responses, so it never re-enters the cache). Without those optimizations the cost would be prohibitive.
\end{itemize}

There is one tension in the guidance that we should call out.
Hadfield et~al.~\cite{hadfield2025multiagent} caution that ``most coding tasks involve fewer truly parallelizable tasks than research, and LLM agents are not yet great at coordinating and delegating to other agents in real time.''
Seven of the eleven systems in this study (eight counting \ocl{}'s protocol-layer variant) nevertheless implement coordinator-worker patterns specifically for coding workflows.
On closer inspection the observed patterns are mostly used for breadth-first exploration phases (parallel codebase research) rather than for parallel implementation, which is consistent with Hadfield et~al.'s caveat once the unit of analysis shifts from the system as a whole to the individual workflow phase.

\subsection{The Minimalism Argument}

\mswe{}'s competitive self-reported benchmark performance (74\%+ on SWE-Bench Verified) with a minimal scaffold and a single bash tool undercuts the assumption that rich tool ecosystems are necessary.
The finding echoes the CodeAct~\cite{wang2024codeact} argument that executable code subsumes discrete tool actions.
For benchmark evaluation alone, most of the scaffolding in the production systems turns out to be overhead.

The counter-argument is that benchmarks measure only a narrow slice of the value proposition.
Production systems (\clcode{}, \cdx{}, \gemini{}, \vibe{}, \oh{}) invest heavily in things that benchmarks do not reward:
\begin{itemize}[itemsep=1pt]
  \item \textbf{Safety}: Preventing destructive actions on user codebases.
  \item \textbf{User experience}: Streaming responses, rich terminal UIs, progress tracking.
  \item \textbf{Extensibility}: Supporting custom tools, MCP servers, plugins.
  \item \textbf{Robustness}: Handling edge cases, stuck detection, error recovery.
  \item \textbf{Cost management}: Prompt caching, context compaction, model selection.
\end{itemize}

These concerns are invisible to SWE-Bench but critical for production adoption.

\subsection{Safety as Architecture}

Safety is the dimension on which the corpus is most spread out.
\cdx{} takes a defense-in-depth approach: OS-level sandboxing plus policy-as-code plus approval workflows, with safety treated as a primary engineering concern from the start rather than as a late addition.
\gemini{} sits at a similar but lighter point on the spectrum: a cross-platform sandbox (reusing OS binaries) plus four explicit approval modes (PLAN/DEFAULT/AUTO\_EDIT/YOLO) and per-mode TOML policies.
\clcode{}'s three-layered system (hooks $\to$ classifier $\to$ dialog) sits between pure automation and human oversight.
\vibe{}'s permission-scope hierarchy (command/file/URL patterns plus agent-profile gates) is more granular than \adr{}'s binary confirmation but has no OS-level enforcement.
\oh{}'s pluggable \texttt{SecurityAnalyzer} framework lets rule-based and LLM-based analysis be combined.

The April edition observed that the corpus's two largest systems (\cdx{}, then at 621\,K lines, and \gemini{} at 568\,K) were also its two cross-platform sandboxers, and read the correlation as structural.
The expanded corpus falsifies the structural reading while preserving the cost claim: \hrm{} and \ocd{} are comparably large with zero OS-level isolation, spending their safety budgets on content-borne threats and permission granularity respectively (Observation~\ref{obs:sandboxcost}).
What remains true is that a native sandbox, where built, is among the most code-expensive components in a harness---\cdx{} \emph{vendors bubblewrap into its tree}, and the meta-harness re-pays the entire bill one layer up (Section~\ref{sec:omnigent}).
The minimal safety surface in \adr{} and \mswe{} is defensible for their target use case (developer tools with trusted users) but limits their applicability in enterprise or automated deployment contexts.

\subsection{The Model--Agent Co-design Thesis}

The coupling between agent scaffolds and foundation models runs in both directions.

\paragraph{The model shapes the agent.}
\cdx{}'s per-model prompts---now server-delivered catalog data spanning GPT-5.2 through the 5.6 family, with per-model tool modes and multi-agent tool generations---make explicit that scaffold design has to evolve with model capabilities, and that the vendor intends to drive that evolution without client releases.
\clcode{}'s extended-thinking integration (budget tokens, thinking blocks) exploits Claude-specific reasoning features.
\gemini{}'s \texttt{ModelRouterService} treats model selection itself as a runtime scheduling decision, dispatching each request to the cheapest sufficient Gemini variant---and absorbing a full model-generation shift (2.5~$\to$~3.x defaults) within one quarter; the same idea is studied formally as Hybrid LLM~\cite{ding2024hybridllm}, RouteLLM~\cite{ong2025routellm}, and the FrugalGPT cascade~\cite{chen2024frugalgpt}.
\vibe{}'s \texttt{reasoning\_effort} mapping and \texttt{ThinkChunk} parsing exploit Mistral's reasoning enum.
\adr{}'s model-aware edit format selection adapts the editing strategy to each model's output patterns, and \ocd{} re-derives the same idea in its tool registry (GPT-family models get a patch DSL; others get string replacement) and its nine-prompt model-family matrix.
\oh{} ships model-specific tool \emph{presets} (default, gemini, gpt5, planning)---co-design reaching even a multi-provider system's tool surface.

\paragraph{The agent shapes model usage.}
\clcode{}'s deferred tool loading reduces prompt tokens, changing the model's effective context window.
\clcode{}'s prompt-caching boundary splits the system prompt to maximize cache hits, structuring the prompt around cache economics as much as around logical organization; Gim et~al.~\cite{gim2024promptcache} provide the systems-level grounding for the modular attention-reuse strategy this implicitly assumes.
\gemini{} keeps extended-thinking traces out of persisted history at response-recording time, so they never leak into future cache lookups (and strips thought signatures on auth switch).
\vibe{}'s middleware pipeline lets per-turn token-budget policies be composed orthogonally to the loop body.
\oh{}'s condensers selectively forget events, shaping what the model ``remembers.''

\paragraph{Implications.}
The four provider-native systems (\clcode{}, \cdx{}, \gemini{}, \vibe{}) can optimize both sides of this coupling at once, evolving prompts, tools, and features in lockstep with model updates---\cdx{} most literally, via its server-delivered model catalog.
The April edition inferred that multi-provider systems must therefore design for the lowest common denominator; the expanded corpus shows that inference was too strong.
\hrm{}, \pib{}, and \ocd{} demonstrate that the full provider-native optimization menu is reachable from a multi-provider substrate by paying the per-provider conditional-code cost deliberately and centrally (Observation~\ref{obs:coupling})---per-model quirk metadata, provider-profile plugins, transform matrices.
What the multi-provider systems cannot replicate is the \emph{update loop}: only the vendor can re-tune scaffold behavior server-side on model-release day.
\vibe{}'s provider-first-with-generic-fallback design and the owned-transports position both remain viable middle paths, differing in who carries the maintenance burden.

\subsection{The Scaffold--Capability Frontier (a Guiding Intuition)}

We close with a guiding intuition rather than a formal hypothesis: across a fixed task distribution and a fixed model, scaffold complexity and observed task success appear to trace a roughly concave curve.
A floor below which the agent cannot operate at all (no loop, no tool, no memory, no result), a steep early-gain region where adding a small set of structural elements (a bash tool, a read/write tool, a Markdown context file) raises success rapidly, and a diminishing-returns plateau where further scaffold work pays off in operational concerns (safety, UX, extensibility) rather than in completion rate.
We label this picture the \emph{scaffold--capability frontier} and treat it as a hypothesis to be tested, not as a proven property of the design space.

Two independent results in the literature are consistent with this picture without confirming it.
\mswe{}'s reported $\sim$74\%+ on SWE-Bench Verified with a deliberately minimal scaffold suggests that, for SWE-Bench-style tasks on a frontier model, the floor is already well below most production scaffolds.
Lin et~al.~\cite{lin2026ahe} report that their \emph{Agentic Harness Engineering} (AHE) system, starting from a bash-only seed comparable to \mswe{} and automatically evolving the harness using observability-driven feedback, reaches 71.9\% on SWE-Bench Verified.
Their component-level ablation finds that \emph{structural} harness elements carry the improvement (tools $+3.3$\,pp, middleware $+2.2$\,pp, long-term memory $+5.6$\,pp), while the system prompt alone \emph{regresses} performance ($-2.3$\,pp); the evolved harness also transfers across four model families.
We read this as suggestive evidence that structural scaffold is more portable than prose-level prompt strategy, consistent with our model--agent co-design discussion at the architectural (rather than weights) level.

What an operational definition would require is at minimum: a fixed task distribution $T$, a fixed model $M$, a metric for scaffold complexity (LoC, tool count, structural feature count, or some weighted combination), and a target success rate $p$.
A ``minimum viable scaffold for $(T, M, p)$'' would then be the lower envelope of scaffold-complexity values whose empirical success rate on $T$ with $M$ exceeds $p$.
We do not produce such a measurement here.
Doing so would require a controlled cross-system run study that this paper deliberately avoids (see Threats to Validity, Section~\ref{sec:discussion}).
We flag the scaffold--capability frontier as a question for follow-on empirical work, and note that the relevant minimum almost certainly varies across:

\begin{itemize}[itemsep=1pt]
  \item \textbf{Task complexity}: Multi-file refactoring may require more sophisticated tool support than single-file bug fixes.
  \item \textbf{Safety requirements}: Enterprise deployments demand scaffolding that benchmarks don't measure.
  \item \textbf{Interaction patterns}: Long-running interactive sessions benefit from memory management that short benchmark runs don't need.
\end{itemize}

\subsection{Threats to Validity}

A study of this shape has real limits, and it is worth being explicit about them.

The analysis rests on source-code reading, not runtime measurement. We do not claim to know how fast any of these systems \emph{are}; we claim to know how they are \emph{structured}. Where the paper reports benchmark numbers, they come from the systems' own documentation, and the last decimal place is not the interesting part.

The qualitative scoring underlying the architectural comparisons involves judgment calls. We tried to anchor each score in a specific implementation detail, and we welcome disagreement from anyone who reads the same codebases and scores them differently. We deliberately avoid line-number citations, and we date all code-size figures to the July-2026 pins: every system in the corpus is actively developed, and finer precision would be obsolete by the time the paper is read. All versions are pinned to exact release tags and commits (Table~\ref{tab:selection}), the April-2026 snapshots are retained for the longitudinal comparison of Section~\ref{sec:evolution_window}, and---a lesson that quarter taught us directly---we now distinguish \emph{inventory} claims, which we date, from \emph{structural} claims, which we expect to endure. Three observations of the April edition were substantively revised in this one on new evidence; the revisions are integrated in place and flagged where they occur.

The \clcode{} analysis is the weakest link on reproducibility. It draws on a publicly circulated source snapshot from March~2026 rather than an official release, and the shipping binary has evolved considerably since (we track its changelog-level evolution but do not treat changelog claims as source evidence). The architectural claims should still be testable against the public \texttt{claude-agent-sdk} and the behavior of the shipping binary, but we acknowledge the asymmetry with the other ten systems, whose source trees are trivially re-derivable from public Git history.

The framework-absence finding (Observation~\ref{obs:absences}) is strong but structurally conservative. We checked dependency manifests and performed import grep across three languages---twice, three months apart, across twelve trees including the meta-harness; we did not trace internal forks, plugins loaded via dynamic \texttt{importlib}/\texttt{require}, or transpiled distributions. A company could in principle use LangChain inside a production SWE agent through a code path we did not inspect. We would be surprised, but not shocked.

The Anthropic-guidance-to-observation mapping (Observation~\ref{obs:guidance}) is suggestive but does not by itself establish causation. The right follow-up is interview work with engineers at each vendor; we did not do that.

We also compared each system independently rather than running all eleven on a shared task set. A head-to-head execution study would turn some of the qualitative scoring into measurable quantities. It would also cost an order of magnitude more effort than this paper, and we chose the tractable scope.

\section{Design Recommendations: A Practitioner's Guide}
\label{sec:recommendations}

The preceding sections are descriptive, cataloging what the eleven production harnesses actually contain.
This section is prescriptive: we distill the observations into actionable recommendations for an engineer designing a new coding agent.
Each recommendation cites the observations that support it, names the systems that implement it, and identifies the trade-off that might justify a different choice.
Where the corpus agrees with the Anthropic \emph{Effective Agents} series~\cite{schluntz2024agents,rajasekaran2025context,aizawa2025tools,hadfield2025multiagent}, we flag the agreement as cross-source validation.
Where the corpus diverges from or extends the series, we note that too.
The recommendations are grouped by subsystem and ordered roughly in the sequence a practitioner would encounter them when bootstrapping an agent.

\subsection{Loop Architecture}

\recommendation{1}{Start with a linear \texttt{while} loop; graduate to a middleware pipeline only when orthogonal turn policies emerge. \emph{Evidence}: Observation~\ref{obs:size} (loop sophistication does not predict benchmark performance); \mswe{}'s 50-line linear loop reports 74\%+ on SWE-Bench Verified (self-reported; see the footnote to Table~\ref{tab:overview}). Upgrade path: once you need three or more independent turn-level policies (turn limits, cost caps, auto-compaction, context-budget warnings, read-only mode), adopt \vibe{}'s middleware pipeline pattern---each policy becomes a composable middleware, not a new branch in the loop body.}

\subsection{Provider Coupling}

\recommendation{2}{If you ship a foundation model, couple tightly to your home provider and expose a generic fallback adapter for portability; if you don't, provider-native optimizations are still on the table---budget for per-model metadata instead. \emph{Evidence}: Observation~\ref{obs:coupling}. \vibe{}'s ``provider-first with generic fallback'' remains the vendor pattern. For multi-provider harnesses, \pib{}'s per-model \texttt{compat} quirk flags, \hrm{}'s declarative provider profiles, and \ocd{}'s transform matrix show that cache breakpoints, thinking tiers, and reasoning effort are reachable from an abstraction---the cost is a maintained per-provider conditioning layer, paid once and centrally.}

\subsection{Tool Design}

\recommendation{3}{Begin with just a \texttt{bash} tool. Add more tools only in response to observed failure modes. \emph{Evidence}: \mswe{}'s single-tool design reports 74\%+ on SWE-Bench with essentially no tool infrastructure (self-reported); Aizawa et~al.~\cite{aizawa2025tools} caution that ``more tools don't always lead to better outcomes.'' A common sequence is: add \texttt{read\_file} and \texttt{write\_file} when bash output truncation becomes a problem; add \texttt{grep}/\texttt{glob} when bash+\texttt{find}/\texttt{rg} feels awkward; add \texttt{search\_replace} when full-file writes waste tokens.}

\recommendation{4}{When your tool count exceeds $\sim$15, adopt deferred tool loading. \emph{Evidence}: \clcode{}'s \texttt{shouldDefer} flag + \texttt{ToolSearchTool} reduces the initial prompt by $\sim$40\%; \cdx{} independently converged with \texttt{defer\_loading} flags plus a BM25-ranked \texttt{tool\_search}; \hrm{} collapses overflowing MCP catalogs into three BM25-searched bridge tools once schemas would exceed 10\% of the context window. Below $\sim$15 tools the indirection is not worth the added complexity; above it, the prompt bloat becomes prohibitive.}

\subsection{File Editing}

\recommendation{5}{Match your edit-tool contract to your model tier: exact unique-substring replacement for frontier models, a fuzzy cascade for open or weaker models---and in either case handle drift \emph{at the tool}, never by line numbers. \emph{Evidence}: Observation~\ref{obs:editing}. The frontier-model camp has converged on exact contracts (\clcode{}; \vibe{} deleted its SEARCH/REPLACE tool and converged on exact matching mid-2026; \pib{} adds only Unicode/whitespace canonicalization with byte-fidelity overlay). The drift-tolerant camp serves broader model ranges (\ocd{}'s nine-stage cascade at Levenshtein~0.65, \hrm{}'s nine-strategy chain, \adr{}'s \texttt{RelativeIndenter}). \gemini{}'s LLM edit-fixer subcall is the emerging third option: repair the match with a model instead of a threshold. Avoid line-number-based edits: models drift on line numbers more than on contextual matching.}

\subsection{Memory and Context}

\recommendation{6}{Auto-discover hierarchical Markdown context files at project, user, and extension scopes---and read your neighbors' filenames too. \emph{Evidence}: all eleven systems that manage repo context converged on the pattern, and the newest ones read \emph{multiple} conventions (\hrm{}: own file, then AGENTS.md, CLAUDE.md, .cursorrules; \ocd{}: AGENTS.md/CLAUDE.md/CONTEXT.md plus remote URLs; \oh{} ingests three ecosystems' files as scoped rules). Inject top-level content near the top of the system prompt, surface \emph{nested} files just-in-time when tools touch their subtree (\vibe{}, \ocd{}, \hrm{}, \gemini{}), and let the model persist durable facts---by direct edits, bounded snapshot files, or a human-reviewed inbox (Observation~\ref{obs:memory}).}

\recommendation{7}{Implement threshold compaction at a fixed buffer below the model's context window; preserve a verbatim recent tail; merge summaries incrementally rather than re-summarizing; and wire the same routine to fire reactively on overflow. \emph{Evidence}: \clcode{} fires below a 13\,K-token buffer and post-restores files; \gemini{} compacts at 50\% preserving the last 30\%; \pib{} and \ocd{} pass the \emph{previous} summary back for merging (anchored/iterative summaries), which preserves early decisions that from-scratch re-summarization loses; \vibe{} re-injects prior user messages into the compacted envelope and reuses the routine on mid-turn \texttt{ContextTooLong} errors; \oh{} routes overflow errors into condensation. Aggressive compaction protects against context rot~\cite{rajasekaran2025context}; conservative preservation keeps recent reasoning coherent.}

\recommendation{8}{Do \emph{not} build RAG over code. Use \texttt{ripgrep}, \texttt{glob}, tree-sitter symbol extraction, and file-system traversal instead. \emph{Evidence}: Observation~\ref{obs:absences}; 0/11 systems use vector embeddings for code retrieval; Rajasekaran et~al.~\cite{rajasekaran2025context} favor JIT retrieval in practice, and all their concrete recommendations point toward deterministic tools. Code has rich deterministic structure (paths, language servers, tree-sitter parses) that semantic similarity cannot replicate, and changes minute-to-minute, making embeddings stale.}

\subsection{Safety Architecture by Deployment Context}

\recommendation{9}{For a \textbf{developer tool (semi-trusted)} context: implement a three-mode approval system (PLAN / DEFAULT / YOLO) with permission-scope patterns. \emph{Evidence}: \gemini{}'s PLAN/DEFAULT/YOLO modes, \vibe{}'s 4-layer permission hierarchy (tool-wide + tool-specific + session rules + interactive callback); matches how a human developer triages risk.}

\recommendation{10}{For \textbf{enterprise / shared / automated} contexts: implement OS-level sandboxing with policy-as-code and per-agent audit trails. \emph{Evidence}: Observation~\ref{obs:sandboxcost}; \cdx{} and \gemini{} ship native cross-platform sandboxes (Bubblewrap on Linux---vendored in \cdx{}'s case---Seatbelt on macOS, restricted tokens on Windows), and \clcode{} wraps Anthropic's reusable \texttt{sandbox-runtime} as an opt-in. \gemini{}'s implementation shows the cost is tolerable when you reuse OS binaries (Node \texttt{child\_process} or equivalent) rather than writing namespace plumbing from scratch.}

\recommendation{11}{Regardless of tier, codify safety rules as data or dedicated policy files rather than as imperative code---and if you support a YOLO mode, keep a floor beneath it. \emph{Evidence}: \cdx{}'s Starlark \texttt{execpolicy} rules with parse-time-validated inline examples, \clcode{}'s \texttt{PreToolUse} hooks, \gemini{}'s per-mode TOML, \ocd{}'s last-match-wins rulesets. \hrm{} adds the floor lesson: its twelve hardline patterns survive \texttt{--yolo}, with the bypass flag frozen at module import so injected content cannot flip it at runtime. Policies-as-code survive refactors and are auditable independently of the loop implementation.}

\subsection{Multi-Agent Orchestration}

\recommendation{12}{Stay single-agent until you can point to a concrete breadth-first exploration phase where parallel context isolation clearly beats serial search. \emph{Evidence}: Hadfield et~al.~\cite{hadfield2025multiagent} caution that multi-agent systems use $\sim$15$\times$ more tokens than a chat baseline and that ``most coding tasks involve fewer truly parallelizable tasks than research''; \vibe{}'s \texttt{task} tool delegates to sub-agents (its prompt encourages launching several in parallel), which is sufficient for most coding workflows.}

\recommendation{13}{Ship an ACP \emph{server}: it is no longer just an editor integration---it makes your harness consumable by hosts and meta-orchestrators. Keep your own sub-agents in-process. \emph{Evidence}: Observation~\ref{obs:protocols}. An ACP server now buys three audiences at once: IDEs (Zed, JetBrains), hosting harnesses (\oh{} runs \clcode{}/\cdx{}/\gemini{} as interchangeable ACP backends), and meta-harnesses (\omni{} drives Goose and Qwen over ACP). Six of eleven corpus systems ship it. For \textbf{agent~$\leftrightarrow$~agent} mesh topologies, \textbf{A2A} remains \gemini{}'s bet alone---defensible, unproven. For \textbf{main-agent~$\leftrightarrow$~sub-agent} coordination within your own runtime, do \emph{not} adopt either (see Recommendation~12): in-process primitives are what production systems use, and even the cross-process exceptions (\pib{}'s JSONL extension, \hrm{}'s SQLite-blackboard swarm) avoid the standard protocols for that role.}

\subsection{Extensibility}

\recommendation{14}{Use \textbf{Skills} for capability templates (workflows, domain knowledge, procedural recipes) and \textbf{MCP} for external integrations (Slack, databases, internal APIs)---in that order of priority. \emph{Evidence}: Observation~\ref{obs:skillsmcp}: skills now lead MCP in adoption (9/11 vs.\ 8/11), discovery is cross-vendor (\ocd{} reads \texttt{\textasciitilde/.claude/skills}), and distribution has registries and provenance. MCP remains the right layer for running external processes; \pib{} demonstrates the defensible minimal position (skills plus CLI tools with READMEs, no MCP). If you install third-party skills, treat them as packages: trust tiers, scanning, and quarantine (\hrm{}) are the emerging baseline~\cite{guo2026skillprobe}.}

\subsection{What Not to Build (Anti-Patterns)}

\recommendation{15}{\textbf{Do not} use LangChain, LangGraph, AutoGen, CrewAI, LlamaIndex, Pydantic AI, Genkit, Google ADK, or Semantic Kernel for the agent runtime. \emph{Evidence}: Observation~\ref{obs:absences}; 0/11 production harnesses in the corpus use these frameworks---\gemini{} skips Google's own. Schluntz \& Zhang~\cite{schluntz2024agents} explicitly warn: ``frameworks often create extra layers of abstraction that can obscure the underlying prompts and responses, making them harder to debug.'' Use raw SDK calls---or, the 2026 corollary (Section~\ref{sec:merger}): if you want a batteries-included starting point, the harness SDKs (Claude Agent SDK, \texttt{openai-codex}, \texttt{openhands-sdk}) \emph{are} the framework layer now, with the corpus's patterns built in.}

\recommendation{16}{\textbf{Do not} build a vector-embedding retrieval layer for code. \emph{Evidence}: Observation~\ref{obs:absences}; 0/11 systems do this. The embeddings that do appear serve \emph{conversation memory} (\ocl{}'s default hybrid search; \hrm{}'s opt-in plugins), and \hrm{} shows lexical SQLite FTS5 sufficing for that role at 642K-line scale. If you believe your domain needs semantic code retrieval, first prove it improves over \texttt{ripgrep}+\texttt{tree-sitter} on a held-out task set before adding the infrastructure.}

\recommendation{17}{\textbf{Do not} wrap every upstream SaaS API as a 1-to-1 tool. \emph{Evidence}: Aizawa et~al.~\cite{aizawa2025tools}; ``a common error we've observed is tools that merely wrap existing software functionality or API endpoints.'' Consolidate instead.}

\recommendation{18}{\textbf{Do not} over-engineer stuck detection---but do ship the cheap caps, which cost a dozen lines. \emph{Evidence}: \clcode{} and \cdx{} still ship no automated stuck detection and deliver production-quality agents. The floor moved, though: even \mswe{} now caps consecutive malformed responses and wall-clock time, \ocd{}'s doom-loop check is a three-identical-calls counter routed to a permission ask, and \hrm{} hashes call signatures but ships the hard stop \emph{disabled}, trusting warnings. \oh{}'s five-scenario \texttt{StuckDetector} and \gemini{}'s hybrid hash+LLM service remain platform-tier investments.}

\subsection{A Minimum Viable Harness}

As a concrete starting point, Listing~\ref{lst:mva} sketches a minimum viable harness in $\sim$90~lines of Python, combining the patterns recommended above: linear loop (\mswe{}), middleware-style policies (\vibe{}), four-tool surface (bash, read, write, search\_replace), hierarchical Markdown context auto-discovery (all four provider-native systems), and threshold compaction (\clcode{}/\allowbreak\gemini{}/\allowbreak\vibe{}).
It is not a drop-in library; it is a scaffold to be copied and specialized.

\begin{lstlisting}[caption={A minimum viable harness, $\sim$90~LoC Python. Illustrative scaffold, not production code.},label={lst:mva}]
from __future__ import annotations
import asyncio, json, pathlib, subprocess
from dataclasses import dataclass, field
from typing import Any, Protocol

# ---- Provider abstraction (Recommendation 2): provider-first fallback ---------
class Model(Protocol):
    async def complete(self, messages: list[dict], tools: list[dict]) -> dict: ...

# ---- Tools (Recommendation 3): bash + 3 file tools --------------------------
def _truncate(s: str, n: int = 25_000) -> str: return s if len(s) <= n else s[:n] + "\n...[truncated]"

def tool_bash(cmd: str) -> str:
    r = subprocess.run(cmd, shell=True, capture_output=True, text=True, timeout=120)
    return _truncate(f"exit={r.returncode}\nstdout:\n{r.stdout}\nstderr:\n{r.stderr}")

def tool_read_file(path: str, offset: int = 0, limit: int = 2000) -> str:
    lines = pathlib.Path(path).read_text().splitlines()
    return _truncate("\n".join(f"{i+1:4}: {ln}" for i, ln in enumerate(lines[offset:offset+limit])))

def tool_write_file(path: str, content: str) -> str:
    pathlib.Path(path).write_text(content); return f"wrote {len(content)} bytes"

def tool_search_replace(path: str, search: str, replace: str) -> str:
    p = pathlib.Path(path); text = p.read_text()
    if text.count(search) != 1: return f"ERROR: search string occurs {text.count(search)}x; must be unique"
    p.write_text(text.replace(search, replace, 1)); return "OK"

TOOLS = {"bash": tool_bash, "read_file": tool_read_file,
         "write_file": tool_write_file, "search_replace": tool_search_replace}

# ---- Hierarchical Markdown context (Recommendation 6) ----------------------
def discover_context(cwd: pathlib.Path = pathlib.Path.cwd()) -> str:
    parts = []
    for p in reversed([cwd, *cwd.parents]):                     # root-to-leaf
        md = p / "AGENTS.md"
        if md.exists(): parts.append(f"<ctx path='{md}'>\n{md.read_text()}\n</ctx>")
    return "\n".join(parts)

# ---- Middleware pipeline (Recommendation 1) ---------------------------------
@dataclass
class Agent:
    model: Model
    messages: list = field(default_factory=list)
    n_turns: int = 0
    cost: float = 0.0
    max_turns: int = 50
    max_cost: float = 5.00
    compact_at_tokens: int = 120_000

    def _token_estimate(self) -> int:
        return sum(len(json.dumps(m)) for m in self.messages) // 4

    async def _check_limits(self):
        if self.n_turns >= self.max_turns: raise StopIteration(f"max_turns={self.max_turns}")
        if self.cost   >= self.max_cost:   raise StopIteration(f"max_cost=${self.max_cost:.2f}")

    async def _maybe_compact(self):                             # Recommendation 7
        if self._token_estimate() < self.compact_at_tokens: return
        summary = await self.model.complete(
            self.messages + [{"role": "user",
                "content": "Summarize the conversation. Preserve decisions and unresolved issues."}], tools=[])
        # Preserve system + last 30% of turns verbatim, drop the middle (Gemini pattern)
        keep = max(4, int(len(self.messages) * 0.30))
        self.messages = [self.messages[0], {"role":"assistant","content":summary["content"]}] + self.messages[-keep:]

    async def run(self, task: str) -> str:
        self.messages = [
            {"role": "system", "content": f"You are a SWE agent.\n\n{discover_context()}"},
            {"role": "user",   "content": task},
        ]
        tool_schemas = [{"name": n, "description": f.__doc__ or n} for n, f in TOOLS.items()]
        while True:
            await self._check_limits(); await self._maybe_compact()
            resp = await self.model.complete(self.messages, tools=tool_schemas)
            self.n_turns += 1; self.cost += resp.get("cost", 0.0)
            self.messages.append(resp)
            if not resp.get("tool_calls"): return resp.get("content", "")
            for call in resp["tool_calls"]:                     # serial execution
                try:    out = TOOLS[call["name"]](**call["args"])
                except Exception as e: out = f"ERROR: {type(e).__name__}: {e}"
                self.messages.append({"role": "tool", "tool_call_id": call["id"], "content": _truncate(str(out))})
\end{lstlisting}

The scaffold above deliberately omits features for which our corpus shows divergence: sandbox (Recommendations~9--10 are deployment-specific), multi-agent (Recommendation~12 defers it), MCP/Skills (Recommendation~14 is extensibility, not core).
Start here, measure~\cite{hadfield2025multiagent}, add the minimum that your observed failure modes demand.

\observation{\label{obs:mva}The 90-line minimum viable agent in Listing~\ref{lst:mva} implements 10 of the 18 recommendations directly and is compatible with the remaining eight. It does so with no framework dependencies, no RAG, no vector store, no multi-agent orchestration, and no sandbox, which lines up with the twin absences (Observation~\ref{obs:absences}) and the ``start simple'' principle of \cite{schluntz2024agents}. We conjecture, without proof, that this scaffold running on a frontier model would match \mswe{}'s SWE-Bench numbers, and that moving beyond those numbers is primarily a model-capability question (Observation~\ref{obs:size}) rather than a scaffold question.}

\section{Conclusion and Future Work}
\label{sec:conclusion}

This paper has laid out a source-code-level anatomy of eleven LLM-powered coding harnesses---one per major commercial LLM provider (\clcode{}/Anthropic, \cdx{}/OpenAI, \gemini{}/Google, \vibe{}/Mistral) plus seven open-source designs---together with a meta-harness contrast point, a longitudinal diff of one quarter's evolution, and an account of the field's turn into platforms.

\subsection{Key Takeaways}

\begin{enumerate}[itemsep=2pt]
  \item \textbf{Architecture matters, but not in the way the current discourse implies.} The sophistication of the agent loop does not predict benchmark performance (\mswe{}'s 50-line loop reports results in the frontier range). It does, however, predict production readiness: safety infrastructure, user experience, extensibility---and increasingly clients and transport, where the largest systems now carry most of their mass.

  \item \textbf{The model--agent relationship is co-evolutionary, and coupling is about the update loop, not capability.} Provider-native optimizations (cache boundaries, thinking budgets, model-specific prompts, routing) are exercised from \emph{both} ends of the coupling spectrum: the vendor systems natively, and \hrm{}, \pib{}, and \ocd{} from multi-provider substrates by paying the per-provider conditioning cost centrally. What only vendors retain is server-side co-evolution---\cdx{} re-tunes prompts, reasoning tiers, and even its multi-agent tooling per model release from a catalog endpoint, no client update required.

  \item \textbf{Safety is architecturally expensive---and a choice, not a consequence of scale.} \cdx{}'s cross-platform sandboxing (now with vendored bubblewrap) remains a substantial investment benchmarks never measure, and the meta-harness re-pays the same bill one layer up. But the April size-implies-sandbox correlation broke: \hrm{} and \ocd{} are among the largest systems with zero OS-level isolation, spending instead on content-borne threat defense and syntax-aware permissioning.

  \item \textbf{Multi-agent orchestration converged on hierarchical patterns, and the protocols found a third role.} Coordinator-worker shapes now appear in seven systems. ACP serves not only the editor~$\leftrightarrow$~agent boundary (six corpus adopters) but a role outside its design brief: \emph{harness hosting}, with \oh{} running \clcode{}, \cdx{}, or \gemini{} as interchangeable backends. A2A remains \gemini{}'s cross-vendor mesh bet. Sub-agent coordination stays in-process in eight of the nine multi-agent systems.

  \item \textbf{The extensibility standards resolved: skills lead.} Skills (9/11 systems) overtook MCP (8/11) when \pib{} broke the tie---implementing agentskills.io while rejecting MCP outright. The skills layer acquired registries, trust tiers, provenance verification, cross-vendor discovery (\ocd{} reads \texttt{\textasciitilde/.claude/skills}), and its first agent authors (\hrm{}'s self-improving loop, \gemini{}'s extraction inbox).

  \item \textbf{The twin absences are real, re-verified, and now historically explained.} Across eleven systems and roughly four million lines, no agent runtime uses a general-purpose agentic framework, and none uses RAG over code---all rely on hand-rolled async loops and deterministic structural retrieval. The absence held through a threefold corpus expansion and a three-month re-audit, corroborating Anthropic's published guidance~\cite{schluntz2024agents,rajasekaran2025context} and the independent AHE result~\cite{lin2026ahe}. Its resolution is the harness--framework merger of Section~\ref{sec:merger}.

  \item \textbf{The Anthropic effective-agents series anticipates the architectures we observe.} The four engineering articles published between December~2024 and September~2025 map closely onto the patterns documented across eleven independently developed systems. Whether this reflects shared empirical reality, guidance-shaped design, or both is a question worth pursuing on its own.

  \item \textbf{The analysis is directly actionable.} Section~\ref{sec:recommendations} distills 18 concrete, evidence-anchored design recommendations for practitioners building new harnesses, together with a 90-line minimum-viable-harness scaffold (Listing~\ref{lst:mva}) that implements ten of them directly. Combined with the source-code audit, this is, as far as we know, the first evidence-based how-to for the discipline that is not a single-vendor whitepaper.

  \item \textbf{The platform turn is no longer a hypothesis.}
  What the April edition advanced cautiously as the CLI-as-framework
  hypothesis, Section~\ref{sec:platformturn} now documents as an
  accomplished turn: the harnesses ship as importable SDKs while the
  framework vendors ship harnesses (the merger of
  Section~\ref{sec:merger}); capability distribution has marketplaces,
  registries, trust tiers~\cite{guo2026skillprobe}, and agent authors;
  vendors write importers for each other's on-disk state and expose
  MDM-grade (mobile-device-management) governance; the agent itself is addressable as a model
  behind an OpenAI-compatible endpoint; and a meta-harness orchestrates
  eleven vendor harnesses---five of the systems studied here among
  them---behind one API (Section~\ref{sec:omnigent}).
  The surrounding ecosystem data points the same way: MCP's 8\,000+
  server ecosystem with app-store curation
  pathologies~\cite{guo2025mcpmeasurement}, skills formalized as
  composable packages~\cite{xu2026agentskills} with npm-like package
  managers~\cite{saha2026skilldex}, 22--29\% of GitHub projects bearing
  agent traces~\cite{robbes2026agenticmuch}, and the ``SE~3.0''
  framing of development as orchestration~\cite{li2025se3}.
  The developer no longer programs \emph{with} the framework but
  \emph{inside} it, and the vendor lock-in of
  Observation~\ref{obs:coupling} is becoming the structural lock-in
  that operating systems and IDEs have always
  produced~\cite{kapoor2025platformagents}---with the meta-layer
  already arbitraging it.
  Whether harnesses consolidate into a few dominant platforms, fragment
  into interoperable runtimes, or commoditize from above is the
  competitive question of the next year; the architectural
  preconditions for all three outcomes are in the source code today.
\end{enumerate}

\subsection{Future Work}

Several directions emerge from our analysis:

\begin{itemize}[itemsep=2pt]
  \item \textbf{Unified evaluation frameworks} that assess safety, user experience, cost efficiency, and extensibility alongside correctness, addressing the gap between benchmark performance and production readiness.

  \item \textbf{Reference architecture specification} that formalizes the patterns we identified (event sourcing, policy-as-code, context forking, deferred tool loading, middleware pipelines, hash-based loop detection, JIT repo context) into a reusable architectural framework.

  \item \textbf{Formal verification of safety policies} for policy-as-code systems, establishing guarantees about what agent actions are permitted under given policy configurations.

  \item \textbf{Empirical study of model--agent co-evolution}, tracking how agent scaffolds change across model generations and quantifying the performance impact of model-specific optimizations. Lin et~al.'s AHE system~\cite{lin2026ahe} demonstrates that automatic harness evolution is already feasible; longitudinal studies could now track how automatically evolved scaffolds compare to hand-crafted ones across model generations.

  \item \textbf{Cross-system benchmark suite} that runs identical tasks across all eleven systems under controlled conditions, enabling direct comparison of the architectural trade-offs we identified.

  \item \textbf{Inter-agent protocol adoption study}, tracking ACP's three roles (editor integration, harness hosting, and---via A2A---the cross-vendor mesh) and measuring the overhead of hosted-harness topologies such as \oh{}'s ACP backends and \omni{}'s adapter fleet.

  \item \textbf{Longitudinal continuation}: the April/July snapshot pair turns this study into a time series; quarterly re-pinning would measure pattern-diffusion half-lives, prompt thinning, and the platformization trajectory quantitatively rather than anecdotally.

  \item \textbf{Meta-harness economics}: whether commoditization-from-above (Section~\ref{sec:omnigent}) captures value durably, and whether the conformance-bench approach to harness capabilities becomes a standard interface.

  \item \textbf{Causation analysis of the Anthropic--industry alignment} via interviews, design-doc archaeology, and commit-history correlation, distinguishing public-guidance influence from convergent engineering.
\end{itemize}

\section*{Acknowledgments and Disclosure of AI Assistance}

This paper was written with substantial assistance from Anthropic's Claude (via the Claude Code CLI), which was used both for the source-code analysis and for drafting the manuscript.
The architectural findings, observations, patterns, and recommendations were verified against the referenced codebases before being kept in the paper.
Any remaining errors are ours.

This disclosure follows the emerging authorship policies of ACL, NeurIPS, ICML, and the IEEE.
The value of the paper, if it has any, lies in what it says about the twelve codebases; the named modules, tools, and version pins cited throughout should let any reader verify the claims independently.

We thank the maintainers and contributors of the twelve studied systems for working in the open.
This paper would not have been possible without their decision to make their source code available for inspection.

\vspace{0.5em}

\appendix
\section{Detailed Comparison Tables}
\label{app:tables}

This appendix collects the densest per-system comparison tables, referenced from the body: the rhetorical content of the eleven system prompts (Table~\ref{tab:prompt-content}), the advanced API-feature matrix (Table~\ref{tab:api_features}), and the head-to-head \clcode{}-versus-\cdx{} pipeline comparison (Table~\ref{tab:pipeline_detail}).

\begin{table}[H]
\centering
\caption{Rhetorical content of the eleven system prompts (July~2026).}
\label{tab:prompt-content}
\scriptsize
\renewcommand{\arraystretch}{1.25}
\begin{tabularx}{\textwidth}{@{}l L{2.5cm} L{2.1cm} L{2.9cm} L{1.7cm} L{2.9cm}@{}}
\toprule
\textbf{System} & \textbf{Verbosity rule} & \textbf{Anti-gold-plate} & \textbf{Commit policy} & \textbf{Emoji rule} & \textbf{Emphasis style} \\
\midrule
\clcode{} & quantified ($\le 25/100$ w; internal A/B) & explicit & no-commit unless asked & conditional & \texttt{IMPORTANT:}/\texttt{NEVER} tags \\
\cdx{}    & qualitative (5.2); persona-driven (5.6) & explicit (5.2); dropped (5.6) & no-commit (5.2); feature-flag (5.6) & none & \texttt{NEVER} tags \\
\gemini{} & via snippet gating & implicit & implicit & none & feature-gated sections \\
\vibe{}   & strict ($<\!150$ w) & explicit & \emph{reversed}: commits expected, signature trailer & strict ban & overridability contract (structural) \\
\oh{}     & none & explicit (file-suffix) & commits taught; push/PR gated & none & XML role tags \\
\adr{}    & ``few short sentences'' & implicit & silent & silent & format examples \\
\mswe{}   & structural (1 cmd/turn) & silent & silent & silent & \texttt{<important>} XML \\
\hrm{}    & qualitative (``lead with the change'') & explicit & no-commit/push/rewrite unless asked & none & \texttt{MUST}/\texttt{NEVER} + exemplar pairs; model-gated XML \\
\pib{}    & single bullet (``Be concise'') & silent (delegated) & silent (delegated) & silent & XML data tags only \\
\ocd{}    & quantified ($<$4 lines; model-dep.) & explicit & no-commit unless asked & conditional; GPT prompt bans & \texttt{IMPORTANT:}/\texttt{NEVER} + \texttt{<system-reminder>} \\
\ocl{}    & per-agent & inherited & inherited & inherited & per-agent \\
\bottomrule
\end{tabularx}
\end{table}

\begin{table}[H]
\centering
\caption{Advanced API features used by each system (July~2026).}
\label{tab:api_features}
\scriptsize
\renewcommand{\arraystretch}{1.25}
\setlength{\tabcolsep}{3.5pt}
\begin{tabularx}{\textwidth}{@{}l L{2.7cm} C{1.25cm} L{2.1cm} L{1.8cm} C{1.1cm} C{0.9cm} L{1.9cm}@{}}
\toprule
\textbf{System} & \textbf{Prompt caching} & \textbf{Ext. thinking} & \textbf{Reasoning effort} & \textbf{Model routing} & \textbf{WebSocket} & \textbf{Vision} & \textbf{Cost tracking} \\
\midrule
\oh{}     & Partial (cache tiers + markers) & \cmark & \cmark & \cmark (RouterLLM) & \xmark & \cmark & Per-model \\
\adr{}    & Partial & \xmark & \xmark & \xmark & \xmark & Opt. & Tok+\$ \\
\clcode{} & Full (Blake2b boundary) & \cmark & \xmark & \xmark & \xmark & \cmark & Tok+\$ \\
\cdx{}    & Partial (\texttt{thread\_id} key) & \xmark & \cmark (to max/ultra) & \xmark & \cmark & \cmark & Per-turn \\
\gemini{} & \xmark (server-side) & \cmark & \xmark & \cmark (router) & \xmark & \cmark & OTel/Tok+\$ \\
\vibe{}   & Partial & \cmark & \cmark (5 levels) & \xmark & \xmark & \cmark & OTel/Per-sess. \\
\mswe{}   & Partial & \xmark & \xmark & \xmark & \xmark & Opt. & litellm \\
\hrm{}    & Full (provider-agnostic markers + prefix normalization) & \cmark & \cmark (5 levels, per-provider translation) & Failover chains only & \xmark & \cmark & Tok+\$, typed provenance \\
\pib{}    & Full (breakpoints + TTL tiers + waste audit) & \cmark (7-level scale) & \cmark (10 dialects) & \xmark (by design) & \cmark (Codex backend) & \cmark & Tok+\$, tiered pricing \\
\ocd{}    & Full (6-dialect fanout + session key) & \cmark & \cmark (date-gated variants) & Small-model selection only & Exp.\ flag & \cmark & Tok+\$ (Decimal) \\
\ocl{}    & Partial & \cmark & \xmark & \xmark & \cmark & \cmark & Per-prov. \\
\bottomrule
\end{tabularx}
\end{table}

\begin{table}[H]
\centering
\caption{Head-to-head pipeline comparison: \clcode{} vs.\ \cdx{}.}
\label{tab:pipeline_detail}
\small
\renewcommand{\arraystretch}{1.2}
\begin{tabularx}{\textwidth}{@{}L{2.5cm} L{5.5cm} L{5.5cm}@{}}
\toprule
\textbf{Stage} & \textbf{\clcode{}} & \textbf{\cdx{}} \\
\midrule
\textbf{Prompt assembly} & 12--15 named sections. Static prefix (identity, rules, tools) cached via Blake2b hash; dynamic suffix (env, memory, MCP) recomputed per-turn. \texttt{SYSTEM\_PROMPT\_DYNAMIC\_BOUNDARY} separates the two. & Per-model prompts as server-delivered catalog data (\texttt{models.json} \texttt{base\_instructions}, remote-refreshed with bundled fallback); personality-templated. AGENTS.md injected via root-to-cwd concatenation (nested content appended last). \\
\addlinespace
\textbf{Tool exposure} & 43 tools, but only a subset loaded initially. Deferred tools discovered on-demand via \texttt{ToolSearchTool} (keyword search + \texttt{select:} syntax). Reduces prompt tokens by $\sim$40\%. & 25--30 built-ins; schemas converted to OpenAI \texttt{Function} format via \texttt{create\_tools\_json\_\allowbreak for\_responses\_api()}. Deferred loading now mirrored: \texttt{defer\_loading} flags (MCP tools by default) + BM25 \texttt{tool\_search}. \\
\addlinespace
\textbf{API protocol} & Anthropic Messages API. SSE streaming with \texttt{content\_block\_\{start,delta,stop\}} events. Extended thinking via \texttt{budget\_tokens}. & OpenAI Responses API. WebSocket primary, SSE fallback. Connection prewarm (\texttt{generate=false}). Turn state via \texttt{x-codex-turn-state} header. Reasoning effort (minimal--xhigh, plus max/ultra; \texttt{ultra} adds automatic delegation). \\
\addlinespace
\textbf{Tool dispatch} & \texttt{partitionToolCalls()}: single-pass reduce partitions by \texttt{isConcurrencySafe()}. Consecutive safe tools batched; unsafe tools solo. Max 10 concurrent via semaphore. & \texttt{FuturesOrdered} through a dedicated \texttt{ToolCallRuntime}; per-tool \texttt{supports\_parallel} flag (default false) gates an execution lock---concurrency-unsafe tools run exclusively. \\
\addlinespace
\textbf{Permission} & 3 layers: (1)~\texttt{PreToolUse} hooks (static pattern match), (2)~LLM risk classifier, (3)~interactive dialog. Sub-agents skip layer~3. & 4 layers: (1)~ExecPolicy Starlark rules, (2)~lifecycle hooks (\clcode{}-compatible vocabulary), (3)~Guardian LLM approval reviewer (fail-closed), (4)~OS-level sandbox enforcement. \\
\addlinespace
\textbf{Execution} & Host execution with an opt-in OS sandbox (Anthropic \texttt{sandbox-runtime}: Bubblewrap/Seatbelt, network restrictions, \texttt{dangerouslyDisableSandbox} escape hatch); git worktrees for branch isolation. & Sandboxed execution. Vendored Bubblewrap (Linux; Landlock legacy fallback), Seatbelt (macOS), restricted tokens (Windows). \\
\addlinespace
\textbf{Compaction} & Triggered when remaining context drops below \texttt{AUTOCOMPACT\_BUFFER\_TOKENS} (13\,K), with an env-overridable percent threshold. Image stripping, message grouping by API round, LLM summarization, \texttt{SystemCompactBoundaryMessage}. Post-cleanup: restore 5 files (50\,K token budget). & \texttt{model\_auto\_compact\_token\_limit} configurable. Local or cloud compaction via \texttt{/responses/compact}; rollout token budgets with turn aborts. \\
\bottomrule
\end{tabularx}
\end{table}

\bibliographystyle{plainnat}

\end{document}